\def\Zqsn{{\cal Z} (q,\sigma,n)}
\def\cZ{{\cal Z}}

\def\cP{{\cal P}}

\def\l{\ell}
\def\esl{s \ell}
\def\fa{f(\alpha)}
\def\tqa{\tau_{{\cal A}}(q)}

\def\sq{\sigma(q)}
\def\sqa{\sigma_{{\cal A}}(q)}
\def\sqq{\sigma_{{\cal Q}}(q)}
\def\sqap{\sigma_{{\cal P}}(q)}
\def\lxc{\langle\negthinspace\langle}
\def\rxc{\rangle\negthinspace\rangle}
\def\lx{\langle}
\def\rx{\rangle}
\def\fm{f_{-}}
\def\fp{f_{+}}
\def\gp{g_{+}}
\def\prop1{\lambda_{1,1} \log n + \lambda_{1,0}}
\def\prop2{\lambda_{2,1} \log^2 n + \lambda_{2,1} \log n + \lambda_{2,0}}
\def\prop3{\lambda_{3,3} \log^3 n + \lambda_{3,2} \log^2 n + \lambda_{3,1} \log n + \lambda_{3,0}}
\def\eq{$$}
\def\en{$$\medskip\noindent}
\def\cline{\centerline}
\def\sect{\bigskip\noindent}
\def\date{Submitted to Journal of Statistical Physics: 12/28/95}
\magnification=\magstep1
\hsize=6.2 true in
\hoffset=.1 true in
\vsize=8.5 true in
\voffset=.2 true in
\baselineskip=10 true pt
\nopagenumbers
\cline{\date}
\bigskip\bigskip\bigskip
\input epsf

\cline{\bf MULTIFRACTAL DIMENSIONS FOR BRANCHED GROWTH} 
\bigskip\bigskip\medskip
\cline{\sl Thomas C. Halsey\footnote{$^1$}{Exxon Research and Engineering, Route 22 East,
Annandale, N.J. 08801 U.S.A.}, Katsuya Honda\footnote{$^2$}{Department of Mathematics, Shinshu University,
Matsumoto 390, Nagano Pref., Japan}, and Bertrand Duplantier\footnote{$^3$}{Service de
Physique Th\' eorique, C.E. Saclay, 91191 Gif-sur-Yvette, France}} \bigskip\bigskip\medskip
\noindent{\bf Abstract}  \baselineskip=20 true pt
\medskip 

A recently proposed theory for diffusion-limited aggregation (DLA), which models this
system as a random branched growth process, is reviewed. Like DLA, this process is
stochastic, and ensemble averaging is needed in order to define multifractal
dimensions. In an earlier work [T.C. Halsey and M. Leibig, Phys. Rev. A {\bf 46}, 7793
(1992)], annealed average dimensions were computed for this model. In this paper, we
compute the quenched average dimensions, which are expected to apply to typical
members of the ensemble. We develop a
perturbative expansion for the average of the logarithm of the multifractal partition
function; the leading and sub-leading divergent terms in this expansion are then
resummed to all orders.  The result is that in the limit where the number of particles $n
\to
\infty$, the quenched and annealed dimensions are {\it identical}; however, the attainment
of this limit requires enormous values of $n$. At smaller, more realistic values of $n$, the
apparent quenched dimensions differ from the annealed dimensions. We interpret these
results to mean that while multifractality as an ensemble property of random branched growth
(and hence of DLA) is quite robust, it subtly fails for typical members of the
ensemble. \vfill 
\line{Keywords: Diffusion-limited aggregation, pattern formation, branched growth
\hfill}

\eject \sect{\bf I. Introduction} \medskip \headline{\hfill \folio \hfill}

Many natural growth processes generate branched structures. Probably the
most celebrated such process is diffusion-limited aggregation (DLA).
A simple algorithm for this type of growth process, introduced by Witten and
Sander, has allowed over ten years of numerical exploration of its
mysterious properties.$^{(1-4)}$ Theoretically, there
has been less progress. A number of studies have, with greater or
lesser success, used real-space renormalization or variants
thereof to study the self-similarity of DLA clusters.$^{(5-7)}$ An alternative
approach has been to try to deepen our understanding of the scaling
structure of DLA clusters by emphasizing the ``multifractal" nature
of the growth of such clusters.$^{(8-10)}$

Recently, a quite broad framework has been proposed to study branched
growth processes.$^{(11)}$ This framework relies upon the dynamics of
competition between neighboring branches in these structures. For
DLA, one of the authors of this work has proposed a method of
determining the universal dynamics underlying this competition, a
method which allows {\it a priori} computation of, e.g., the overall fractal
dimension of DLA clusters.$^{(12)}$ Results of this approach are in good
agreement with numerical results for DLA.

In this work, we turn to the implications of this approach for the
multifractal properties of DLA clusters. Our principal result is that
fluctuations in the ensemble of possible DLA clusters seem to play a
more important role than previously expected in determining these
multifractal properties. In fact, while multifractality is well
defined as an {\it ensemble} property of DLA, there appears to be a
quite precise sense in which {\it typical} DLA clusters are not
multifractals at all!

This result should be placed in the context of a number of studies
claiming deviations from perfect fractal or multifractal behavior for DLA. A number
of groups have reported anomalies in the
scaling behavior of the very weak growth sites in DLA.$^{(13)}$  More recently,
Mandelbrot and collaborators have proposed that the radius of gyration scaling
of large DLA clusters may be more complicated than previously believed.$^{(14)}$
However, in our case, the scaling anomalies correspond to the most active,
strongly growing regions of the cluster, and the form of the anomalies can be
computed. \medskip \noindent{\bf Diffusion-limited aggregation} \smallskip

The Witten-Sander algorithm is based on modelling growth as an
aggregation of random walkers. Consider a cluster composed of $n$
particles. To generate the $n+1$ particle cluster, introduce a random
walker at a great distance from the cluster. This random walker will
either escape to infinity without encountering the cluster, or else
will eventually encounter the cluster. In the
Witten-Sander algorithm, the particle sticks at the point of its
first contact with the cluster, thereby forming the $n+1$ particle
cluster.  The procedure is then iterated to create arbitrarily large
clusters (the current record in two dimensions, with off-lattice
random walks, is in the neighborhood of $10^8$ particles!)$^{(15)}$

Structures generated by this algorithm are highly branched and
ramified (Figure 1). Fractal dimensions, defined by the scaling of the
average cluster radius of gyration $r_g$ with particle number $n$, $n
\propto r_g^{D_f}$, are $D_f \approx 1.71$ in spatial dimensionality
$d=2$ and $D_f \approx 2.49$ in spatial dimensionality $d=3$. In higher
dimensionalities, fractal dimensions are given approximately by the
phenomenological formula $D_f \approx (d^2 + 1)/(d+1)$.$^{(16)}$

DLA is observed to be a good model
for a variety of natural growth processes, including
electrodeposition, viscous fingering, and solidification.$^{(17)}$ Variants of
the model have been proposed to account for colloidal aggregation,
dielectric breakdown, and the fracture of brittle media.$^{(18)}$

Note that DLA is a stochastic process, due to the
underlying stochastic nature of the random walks. Properly speaking,
for fixed particle number $n$ there exists a large ensemble of
clusters, each with a certain probability of appearing. If we (naively) suppose
that each arriving particle can attach to any of the pre-existing
particles, then there are $(n-1)!$ different ``genealogically distinct"
(i.e., with respect to particle attachments) members of this ensemble.

This implies that quantities such as $D_f$ must be defined as ensemble
averages, since not all possible members of the ensemble will have
the same $D_f$.  For instance, a possible, though unlikely, ensemble
member consists of a long straight chain of particles, which has $D_f
=1$. Unfortunately, researchers have rarely paid much attention to
this aspect of the problem, and few numerical studies even carefully report the
method of ensemble averaging used.
\medskip
\noindent{\bf Multifractality}
\smallskip

In a particular cluster of $n$ particles, there is a well-defined
probability that the $n+1$'st particle will stick to any particular
pre-existing particle. Indexing the particles by $i$, we write
the probabilities of attachment $\{ p_i \}$; $p_i$ will equal zero for any particle
inacessible to an approaching random walker. These quantities
are very broadly distributed, being relatively large at the tips of clusters, where
arriving particles are quite likely to attach, and extremely small deep inside
the ``fjords" of a cluster, where random walkers will almost never
successfully arrive without having previously contacted the cluster.
If the distribution of $\{ p_i \}$ is multifractal, then we expect
that

\eq
\sum_{i=1}^n p_i^q = n^{-\sigma(q)} ,
\eqno(1.1)
\en
where the exponent function $\sigma(q)$, which is defined by this
relation, should be asymptotically independent of particle number
$n$.$^{(19),(20)}$ (Note that usually multifractality is defined in terms of
the scaling of moments of $\{ p_i \}$ with respect to the length
scale $r_g$, and not particle number, defining an exponent function
$\tau(q)$. We use $\sigma(q)$ because for our purposes particle
number is a more convenient basis for defining the exponent
function. We expect $\tau(q) = D_f \sigma(q)$.)

These exponent functions have well-known interpretations in terms
of the distribution of $\{ p_i \}$. We will temporarily follow the literature and
use $\tau(q)$ rather than $\sigma(q)$. Suppose that for any particular
$p_i$ a corresponding $\alpha_i$ is defined by $p_i =
(a/r_g)^{\alpha_i}$, where $a$ is the particle diameter, and $r_g$ is the cluster radius of
gyration. Then the total number of particles with $p_i$ in a range $d \alpha$, $N(\alpha
) d \alpha$, defines a quantity $\fa$ via $N(\alpha) =
(a/r_g)^{-\fa}$; $\fa$ is independent of $r_g$ for multifractals.$^{(20)}$
The quantity $\fa$ can be obtained by Legendre transformation
of $\tau (q)$,

\eq\eqalign{
\alpha(q) = &{d \tau (q) \over dq} ;\cr
f(\alpha (q) ) = &q \alpha(q) -\tau(q) .\cr} 
\eqno(1.2)
\en
The values of $f$ and $\alpha$ at a particular value of $q$ satisfy
the tangent condition

\eq
{d \fa \over d \alpha} = q .
\eqno(1.3)
\en
By definition, $\fa$ is an
intrinsically positive quantity, since $N(\alpha) \ge 1$.

Now we must modify these standard relations to account for the
stochastic nature of DLA. One approach is to define the exponent
function $\sigma (q)$ by a simple average of the moments; by analogy
with statistical mechanics, we term this ``annealed" averaging, and
the resulting dimensions $\sqa$ ``annealed" dimensions,$^{(21)}$

\eq
\lx \sum_{i=1}^n p_i^q \rx = n^{-\sqa} ,
\eqno(1.4)
\en
where the brackets $\lx \cdots \rx$ denote, here and elsewhere, averaging
over the ensemble of DLA clusters. We can similarly define $\tqa$
in terms of the scaling with respect to the
cluster radius\footnote{$^1$}{Unfortunately,
$r_g$ is also a stochastic quantity. The ambiguity this introduces is one argument for
using $\sq$, defined in terms of $n$, for which there is no such ambiguity.} $r_g$ by $\lx \sum_{i=1}^n p_i^q \rx =
(a/r_g)^{-\tqa}$. 

One advantage of this procedure is that the
corresponding function $\fa$ retains a simple interpretation. Suppose
that the expectation value of the number of particles with
probabilites $p_i$ corresponding to the exponent $\alpha$ is $\lx
N(\alpha) \rx$. Direct application of Legendre transformation to
Eq.~(1.4) gives $\lx N(\alpha) \rx \propto (a/r_g)^{-\fa}$, where $\fa$
obeys Eqs.~(1.2-3), and can be thereby generated from $\tqa \equiv
D_f \sqa$. One difference with the non-stochastic case is that, since
the expectation value of a number can take any non-negative value,
$\fa$ can now be negative. Figure 2 shows characteristic $\fa$
functions for toy models, in the non-stochastic case and for
stochastic models with annealed averaging. Typically in toy models,
negative values of $f$ appear for $\alpha$ sufficiently far from the
value of $\alpha$, $\alpha_m$, which corresponds to the maximum value of $f$,
($\alpha_m$ represents the most frequently occuring value of the growth probability.)
In terms of $\tqa$ or $\sqa$, negative values of $f$ thus correspond to large absolute
values of $q$.

The great drawback of annealed averaging arises if the quantity
$\Omega(q) = \sum_i p_i^q$ is subject to large fluctuations between
different ensemble members. Suppose that these fluctuations are
log-normal, so that the probability $\cP (\Omega) d \Omega$ that a
particular ensemble member has $\sum_i p_i^q$ in the range $[\Omega,
\Omega + d \Omega]$ is

\eq
\cP (\Omega) d \Omega = P_0 \exp \left ( {(\log \Omega- \log \bar
\Omega)^2 \over \bar \Sigma} \right ) d  \log \Omega .
\eqno(1.5)
\en
Clearly, the most
likely value of $\log \Omega$, $\log \bar \Omega$, is not equal to $\log \lx \Omega
\rx$.$^{(22)}$ This raises the possibility that observation of the scaling function
$\sigma(q)$ for a typical DLA cluster may lead to a different result
than $\sqa$. However, Eq.~(1.5) shows that if the fluctuations of $\Omega$ are
log-normal, then $\lx \log \Omega \rx = \log \bar \Omega$. This suggests that
dimensions $\sqq$ defined through ``quenched" averaging,$^{(21)}$

\eq
\lx \log \sum_{i=1}^n p_i^q \rx \equiv - \sqq \log n ,
\eqno(1.6)
\en
may be closer to the typical result of an observation on, for
instance, a numerically generated cluster. Of course, it is not
necessarily the case that the fluctuations of $\Omega$ for DLA are
log-normal.  However, whatever the nature of these fluctuations,
quenched averaging reduces the impact of rare ensemble members, and
thus should yield results closer to the behavior of typical ensemble
members.

If $\sqq$ gives results characteristic of a typical member of the
ensemble, then the corresponding $\fa$ cannot include negative values
of $f$, which can only be interpreted as ensemble average quantities.$^{(23)}$
Thus if the fluctuations of the ensemble are strong enough to lead to
negative values of $f$ in $\sqa$, then we also expect that $\sqa \ne
\sqq$.

This does not imply that $f < 0$ are necessarily unobservable. Suppose
at some value of $q$, $f(\alpha (q))$ (annealed) is $< 0$. Then in
order to see this value of $\alpha$, one must average over an
ensemble of at least $N \sim (r_g / a)^{\vert f \vert}$ clusters. This will become
difficult as sizes increase, but for small absolute values of $f$,
negative values of $f$ should be observable.

It is convenient to introduce the ``partition sum" $\Zqsn$ which
is defined for a specific cluster by

\eq
\Zqsn = n^{\sigma} \sum_{i=1}^n p_i^q .
\eqno(1.7)
\en

The definitions above are equivalent to requiring that the annealed
dimensions satisfy

\eq
\lx \cZ (q, \sqa ; n) \rx = 1 ,
\eqno(1.8)
\en
and that the quenched dimensions satisfy

\eq
\lx \log \cZ (q, \sqq; n) \rx = 0 ,
\eqno(1.9)
\en
in the limit of large $n$. By definition, $\lx \cZ (q, \sigma; n) \rx = n^{\sigma}
Z(q,n)$, so that

\eq
\lx \log \cZ (q,\sigma; n) \rx = \sigma \log n + F (q,n) ,
\eqno(1.10)
\en
where $F$ is independent of $\sigma$.

\medskip
\noindent{\bf Branched growth model}
\smallskip

The branched growth model introduced in Ref.~11 is based upon an
analysis of the dynamics of competition between neighboring branches.
In standard off-lattice DLA algorithms, any particle that aggregates to a
cluster has a unique ``parent" particle, to which it attaches. Since
no particle has two parents, no loops can develop in the structure.
The position of any particular particle can be uniquely identified by
specifying which branch it lies upon, on a succession of increasing
length scales. 

Consider a branch point, which is any particle that is parent to
more than one succeeding particle. At a particular point in the
development of the cluster, the two (higher numbers are
possible, but are both unlikely and irrelevant) sub-branches coming off
of this particle have respectively total masses $n_1$, $n_2$ and total
growth probabilities $p_1$, $p_2$. These growth probabilities represent the total
probability that the next particle will stick to any constituent of that particular
sub-branch.  We define relative masses and relative growth probabilities by

\eq\eqalign{
x=&{p_1 \over p_1 + p_2}\equiv {p_1 \over p_b}\cr
y=&{n_1 \over n_1 + n_2}\equiv {n_1 \over n_b}\cr}
\eqno(1.11)
\en
where we have also defined total branch mass and growth probability
$n_b$, $p_b$, each made up of a contribution from the two
sub-branches. 

Simple kinematics shows that, neglecting fluctuations in the
numbers of particles arriving at this branch, we have$^{(11)}$

\eq
{d y \over d \log n_b} = x - y .
\eqno(1.12)
\en
If $dx / d \log n_b$ can be expressed as a function of $x$ and $y$
alone, then there is a closed dynamics for branch competition as a
function of these two variables. In Ref.~12, it was shown that by
averaging over the intrinsic stochasticity of the DLA problem, a
function $g(x,y)$ can be determined (at least in $d=2$) such that

\eq
{d x \over d  \log n_b} = g(x,y) ,
\eqno(1.13)
\en
with $g(x,y)$ a calculable function of $x$ and $y$ alone. By
symmetry, $g(x,y) = -g(1-x,1-y)$, so that there is a fixed point of
the dynamics at $(x,y)=(1/2,1/2)$. For the computed function
$g(x,y)$, this is a hyperbolically unstable fixed point, with one
stable and one unstable direction. (Actually, this qualitative
feature can also be predicted on general grounds, without
computation). The unstable manifold of the fixed point leads to two
other (stable) fixed points, at $(x,y)=(0,0)$ and $(x,y)=(1,1)$.
Thus, sub-branch pairs, which are born in a tip-splitting process,
compete in an unstable fashion, so that asymptotically one of
the two sub-branches possesses virtually all of the mass and all of the
growth probability of the pair.

Although the dynamics of a branch pair in the $x-y$ plane is thus
deterministic, the stochasticity of the DLA problem is retained since
the original position of a branch pair in this plane, i.e., when $n_b
\sim 1$, and a tip-splitting event creates the pair, is a random
function. In fact, sub-branch pairs born in most sections of the $x-y$
plane will be quickly driven to one or the other of the stable fixed
points, at which one of the two sub-branches completely
dominates the other.  In Eqs.~(1.12-13), we see that for general values of $x$, $y$, the
fundamental scale over which
$x$ and $y$ will change their values is $\log n_b \sim 1$, i.e., while the
branch as a whole is still of microscopic size. 

However, if the sub-branch pair is originally created quite
close to the central, unstable fixed point, then it will remain
in the vicinity of that fixed point up to larger values of $\log
n_b$. Such pairs correspond to the large, relatively equal branch pairs seen in a
DLA cluster. Thus, the large scale cluster structure is
sensitive only to the distribution of birth probability in the
immediate vicinity of the unstable fixed point. This probability
distribution will be determined by the microscopic dynamics of
tip-splitting, which do not recognize any special role of the
unstable fixed point. Thus we expect this probability
distribution to be only slowly varying near that point; it can
be approximated for our purposes by a constant.

This criterion, and the specific form of $g(x,y)$, defines the random
branched growth model. Computations with this model are much
simplified by the fact that the branch competition trajectories born quite close to the
stable manifold, which correspond to the large-scale structure, will
be quickly drawn onto the unstable manifold. This manifold can thus
be parameterized by $x$ and $y$ as functions of $\log n_b$, or of
$n_b$. Of course, the manifold is symmetric about the unstable fixed
point. It is convenient to introduce a parameter $\epsilon$, which
parameterizes the original distance (when $n_b \sim 1$) of the branch pair
from the stable manifold in the $x - y$ plane. If this distance is $\epsilon^{\nu}$, where
$\nu$ is the eigenvalue along the unstable manifold at
the unstable fixed point of the dynamics defined by Eqs.~(1.12-13), it can be shown that
$x$ and $y$ are functions of the combination $\epsilon n_b$, and that $\epsilon$ plays no
further role in the dynamics.$^{(11)}$

For practical purposes, we must thus specify $x(\epsilon n_b)$,
$y(\epsilon n_b)$ in order to compute with our model. A useful toy
model, ``Model Z" is displayed in Figure 3. In this model, the
unstable manifold follows a straight line away from the fixed
point until it strikes the line $x=0$ (or $x=1$). It then moves
vertically into the stable fixed points. The angle of the unstable
manifold at the fixed point is the one adjustable parameter for this
type of trajectory. 

For more realistic purposes, we can use the unstable manifold
computed for DLA in two dimensions in Ref.~12.$^{(24)}$  This manifold is
displayed in Figure 4.

\medskip
\noindent{\bf Summary of results}
\smallskip

In Ref.~11, it was shown that the annealed dimensions $\sqa$ for a
branched growth model are determined by the simple implicit formula

\eq
\int_0^{\infty} d \eta \; \eta^{\nu-1} \left \{ {x^q (\eta) \over y^{\sigma}
(\eta)} + {[1 - x(\eta) ]^q \over [1-y(\eta)]^{\sigma}} - 1 \right \}
\equiv \int_0^{\infty} d \eta \; \eta^{\nu-1} \psi(\eta; q,\sigma)  = 0 ,
\eqno(1.14)
\en
where $\eta \equiv \epsilon n_b$ parameterizes position on the
unstable manifold described in the above discussion. We have also
defined the quantity $\psi(\eta; q,\sigma)$ by

\eq
\psi(\eta; q, \sigma) \equiv {x^q (\eta) \over y^{\sigma}
(\eta)} + {[1 - x(\eta) ]^q \over [1-y(\eta)]^{\sigma}} - 1 .
\eqno(1.15)
\en
This quantity will appear frequently
in our results. For model Z or for the computed DLA trajectories,
Eq.~(1.14)  leads for $q > 0$ to an annealed multifractal spectrum with
the expected properties, such as negative values of $\fa$ for large
values of $q$. For $q<0$, annealed multifractal dimensions are
typically not well-defined, due to a divergence of the left-hand side
in Eq.~(1.14).

In this work, we study the structure of $\lx \log \cZ \rx$ for
the random branched growth model, in order to determine the quenched
multifractal dimensions for these models.  We develop a diagrammatic
perturbation expansion for this quantity, and we show
how sets of terms in this expansion can be resummed to all orders.
The form of the resulting expression is

\eq\eqalign{
\lx \log \Zqsn \rx = \left ( \int d \eta \eta^{\nu - 1} \psi(\eta; q,
\sigma) \right ) &\Gamma_0 (q, \sigma) \log n \cr  + 
& \Gamma_1 (q, \sigma) (1 -
n^{-\Delta(q)}) +  \cdots ,\cr}
\eqno(1.16)
\en
where the exponent $\Delta (q)$ governing the finite-size corrections is positive and
calculable. Because
$\lx \log \cZ \rx = \sigma \log n + F(q,n)$, the power law terms should be independent of
$\sigma$; we retain an {\it ersatz} dependence on $\sigma$ in Eq.~(1.14) because this
independence is difficult to see directly within our perturbative method.  In the limit $n
\to \infty$, the power-law and constant terms can be disregarded; thus it appears that the
criterion $\int d \eta \; \eta^{\nu -1} \psi(\eta; q,\sigma) = 0$ is sufficient to determine
$\sqq$. Since this is the same criterion as that for the annealed dimensions $\sqa$, this
implies that $\sqq=\sqa$. This result is somewhat surprising, and also undermines the
interpretation of $\sqq$ as a property of a typical cluster, since we expect $\sqa$ to
include regions of negative $\fa$.

This paradox is resolved by a further
feature of this result. It concerns the order of the limits $q \to \infty$, $n \to\infty$. 
Negative values of
$f$ appear for large values of $q$; however, $\lim_{q \to \infty} \Delta (q) = 0$. Thus, for
large $q$, the power-law terms in Eq.~(1.16) do contribute logarithms to
the result, provided that $n$ is not too large. Thus for moderate
values of $n$, a different result is obtained for $\sqq$, which proves not to include
negative values of $f$. At large $n$, one will eventually cross over to the asympotic
regime in which $\sqq = \sqa$, but this value of $n$, which we term $n_c$,
grows very quickly with $q$, as $n_c (q) \sim \exp(ae^{bq})$, with $a$
and $b$ positive constants, as shown below explicitly in section 4 for a specific model,
model Z. Thus for practically accessible values of
$n$, we expect never to see negative values of $f$.

Of course, this resolution also implies that any typical cluster of
finite size $n$ should probably not be viewed as a true multifractal, as
its exponent function $\sqq$ actually represents different scaling
behaviors above and below the value of $q$ for which $n = n_c (q)$.
One conclusion of our study is thus that multifractality should be
properly viewed as an ensemble property of DLA, and not a property of
individual DLA clusters.

This paper is comprised of five sections and five appendices. In
section 2, we develop the perturbative expansion of $\lx \log \cZ\rx$. In section 3, we show how
families of terms within this expansion can be resummed to all orders. In section 4, we discuss the
implications of these results for the multifractality of DLA
clusters, and we compare our analytical results with numerical
results for random branched growth. In section 5, we conclude and
summarize.  In appendix A, we show how to perform
sums that play the role of ``propagators" in our perturbative
expansion. In appendix B, we list perturbation theory terms, with their diagrammatic
expressions, up to fourth order. In appendix C, we list some mathematical
identities that are useful in the resummation of section 3. In appendix D
we show a simpler method of obtaining some of the terms in the final expression for
$\lx \log \cZ \rx$. Finally, in appendix E, we show how the expansion of section 2 can be
recast as a perturbation theory in $q-1$.

\bigskip
\sect{\bf 2. Perturbative expansion for $\lx \log \cZ \rx$}
\medskip

Formally, we can write

\eq
\lx \log \Zqsn \rx = \sum_{N=1}^{\infty} { (-1)^{N-1} \over N}\left
\lx ( \cZ - 1 )^N \right \rx ,
\eqno(2.1)
\en
so that our problem is one of computing expectation values of powers
of $\cZ - 1$. We know that

\eq
\Zqsn = n^{\sigma} \sum_{i=1}^n p_i^q = \sum_{i=1}^n {p_i^q \over
(1/n)^{\sigma}} , \eqno(2.2)
\en
where we recall that the index $i$ labels the particles. 

In our model, the only sites which are
allowed to grow are at the end-points of the branching process. Thus we use Eq.~(2.2), substituting
a sum over these ``elementary" sub-branches for the sum over particles $i$ in (2.2). Consider a
particular elementary sub-branch of the cluster, of total growth probability $p_e$ and total mass
$n_e$ (Figure 5). This sub-branch branches off from its sibling at a
particular node, indexed by J; the total number of particles in the
elementary sub-branch and in its sibling taken together is $n_J$. If
the elementary sub-branch is the weaker (stronger) of the two
sub-branches, then it has a proportion $x(\epsilon_J n_J)$ (or $1-
x(\epsilon_J n_J) $) of the total growth probability of the two
sub-branches taken together, and a proportion $y(\epsilon_J n_J)$ (or
$1-y(\epsilon_J n_J)$) of the total mass of the two sub-branches taken
together. Here $\epsilon_J$ is the random variable specifying the
original state for the branch point indexed by $J$ when the total number of
descendants of that branch point was $n_J = 1$. We will sometimes refer to the path that leads
through the stronger of the two sub-branches as the ``main branch", the weaker sub-branch
corresponds to a ``side-branching."

Since $\cZ$ can be expressed in such a way that $x$ and $y$ always appear together, it is convenient
to define quantities $f_{\pm} (\epsilon n)$ by

\eq\eqalign{
\fm(\eta) \equiv & {x^q(\eta) \over y^{\sigma} (\eta) } ;\cr
\fp(\eta) \equiv & { (1-x(\eta) )^q \over (1-y(\eta))^{\sigma}} .\cr}
\eqno(2.3)
\en

The contribution of our particular sub-branch to $\cZ$ can be
written as

\eq
{p_e^q \over (n_e/n)^{\sigma}} = \prod_{j=1}^J f_{\mu_j} (\epsilon_j n_j) ,
\eqno(2.4)
\en
where $\mu_j = \pm$, depending on whether the stronger or weaker
branch is taken at the $j$'th node. The index $j$ indexes the nodes between the root of
the cluster, $j=1$, and the elementary sub-branch, for which we have taken $j=J$. Note
that $n_1 \equiv n$, the total number of particles in the cluster. Note also that the index
$j$ measures position upon a particular path from the root to an elementary sub-branch, so
its meaning depends upon the exact sequence of $\{
\mu_j \}$ chosen. Comparing with Eq.~(2.2), we see that in Eq.~(2.4) we are introducing a slightly
different cut-off procedure than that used in Eq.~(2.2), where $n_e = 1$. 

Averaging the quantity appearing on the right-hand
side in Eq.~(2.4) is difficult because $n_j$ is actually a function
of all $\epsilon_k$ with $k<j$, since either $n_j =
y(\epsilon_{j-1} n_{j-1} ) n_{j-1}$, if at the $j$'th node we take
the weaker of the two sub-branches originating at that node, or else $n_j = (1 -
y(\epsilon_{j-1} n_{j-1} ) ) n_{j-1}$, if we take the stronger of the
two sub-branches. Thus the individual terms in the product appearing
in Eq.~(2.4) cannot be averaged independently of one another.

Nevertheless, it is instructive to investigate how a single one of
these terms averages. Consider $\fm(\epsilon_j n_j)$. The average of
this over the random variable $\epsilon_j$ is given by

\eq
\lx \fm (\epsilon_j n_j ) \rx_{\epsilon_j} = \int_0^{\infty} d \epsilon_j
\rho(\epsilon_j) \fm (\epsilon_j n_j ) ,
\eqno(2.5)
\en
where $\lx \rx_{\epsilon_j}$ denotes averaging over $\epsilon_j$, and
$\rho (\epsilon_j )$ is the probability distribution for
$\epsilon_j$. In the introduction, we stated that the
initial distance in the $x-y$ plane from the unstable manifold is
proportional to $\epsilon^{\nu}$. In addition, we stated that the probability
distribution with respect to the distance measure $d \epsilon^{\nu}$ should be constant
close to the unstable manifold. Since $d \epsilon^{\nu} \propto \epsilon^{\nu-1} d
\epsilon$,  we see that 

\eq
\lim_{\epsilon \to 0} \rho(\epsilon ) = \rho_0 \epsilon^{\nu - 1} ,
\eqno(2.6)
\en
where $\rho_0$ is a constant. (The reader may consult Ref.~11 for a more detailed
discussion of this point.) Now suppose that $n_j \gg 1$. The quantities $x(\eta)$,
$y(\eta) \to 0$ for $\eta \gg 1$, (see Figures 3 and 4), and for physical values of
$\sigma$, we thus expect $\fm(\eta ) \to 0$ for $\eta \gg 1$. Thus we have

\eq
\int_0^{\infty} d \epsilon_j \rho(\epsilon_j) \fm(\epsilon_j n_j) \approx {\rho_0
\over n_j^{\nu} } \int_0^{\infty} d \eta \; \eta^{\nu - 1} \fm( \eta ) \equiv {\rho_0
\over n_j^{\nu} } \lxc \fm \rxc ,
\eqno(2.7)
\en
where we have defined a normalized
expectation value $\lxc \cdots \rxc$ by

\eq
\lxc f(\eta) \rxc \equiv \int_0^{\infty} d \eta \; \eta^{\nu-1} f(\eta).
\eqno(2.8)
\en
This expectation value does not depend on
$n_j$, nor on anything else except the form of the unstable manifold. The
corrections to Eq.~(2.7) are of higher order in $n_j^{-1}$. Ignoring these
corrections, we write

\eq
\lx \fm (\epsilon_j n_j ) \rx =  \lx {\rho_0 \over n_j^{\nu} }\rx
\lxc \fm \rxc
\eqno(2.9)
\en
where the remaining true expectation value $\lx n_j^{-\nu} \rx$ depends
only upon random variables $\epsilon_k$ with $k < j$.

The averaging of $\fp$ is performed rather differently, because $\lim_{\eta \to \infty}
\fp(\eta) = 1$. Thus

\eq\eqalign{
\int_0^{\infty} d \epsilon_j \rho(\epsilon_j) \fp (\epsilon_j n_j ) = &
\int_0^{\infty} d \epsilon_j \rho(\epsilon_j) \left [ 1 + (\fp (\epsilon_j n_j ) -1 )
\right ] \cr = &1 + \int_0^{\infty} d \epsilon_j \rho(\epsilon_j)  (\fp
(\epsilon_j n_j ) -1 ) \cr \approx & 1 + {\rho_0 \over n_j^{\nu} } \lxc
(\fp - 1 ) \rxc \cr \equiv & 1 + {\rho_0 \over n_j^{\nu} } \lxc
\gp \rxc} ,
\eqno(2.10)
\en
where we have used $\int_0^{\infty} d \epsilon \rho(\epsilon) = 1$, and have introduced

\eq
\gp(\eta) \equiv \fp(\eta) - 1 .
\eqno(2.11)
\en 

It is now natural to expand $\lx \log \cZ \rx$ perturbatively in
$\fm$ and $\gp$. Formally, we can associate a dummy parameter
$\delta$ to each factor of $\fm$ or $\gp$, expand order by order in
$\delta$, and set $\delta =1$ at the end of the computation.

\medskip
\noindent{\bf Zeroth order in $\delta$}
\smallskip

At zeroth order in $\delta$, only one elementary sub-branch in the
cluster contributes, that in the stronger sub-branch at each node,
starting at the root (Figure 6). The contribution
$p_e^q/(n_e/n)^{\sigma}$ from this sub-branch is given by

\eq
{p_e^q \over (n_e/n)^{\sigma}} = \prod_{j=1}^J \fp(\epsilon_j n_j) =
\prod_{j=1}^J \left (1 + \gp(\epsilon_j n_j) \right ) 
\eqno(2.12)
\en
where the index $j$ now denotes nodes along this main branch of the
cluster. To zeroth order in $\gp$, this is simply
$p_e^q/(n_e/n)^{\sigma}=1$, so that $\cZ - 1 = O(\delta)$, and
to zeroth order in $\delta$, $\lx \log \cZ \rx = 0$.

\medskip
\noindent{\bf First order in $\delta$}
\smallskip

To first order in $\delta$, we can approximate

\eq
\lx \log {\cal Z} \rx \approx \lx \cZ - 1 \rx ,
\eqno(2.13)
\en
because the other terms in the expansion of the logarithm are at
least of $O(\delta^2)$. The terms with only one factor of $\fm$
represent elementary branches removed at some point from the main
branch by the choice of the weaker branch at only one node. If there
are $J$ nodes in the main branch, there are $J$ such ``first-order"
sidebranches. Since we are interested in terms of first order in
$\delta$ only, for these terms the factors of $\fp$ appearing in the partition
function can be replaced by $1$, because $\fp = 1 + \gp =1 +
O(\delta)$. Thus the term of $O(\delta )$ that is proportional to
$\fm $ is

\eq
\lx \log Z \rx_{1, \propto  \fm } =  \sum_{j\ge 1} \left \lx {\rho_0 \over
n_j^{\nu}} \right \rx \lxc \fm \rxc ,
\eqno(2.14)
\en
where the subscript on the left-hand side represents terms of first
order in $\delta$ that are also proportional to $\lxc \fm \rxc$. The
index $j$ represents nodes on the main branch of the cluster, starting at the seed (or root) $j=1$. 

There are
also terms of first order in $\delta$ that are proportional to $\lxc \gp \rxc$. These terms include
no factors of $\fm$, and refer to the elementary sub-branch at the end of the main branch.
The actual partition sum contribution from this sub-branch is represented as a product of
factors of $\fp$, one at each node. To first order in
$\delta$, one chooses one of these nodes to contribute a factor of
$\gp$, and all others contribute a factor of $1$. Thus

\eq
\lx \log Z \rx_{1, \propto  \gp} =  \sum_{j \ge 1} \left \lx {\rho_0 \over
n_j^{\nu}} \right \rx \lxc \gp \rxc ,
\eqno(2.15)
\en
where, again, the index $j$ refers to nodes on the main branch of the cluster.

Although the term $\propto  \gp $ is thus identical in form to that
$\propto \fm$, Eq.~(2.14), the origin of the two terms is quite
different. Equation (2.14) gives the $O ( \delta )$ partition sum
contribution of all of the elementary sub-branches along the ``main
line" of the sidebranches of the main branch, while Eq.~(2.15) gives
the $O(\delta )$ contribution from the elementary sub-branch at
the end of the main branch itself (Figure 7).

We can also represent these terms graphically. We represent a $\gp$
``vertex" by an open circle, and an $\fm$ vertex by a solid circle.
Their sum 

\eq
\psi \equiv \fm + \gp
\eqno(2.16)
\en
(from Eq. (1.15)) is indicated by two circles connected by a short vertical
line. A set of nodes on a branch over which $n_j^{-\nu}$ (and later
more complicated functions of $n_j$) is summed is indicated by a
solid horizontal line. The left side of a diagram indicates the root,
and the right hand side represents structure successively further
down the branching tree. Thus the $\lx \log {\cal Z} \rx_1$ terms
that we have written in Eqs.~(2.14-15) above are indicated by the diagrams
in Figure 8. 

The perturbation expansion we are developing is
analogous to field theoretic perturbation series, with the factors
of $ \fm $ and $ \gp $ playing the role of vertices,
and the sums of $n^{-\nu}$ playing the role of propagators. We shall
see below that at higher order, both of these objects become more
complicated. One difference with field theory is that topologically,
the diagrams always have the shape of branched trees, with no loops
in the structure.

To evaluate these terms, we must evaluate $\lx \sum_j n_j^{- \nu
} \rx$, where the sum is along the nodes of the main branch. We shall evaluate this sum by a
somewhat roundabout path. Consider the quantity $\log (n/n_e)$, where $n_e$ is the number of particles
in the elementary branch at the end of the main branch. By the definition of the $y$ parameters
(Eq.~(1.10)), we can write the identity

\eq
\log (n/n_e) = - \log \prod_{j=1}^J \left (1-y(\epsilon_j n_j) \right) ,
\eqno(2.17)
\en
where the index $j$ ranges over all of the nodes on the main branch, of which $J$ is the
last. Taking the expectation value of the right-hand side, and using the methods we have
developed above for computing these expectation values, we see (referring to Eqs.~(2.7-8))
that

\eq
\log n - \log n_e = -\sum_{j=1}^J  \left \langle{\rho_0 \over n_j^{\nu}} \right \rangle
\lxc
\log (1 - y)
\rxc ,
\eqno(2.18) 
\en
so that

\eq
\sum_{j=1}^J  \left \langle {\rho_0 \over n_j^{\nu}} \right \rangle = \lambda \log n + a_0 ,
\eqno(2.19)
\en
with $\lambda = -\lxc \log (1-y) \rxc^{-1}$ and $a_0 =  \log n_e / \lxc \log (1-y) \rxc$.  
The parameter $\lambda$ is thus a function only of the unstable manifold in the
$x-y$ plane. This is not the case with $a_0$, which is cutoff dependent, and thus
sensitive to small-scale details of the theory. We shall see
below that such non-universal constants do not affect results for
multifractal dimensions.

Thus our final result is that

\eq
\lx \log {\cal Z} \rx_1 = \lxc \fm + \gp \rxc (\lambda \log n + a_0) .
\eqno(2.20)
\en
We should remark that the non-universal constant $\rho_0$, and the exponent
$\nu$, no longer appear explicitly in our formulae, although they do appear in
intermediate steps in this computational method, and $\nu$ appears in the definition
of the average $\lxc \cdots \rxc$. These quantities do not appear explicitly in final results
at any order in $\delta$. 

In Eqs. (1.15) and (2.16), we defined the quantity $\fm + \fp - 1 = \fm + \gp
\equiv \psi$, and stated that the criterion $\lxc \psi \rxc = 0$
determined the annealed multifractal dimensions $\sqa$. (Recall that
$\psi$, like $\fm$ and $\gp$, is a function of $q$ and $\sigma$.)
Thus, in the limit $n \to \infty$, our $O(\delta)$ result for the
quenched multifractal dimensions, determined by the criterion $\lx
\log {\cal Z} \rx_1 = 0$, is identical to our result for the annealed
multifractal dimensions, so that $\sqq = \sqa + O (\delta^2 )$.

\medskip
\noindent{\bf Second order in $\delta$}
\smallskip

Although similar in spirit to the computation of $\lx \log {\cal Z}
\rx_1$, the computation of $\lx \log {\cal Z} \rx_2$ introduces some
new complications. Let us start by considering the term arising from
$\lx {\cal Z} - 1 \rx_{2,\propto   \gp^2 }$. This term arises from the
elementary sub-branch at the end of the main branch; in this case we
are computing the $O(\delta^2)$ term in its contribution to the
partition sum.

Specifically, we must compute

\eq
\lx {\cal Z} - 1 \rx_{2,\propto   \gp^2 }= \lx \sum_{k \ge 1} \sum_{j>k}
\gp(\epsilon_k n_k) \gp(\epsilon_j n_j) \rx .
\eqno(2.21)
\en
Graphically, this term is represented by a horizontal bar interrupted
by two open circles, representing the two factors of $\gp$, one
downstream from the other on the main branch. Note the presence of
two horizontal segments, corresponding to the two sums in Eq.~(2.21) (Figure 9).

In Eq.~(2.21), we start by averaging over $\epsilon_j$, thereby obtaining

\eq
\lx \sum_{k \ge 1} \sum_{j>k}
\gp(\epsilon_k n_k) \gp(\epsilon_j n_j) \rx = \lx \sum_{k \ge 1}
\sum_{j>k} \gp(\epsilon_k n_k) {\rho_0 \over n_j^{\nu}} \rx \lxc \gp
\rxc . \eqno(2.22)
\en

Using the result from Eq. (2.19) for $\sum_{j \ge 1} n_j^{-\nu}$, we see that

\eq
\lx \sum_{j>k} {\rho_0 \over n_j^{\nu}} \rx =  \lx \lambda
\log n_{k+1} + a_0 \rx =  \lx \lambda [ \log n_k + \log (
1 - y(\epsilon_k n_k ) ) ] + a_0  \rx .
\eqno(2.23)
\en
where $n_{k+1}$ refers to the number of particles in the stronger branch at the $k'th$
node on the main branch. Then, by summing over $j$ and averaging over $\epsilon_k$, we
obtain

\eq
\lx {\cal Z} - 1 \rx_{2,\propto  \gp^2 } = \Bigg \lx \sum_{k \ge 1}{\rho_0 \over
n_k^{\nu}} \Big \{ \lambda \big [ \lxc \gp \rxc \log n_k + \lxc \gp \log (1
- y) \rxc \big ] + a_0 \lxc \gp \rxc \Big \} \Bigg \rx \lxc \gp \rxc .
\eqno(2.24)
\en

In order to perform the final sum, we must compute $\lx \sum_{k}\log
n_k/n_k^{\nu} \rx$. This can be done by an iterative procedure based upon our result
above for $\sum_{j} n_j^{-\nu}$, Eq.~(2.19)
(details are given in appendix A.) The result is

\eq
\lx  \sum_{j  \ge 1} { \rho_0 \log n_j \over n_j^{\nu} } \rx =  \left (
\lambda_{2,2} \log^2 n + \lambda_{2,1} \log n + \lambda_{2,0} \right ) ,
\eqno(2.25)
\en
where $n \equiv n_1$ is again the total number of particles in the
entire cluster, and $\lambda_{2,0}$ is a non-universal constant. By contrast, $\lambda_{2,2}$ and
$\lambda_{2,1}$ are functions only of the form of the unstable manifold. The generalization of
Eq.~(2.25) is straightforward

\eq
\lx \sum_{j \ge 1}  {\rho_0 \log^{N-1} n_j \over n_j^{\nu} } \rx =
\sum_{M=0}^{N} \lambda_{N,M} \log^M n .
\eqno(2.26)
\en

In this notation, all of the $\{ \lambda_{N,0} \}$ are non-universal; the first of
these is $\lambda_{1,0} \equiv a_0$. Furthermore,  $\lambda \equiv \lambda_{1,1}$.
In appendix A, an iterative method which can be used to compute arbitrary numbers of
these coefficients (except, of course, for the non-universal ones) is
demonstrated. For the purposes of this work, we need only $\lambda = \lxc \log(1-y
\rxc^{-1}$, $\lambda_{2,2} = \lambda/2$, and $\lambda_{2,1}$, which is given by

\eq
\lambda_{2,1} =  { \lxc \log^2 (1-y ) \rxc \over 2 \lxc
\log(1-y) \rxc^2 } \equiv \lambda^{\prime}
\eqno(2.27)
\en

Applying these results to the computation of the right-hand side of
Eq.~(2.24), we obtain

\eq\eqalign{
\lx {\cal Z} - 1 \rx_{2,\propto  \gp^2  }  = & ( \lambda_{2,2} \log^2
n + \lambda_{2,1} \log n + \lambda_{2,0} ) \lambda \lxc \gp \rxc^2 \cr
&\quad + (\lambda \log n + a_0 ) \lambda \lxc \gp \log (1-y) \rxc \lxc \gp
\rxc \cr
&+ (\lambda \log n + a_0 ) a_0 \lxc \gp \rxc^2 \cr} .
\eqno(2.28)
\en

Another term, $\lx \cZ - 1 \rx_{2, \propto  \fm^2 }$, corresponds to
partition sum contributions from elementary sub-branches with two
weak ancestor nodes. In this case, since we wish to consider only
terms of $O(\delta^2)$, we take no factors of $\gp$ whatsoever. The
diagrammatic representation of this term is shown in Figure 10. A short
vertical bar is added before the second horizontal line, to indicate
that one of the summations of $n^{-\nu}$ takes place off of the main
branch.  The calculation is entirely analogous to that leading to
Eq.~(2.28), and the result is

\eq\eqalign{
\lx {\cal Z} - 1 \rx_{2,\propto   \fm^2 } = & ( \lambda_{2,2} \log^2
n + \lambda_{2,1} \log n + \lambda_{2,0} ) \lambda \lxc \fm \rxc^2 \cr
&\quad + (\lambda \log n + a_0 ) \lambda \lxc \fm \log y \rxc \lxc \fm
\rxc \cr
&+ (\lambda \log n + a_0 ) a_0 \lxc \fm \rxc^2 \cr} .
\eqno(2.29)
\en

The mixed terms (Figure 11), which contain one factor of $\gp$, and one factor of
$\fm$ (in either order), have the same structure,

\eq\eqalign{
\lx {\cal Z} - 1 \rx_{2,\propto   \fm \cdot \gp} = & ( \lambda_{2,2}
\log^2 n + \lambda_{2,1} \log n + \lambda_{2,0} ) 2 \lambda \lxc \fm
\rxc \lxc \gp \rxc  \cr &+ (\lambda \log n + a_0 ) \lambda [ \lxc \fm
\log y \rxc \lxc \gp \rxc +  \lxc \gp \log (1-y)\rxc
\lxc \fm \rxc ] \cr
&+ (\lambda \log n + a_0 ) 2 a_0 \lxc \fm \rxc \lxc \gp \rxc \cr} .
\eqno(2.30)
\en

There are also contributions to $\lx \log \cZ \rx_2$ from the
$-(1/2) \lx (\cZ - 1 )^2 \rx$ term in the expansion of the
logarithm. These reflect cross-products between two elementary
sub-branches, with the partition sum contribution of each being
taken to $O (\delta)$. The diagrams appearing are shown in
Figure 12. Note that symmetry factors appear multiplying some of
the diagrams. Also, new vertices of the type $\lxc \gp^2 \rxc$, $\lxc
\fm^2 \rxc$ appear, arising from cases in which $\gp$ or $\fm$
factors are taken at the same vertices in the two different
``replicas" of $\cZ - 1$. Such vertices are indicated by adjoining circles, displaced
perpendicularly. The contribution $\propto \gp^2 $ is

\eq\eqalign{
-{1 \over 2 } \lx (\cZ - 1)^2 \rx_{2,\propto  \gp^2 } = - & ( \lambda_{2,2} \log^2
n + \lambda_{2,1} \log n + \lambda_{2,0} ) \lambda \lxc \gp \rxc^2 \cr
&\quad - (\lambda \log n + a_0 ) \lambda \lxc \gp \log (1-y) \rxc \lxc \gp
\rxc \cr
&- (\lambda \log n + a_0 ) a_0 \lxc \gp \rxc^2 \cr
&\quad -{1 \over 2} (\lambda \log n + a_0 )
 \lxc \gp^2 \rxc, \cr}
\eqno(2.31)
\en
while that $\propto  \fm^2 $ is

\eq\eqalign{
-{1 \over 2 } \lx (\cZ - 1)^2 \rx_{2,\propto   \fm^2 } = - & (
\lambda_{2,2} \log^2 n + \lambda_{2,1} \log n + \lambda_{2,0} )
\lambda \lxc \fm \rxc^2 \cr &\quad - (\lambda \log n + a_0 ) \lambda \lxc
\fm \log (1-y) \rxc \lxc \fm \rxc \cr
&- (\lambda \log n + a_0 ) a_0 \lxc \fm \rxc^2 \cr
&\quad -{1 \over 2} (\lambda \log n + a_0 )
 \lxc \fm^2 \rxc . \cr} 
\eqno(2.32)
\en
Note that the vertex $\lxc \fm \log (1-y) \rxc$ appears in this term,
rather than $\lxc \fm \log y \rxc$, as in Eq.~(2.29). Finally, the mixed
term is

\eq\eqalign{
-{1 \over 2 } \lx (\cZ - 1)^2 \rx_{2,\propto  \fm\cdot\gp } = - & (
\lambda_{2,2} \log^2 n + \lambda_{2,1} \log n + \lambda_{2,0} )
2 \lambda \lxc \fm \rxc \lxc \gp \rxc \cr &- (\lambda \log n + a_0 )
[\lambda \lxc \fm \log (1-y) \rxc \lxc \gp \rxc \cr & \quad\quad + \lambda \lxc
\fm \rxc \lxc \gp \log(1-y) \rxc ]\cr &- (\lambda \log n + a_0 ) 2 a_0
\lxc \fm \rxc \lxc \gp \rxc \cr &\quad -(\lambda \log n + a_0 )
 \lxc \fm \cdot \gp \rxc . \cr} 
\eqno(2.33)
\en

Collecting terms from Eqs.~(2.28-33), we finally obtain

\eq
\lx \log \cZ \rx_2 = \left [ \lambda \lxc \psi \rxc \lxc \fm \log
 ( {y \over 1-y } ) \rxc -{1 \over 2} \lxc \psi^2 \rxc
\right ] ( \lambda \log n + a_0 )
\eqno(2.34)
\en
where, again, $\psi=\fm + \gp$. Note that the $\log^2 n$ term has
cancelled out, as have the non-universal contributions to the $\log
n$ term ($\propto a_0 \lambda \log n$ in the above Eqs.~(2.28-33)). 

\medskip
\noindent{\bf Structure of $\delta$ expansion}
\smallskip

In appendix B, we continue this expansion, deriving $\lx \log \cZ
\rx_3$ and $\lx \log \cZ \rx_4$. In general, we expect that the
$O(\delta^N)$ term in the expansion will have the form

\eq
\lx \log \cZ \rx_N = \sum_{M=0}^N \beta_{N,M} \log^M n ,
\eqno(2.35)
\en
where the coefficients $\beta_{N,M}$ will include ``universal"
(independent of the $\{ \lambda_{N,0} \}$ of Eq.~(2.26)) and non-universal terms. It is
clear from the structure of the expansion that the leading coefficients
$\beta_{N,N}$ do not contain any non-universal terms, as the leading
coefficient $\lambda_{N,N}$ in Eq.~(2.26) for any sum $\sum_j \log^{N-1} n_j / n_j^{\nu}$ is
universal (see Eq.~(A.8) below). We also note that with the exception of the $O(\delta )$
term, we have $\beta_{N,N} = 0$ to fourth order. Below we will show that this
is true to all orders.  The situation is summarized in Table I, which
shows, as a function of order in $\delta$ and power of $\log n$, the
terms appearing in this series.

Unfortunately, this series does not allow us to directly determine
the quenched dimensions $\sqq$. Recall that these are determined by
the requirement that $\lx \log \cZ \rx \to 0$ in the limit $n \to
\infty$ (more precisely, that it be bounded above and below by
quantities which go to neither $\pm \infty$, see Ref.~20 for a
discussion). Since each individual term $\log^M n$ diverges as $n \to
\infty$, and since there is no obvious relation between the
coefficients of different powers of $\log n$ in the series for $\lx
\log \cZ \rx$, we cannot extract unique results for $\sqq$ from this
series.

For this reason, we turn in section 3 to the resummation to all
orders in $\delta$ of the leading and sub-leading divergent terms in this series. We
shall see that the annoying higher order logarithms can be safely
resummed to decaying power laws, allowing unambiguous determination
of $\sqq$ in the limit $n \to \infty$.

\medskip
\noindent{\bf Resummation of ``one-propagator" series}
\smallskip

Before turning to the general re-summation of the series, we would
like to remark on a simpler resummation, which is analogous to loop
expansions in ordinary field theory. Consider all terms in the series
with only one summation over $n^{-\nu}$. These terms can be of any
order in $\delta$, as vertices can be multiplied together at the same
node, as in the $O (\delta^2 )$ terms above. The associated diagrams are shown in Figure
13. This ``one-propagator" result can be easily derived:

\eq
\lx \log \cZ \rx_{1p} = \sum_{N=1}^{\infty} {(-1)^{N-1} \over N} \lxc
\psi^N \rxc (\lambda \log n + a_0 ) .
\eqno(2.36)
\en
We have used the subscript $1p$ to indicate that this is the
one-propagator term. 
Taking the sum inside the brackets $\lxc \rxc$ yields

\eq
\lx \log \cZ \rx_{1p} = \lxc \log ( 1 + \psi ) \rxc (\lambda \log n 
+ a_0) .
\eqno(2.37)
\en

This result has two appealing properties--in the first place it leads to a result for
$\sqq$ in which no negative values of $f$ appear. It is thus an appealing
approximate formula for the dimensions of a ``typical" cluster. Also,
as we shall see in section 4, the $\sqq$ resulting from setting $\lx
\log \cZ \rx_{1p} = 0$ is quite close to numerical results in the case of our
toy model, model Z. 

However, if one tries to extend this approach by resumming the
``two-propagator" terms, and so forth, one encounters the same
problem as with the $\delta$-series, that higher orders in $\log n$
are also generated. Thus this expansion technique suffers from the same difficulty
as the $\delta$ expansion, that unique values of
$\sqq$ are impossible to obtain without resummation. 

\bigskip
\noindent{\bf 3. Resummation of the $\delta$-series}
\medskip

We have shown that the form of the $\delta$-series is:

\eq
\lx \log \cZ \rx = \sum_{N=1}^{\infty} \delta^N \lx \log \cZ \rx_N = \sum_{N=1}^{\infty} \sum_{M=0}^N
\beta_{N,M}
\log^M n ,
\eqno(3.1)
\en
where each $\beta_{N,M}$ is of $O(\delta^N )$. Our procedure is now to resum
this series by summing first the most divergent terms $M=N$ at all orders in
$\delta$, then the second-most divergent terms $M=N-1$, and so on. Introducing
yet another subscript, $\ell$ (leading), to indicate the most divergent sum,
and the subscript $s\ell$ (sub-leading), to indicate the next most divergent sum, we
have

\eq
\lx \log \cZ \rx_{\ell} = \sum_{N=1}^{\infty} \beta_{N,N} \log^N n ,
\eqno(3.2)
\en
and
\eq
\lx \log \cZ \rx_{s \ell} = \sum_{N=2}^{\infty} \beta_{N,N-1}
\log^{N-1} n .
\eqno(3.3)
\en

In principle, we could continue with sub-sub leading terms, but in
practice we will restrict ourselves to computing only these two terms.
Note that our resummation procedure corresponds to summing down the
diagonals of Table I.

\medskip
\noindent{\bf Summation of leading terms}
\smallskip

The leading logarithm at each order in $\delta$ will be universal, as
pointed out in section 2 above. Let us start by computing $\lx Z - 1
\rx_{\ell}$. In order to compute this, we will need the result from
appendix A: $\lambda_{N,N} = {\lambda / N}$, where the reader might recall the definition of
$\lambda_{N,M}$

\eq
\lx  \sum_{j \ge 1} {\rho_0 \log^{N-1} n_j \over n_j^{\nu}} \rx = \sum_{M=0}^N
\lambda_{N,M} \log^M n .
\eqno(3.4)
\en
Here the sum is along a horizontal bar of some diagram.  Since we are taking the leading order in
$\log n$ in the entire diagram, we wish to take the leading order in each
propagator. No factors of $1-y$ or $y$ will appear in the
leading order, since each such factor would take the place of a factor of
$\log n$. Also, we take the leading logarithm of each term of the form of Eq.~(3.4), which
will have as a coefficient one of the $\{ \lambda_{N,N} \}$. 

Consider a
diagram contributing to $\lx \cZ - 1 \rx_{\ell}$ consisting of $n_1$
factors of $\fm$ and $n_2$ factors of $\gp$. There will be a total of
$N = n_1 + n_2$ horizontal bars, or propagator sums, in such a
diagram (see Figure 14). If we
call the contribution of this diagram ${\cal C}$, then

\eq
{\cal C} = {\lambda^N \over N!}  \lxc \fm \rxc^{n_1} \lxc \gp 
\rxc^{n_2} \log^N n ,
\eqno(3.5)
\en
because $\lambda^N/N! = \prod_{N^{\prime}=1}^{N} \lambda_{N^{\prime},N^{\prime}}$. 

Since any vertex can be taken to be either $\fm$ or $\gp$ without
changing other features of the term, we immediately obtain

\eq
\lx \cZ - 1 \rx_{\ell} = \sum_{N=1}^{\infty} {\lambda^N \over
N!} \lxc \psi \rxc^{N} \log^N n = \exp ( \lambda \lxc \psi \rxc \log n
) - 1 .
\eqno(3.6)
\en

Now let us turn to ${-1/2} \lx (\cZ - 1)^2 \rx_{\ell}$ from the expansion of $\lx \log \cZ \rx$ in
Eq.~(2.1). For the sake of argument, let us consider terms in which each factor of $\cZ - 1$
has at least one factor of $\fm$. We will say that the first factor
of $\cZ - 1$ has $l_1$ factors of $\gp$ before the first appearance of
$\fm$, and $m_1$ factors of either $\fm$ or $\gp$ after the first appearance of $\fm$.
The second factor of $\cZ - 1$ has $l_2$ factors of $\gp$ before the
first factor of $\fm$, and $m_2$ factors of either $\fm$ or $\gp$ after the first
factor of $\fm$ (see Figure 15). 

Now we average these two factors
together. Note that two factors of $\fm$ appearing in the main line
cannot appear at the same node; were they to be averaged together, the
diagram overall would lose one propagator, and thus one power of $\log
n$, without losing a factor of $\delta$, which counts the total number of factors of $\fm$ and
$\gp$. Such diagrams first contribute at sub-leading order. Similarly, two factors of $\gp$
cannot appear at the same node. 

Suppose that the factor of $\fm$
coming from the second $\cZ-1$ is further down the main line than
that coming from the first $\cZ-1$. Then in any diagram resulting
from this product, there will be a number $l \le l_2$ of $\gp$ vertices
between the two $\fm$ vertices, and $L-l$ to the left (further
up the main branch) of the first $\fm$ vertex, where $L=l_1 +
l_2$ (see Figure 16). Now we must consider the vertices off the main
line. The structure of the diagram does not depend on whether these
are $\fm$ or $\gp$ vertices, since we average no factors of $\log(y)$
or $\log(1-y)$ with these vertices at leading order. Thus we will sum
together these two types of off-main line vertices, and attach $m_1$
vertices $\psi$ to the first factor of $\fm$, and $m_2$ such vertices
to the second factor of $\fm$.

In order to compute the diagram, we must keep track of the number
of factors of $\log n$ appearing for each propagator; at
leading order we take only the propagator terms multiplying one of elements of the set $\{
\lambda_{N,N}\}$. The total number of propagators to the right (on the main line) of
the ``upstream" factor of $\fm$ is $m_2 + l + 1$, while the total number
downstream of the first factor of $\gp$ to the left of this $\fm$ vertex is
$m_1 + m_2 + l + 2$. (Note that we use ``upstream" to mean closer to the root, not
closer to the elementary sub-branches.) A simple computation, keeping careful track of
these factors, gives the contribution ${\cal W}_{l}$ of this particular diagram as:

\eq\eqalign{
{\cal W}_{l} = & {1 \over (m_2 + l + 1)!} {1 \over m_1 !} {(M + l + 1)! \over (M + L + 2)!} \cr &
\quad \cdot \lxc \fm
\rxc^2 \lxc \gp \rxc^{L} \lxc \psi \rxc^{M} (\lambda \log n
)^{M + L + 2}, \cr}
\eqno(3.7)
\en
where $M=m_1+m_2$ and we defined above $L=l_1+l_2$.

Of course, the two original factors of $\cZ - 1$ can be multiplied
together in various distinguishable ways to make this diagram,
corresponding to the different origins of the $L-l$ factors of $\gp$
on the left hand side of the diagram. Since $l_1$ of these come from
the first factor of $\cZ-1$, and $L-l-l_1$ come from the second
factor of $\cZ-1$, the total number of relative permutations $N_p$ is

\eq
N_p = \left ( {L-l \atop l_1} \right ) \equiv {(L-l)! \over (L-l-l_1)!
l_1!} . 
\eqno(3.8)
\en

Finally, to obtain the expectation value of the product of the factor
$\cZ-1$ containing $l_1$ factors of $\gp$ in the main line and $m_1$
factors of $\psi$ off the main line with the other factor of $\cZ-1$, which contains $l_2$
factors of $\gp$ on the main line and $m_2$ factors of $\psi$ off the
main line, we must sum over $l \le l_2$ and permute the choice of whose main
line factor of $\fm$ is upstream (see Figure 16). If we are interested in
contributions to $-(1/2) \lx (\cZ - 1)^2 \rx$ with at least two factors
of $\lxc \fm \rxc$, we must then sum over $m_1$, $m_2$, $l_1$, and
$l_2$. This yields

\eq\eqalign{
-{1 \over 2}\lx (\cZ - 1)^2 \rx_{\ell, \propto \lxc \fm \rxc^2} = &-
\sum_{m_1=0}^{\infty}
\sum_{m_2=0}^{\infty}
\sum_{l_1=0}^{\infty}
\sum_{l_2=0}^{\infty}
\sum_{l=0}^{l_2} \cr & \quad \cdot
\left ( {L-l \atop l_1} \right ) {1
\over (m_2 + l + 1)!} {1 \over m_1 !} {(M + l + 1)! \over (M + L + 2)!} \cr & \quad \cdot\lxc \fm
\rxc^2 \lxc \gp \rxc^{L}
\lxc \psi \rxc^{M} (\lambda \log n )^{M + L + 2} . \cr}
\eqno(3.9)
\en
In addition, the term including only one factor of $\fm$ is easily
written as

\eq
\eqalign{
-{1 \over 2}\lx (\cZ - 1)^2 \rx_{\ell, \propto \lxc \fm \rxc} = &-
\sum_{m_1=0}^{\infty}
\sum_{l_1=0}^{\infty}
\sum_{l_2=1}^{\infty}
\sum_{l=0}^{l_2} \cr & \quad \cdot
\left ( {L-l \atop l_1} \right ) {1 \over m_1 !} {1 \over l !} {(m_1 + l )! \over (m_1
+ L + 1)!}  \cr & \quad \cdot\lxc \fm \rxc \lxc \gp \rxc^{L}
\lxc \psi \rxc^{m_1} (\lambda \log n )^{m_1  + L + 1} ,\cr}
\eqno(3.10)
\en
while that with no factors of $\fm$ at all is

\eq
\eqalign{
-{1 \over 2}\lx (\cZ - 1)^2 \rx_{\ell, \propto \lxc \fm \rxc^0} = &-{1
\over 2}
\sum_{l_1=1}^{\infty}
\sum_{l_2=1}^{\infty} 
\left ( {L \atop l_1} \right ) {1 \over L!} \lxc \gp \rxc^{L}
(\lambda \log n )^{L} \cr} \eqno(3.11)
\en
The sums are tedious but elementary. In appendix C we illustrate the
performance of the types of sums appearing in this work. Performing
the sums and adding Eqs.~(3.9-11) we obtain the quite simple result,

\eq
-{1 \over 2}\lx (\cZ - 1)^2 \rx_{\ell} = -{1 \over 2} \left [ \exp(
\lambda \lxc \psi \rxc \log n ) - 1 \right ]^2= -{1 \over 2} [\lx \cZ - 1
\rx_{\ell} ]^2 . \eqno(3.12) \en

Remarkably, at leading order, the process of taking the expectation
value of $\cZ - 1$ commutes with taking the square of this quantity.
We will now address the origin of this identity.

Let us first consider a simpler case, with only one factor of $\cZ-1$. The leading order
${\cal D}$ of a diagram with $l$ factors of $\gp$ on the main line,
followed by one factor of $\fm$, followed by $m$ factors of $\psi$, can be written as

\eq
{\cal D} = {\cal N} (\lambda \lxc \fm \rxc \log n) (\lambda \lxc \gp \rxc \log
n)^l (\lambda \lxc \psi \rxc \log n)^m ,
 \eqno(3.13)
\en
where the factor ${\cal N} = 1/(m+l+1)!$. The factor ${\cal N}$ can be written as a
nested integral over $l+1$ variables $x_i$:

\eq
{\cal N} = \left (\prod_{i=1}^l \int_0^{x_{i-1}} d x_i \right )\int_0^{x_l} dx_{l+1} {x_{l+1}^m
\over m!} = {1 \over (m+l+1)!} ,
\eqno(3.14)
\en
with $x_0 \equiv 1$. In this product, the $l$ integrals on the left correspond to $\gp$
vertices, while the last integral corresponds to the $\fm$ factor and the accompanying
insertion of $m$ additional factors of $\log n$ associated with the $\psi$ vertices. 

Now consider a term arising from the product of a factor of $\cZ-1$ containing $l_1$ factors of $\gp$
in the main line and $m_1$ factors of $\psi$ off the main line with another factor of
$\cZ-1$ containing
$l_2$ factors of $\gp$ and $m_2$ factors of $\psi$. Again, we choose $L=l_1 + l_2$ and
$M=m_1+m_2$. Each factor of $\cZ-1$ in addition has one $\fm$ vertex on the main line. At leading
order, any diagram ${\cal D}$ resulting from this product will have the form

\eq
{\cal D} = {\cal N} (\lambda \lxc \fm \rxc \log n)^2 (\lambda \lxc \gp \rxc \log n)^L
(\lambda \lxc \psi \rxc \log n)^M, \eqno(3.15)
\en
where the combinatorial factor ${\cal N}$ depends on the precise way in which the diagrams
are multiplied together. It is simple to write a formula for ${\cal N}$. Let us assign variables
$x^{(1)}_i$ to the main line vertices arising from the first factor of $\cZ-1$, and variables
$x^{(2)}_j$ to main line vertices arising from the second factor of $\cZ-1$. Then the factor ${\cal
N}$ for any particular diagram can be written analogously to Eq.~(3.14). For instance,
if all of the factors of $\gp$ coming from the second factor of $\cZ-1$ are further
down the main line (to the right of) the factor of $\fm$ arising from the first factor
of $\cZ-1$, then we immediately see that

\eq\eqalign{
{\cal N} = & \left (\prod_{i=1}^{l_1} \int_0^{x^{(1)}_{i-1}} d x^{(1)}_i \right
)\int_0^{x^{(1)}_{l_1}} dx^{(1)}_{l_1+1} {(x^{(1)}_{l_1+1})^{m_1} \over m_1!}
\cr&\quad\cdot\left (\prod_{j=1}^{l_2} \int_0^{x^{(2)}_{j-1}} d x^{(2)}_j \right
)\int_0^{x_{l_2}} dx_{{l_2}+1} { (x^{(2)}_{l_2+1})^{m_2} \over m_2!} \cr = &{(M+l_2+1)!
\over (M+L+2)!}  {1 \over m_1!} {1 \over (m_2 + l_2 +1)!}, \cr}
\eqno(3.16)
\en
with $x^{(2)}_0 = x^{(1)}_{l_1+1}$ and $x^{(1)}_0=1$. Now the allowed permutations of the
vertices correspond to all permutations of orderings of $\{x^{(1)}_i\}$, $\{x^{(2)}_j\}$
within the multiple integral that preserve the orderings $1>x^{(1)}_1>x^{(1)}_2\cdots$
and $1>x^{(2)}_1>x^{(2)}_2\cdots$. But this allows one to factor the two nested
integrals over the points $\{ x^{(1)} \}$ and $\{ x^{(2)} \}$ completely, so that the sum
over permutations
$P$ of
${\cal N}$ can be expressed as

\eq\eqalign{
\sum_P {\cal N} = & \left (\prod_{i=1}^{l_1} \int_0^{x^{(1)}_{i-1}} d x^{(1)}_i \right
)\int_0^{x^{(1)}_{l_1}} dx^{(1)}_{l_1+1} {(x^{(1)}_{l_1+1})^{m_1} \over m_1!} \cr&\quad\cdot\left
(\prod_{j=1}^{l_2} \int_0^{x^{(2)}_{j-1}} d x^{(2)}_j \right )\int_0^{x^{(2)}_{l_2}}
dx^{(2)}_{{l_2}+1} { (x^{(2)}_{l_2+1})^{m_2} \over m_2!} \cr = &{1 \over (m_1+l_1+1)!}   {1
\over (m_2 + l_2 +1)!} ,\cr}
\eqno(3.17)
\en
where $x^{(2)}_0=1$, which deconvolves the two nested integrals, allowing their factorization
into terms of the form (3.14). This provides a direct proof of the identity (3.12).

Using the same argument, we can easily show that at leading order,

\eq
\lx (\cZ-1)^p \rx_{\ell} = [ \lx \cZ-1 \rx_{\ell} ]^p=
[ \exp(\lambda \lxc \psi \rxc \log n) - 1]^p,
\eqno(3.18)
\en
so that the leading order contribution to $\lx \log \cZ \rx$ is

\eq\eqalign{
\lx \log \cZ \rx_{\ell} & = \sum_{p=1}^{\infty} {(-1)^{p-1} \over p} (\lx \cZ-1 \rx_{\ell})^p
\cr &=\lambda \lxc \psi \rxc \log n , \cr}
\eqno(3.19)
\en
We therefore arrive at the result

\eq
\lx \log \cZ \rx_{\ell} = \log( 1 + \lx Z - 1
\rx_{\ell}).
\eqno(3.20)
\en
We shall see that this is true also at sub-leading order, provided we
restrict ourselves to terms diverging logarithmically with $n$. 

\medskip
\noindent{\bf Summation of sub-leading terms: Universal}
\smallskip

Sub-leading terms are those with one less power of $\log n$ than $\delta$. There are
two ways in which such terms can arise:

\item{1.} In a term arising from $\lx (\cZ-1)^p \rx$ with $p>1$, more than one factor of
$\fm$ or $\gp$ may appear at the same vertex. Since these two or more factors are integrated
together, the net result is that the order of the term in $\log n$ is reduced with respect to the
order in $\delta$. If more than two factors of $\fm$ or $\gp$ appear at the same vertex, or
if more than one vertex contains more than one factor, then at least two factors of $\log n$ will
be lost with respect to factors of $\delta$, so the term will be at most of sub-sub-leading
order. Thus we need only consider diagrams with one vertex possessing two factors of $\fm$ or
$\gp$, with all other vertices containing at most one factor.

\item{2.} Recall that the form of the propagator sum along a particular sub-branch up to a given
vertex is
$\sum_{M=0}^N
\lambda_{N,M}
\log^M n^{\prime}$, where $n^{\prime}$ is the number of particles below the vertex in question,
either on the weak or the strong side (See Eq.~(2.26)). Thus either $n^{\prime}=y(\epsilon n) n$, or
$n^{\prime} = (1-y(\epsilon n) ) n$, where $\epsilon$ is the random
variable at the vertex being explicitly considered, and $n$ is the total number of descendants
below that vertex. There are two ways in which the order of the propagator in $\log n$ may be
reduced by one. The first way is for a factor of $\log y$ or $\log(1-y)$ to be taken in the highest
order in $\log n^{\prime}$ term ($M=N$) in the propagator. The second way is
for the second highest term in
$\log n^{\prime}$ ($M=N-1$) to be taken, but without including any factors of $\log(1-y)$ or $\log
y$.

\smallskip\noindent
{\it Terms of type 1}
\smallskip

An example of a term of type 1. is provided by $\lx \log \cZ \rx_{s\ell,\propto \lxc \fm^2
\rxc}$. There is no contribution from $\lx \cZ - 1 \rx$ to this term; the first contribution
arises from $-(1/2) \lx (\cZ-1)^2 \rx$ (see Figure 17). It is easy to see that

\eq\eqalign{
\lx (\cZ-1)^2 \rx_{s\ell,\propto \lxc \fm^2 \rxc} = &\sum_{l_1=0}^{\infty}
\sum_{l_2=0}^{\infty} \sum_{m_1=0}^{\infty} \sum_{m_2=0}^{\infty} \left ( { L \atop l_1 }
\right ) {1 \over m_1!} {1 \over m_2!}{M! \over (M+L+1)!} \cr&\quad\cdot(\lambda \lxc \gp \rxc \log
n)^L (\lambda \lxc \psi \rxc \log n)^M (\lambda \lxc \fm^2 \rxc \log n) .\cr}
\eqno(3.21)
\en
As in the above, $M=m_1 + m_2$, and $L=l_1+l_2$. The sums are straightforward, and yield

\eq
\lx (\cZ-1)^2 \rx_{s\ell,\propto \lxc \fm^2 \rxc} = {\lxc \fm^2 \rxc \over 2 \lxc \fm \rxc }
\left [ \exp(2 \lambda \lxc \psi \rxc \log n ) - \exp(2 \lambda \lxc \gp \rxc \log n )
\right ] .
\eqno(3.22)
\en

The sub-leading terms proportional to $\lxc \gp^2 \rxc$ and $\lxc \fm \gp \rxc$ are computed in
exactly the same way, and combine with the term proportional to $\lxc \fm^2 \rxc$ to yield

\eq
\lx (\cZ-1)^2 \rx_{s\ell,\propto \lxc \psi^2 \rxc} = {\lxc \psi^2 \rxc \over 2 \lxc \fm \rxc
} \left [ \exp(2 \lambda \lxc \psi \rxc \log n ) - \exp(2 \lambda \lxc \gp \rxc \log n )
\right ] .
\eqno(3.23)
\en

Now we must compute the sub-leading contribution $\propto \lxc \psi^2 \rxc$ arising from
$\lx (\cZ-1)^p \rx$, with $p > 2$. A typical diagram is shown in Figure 18. There are a
total of
$p(p-1)/2$ choices of the two factors of $\cZ-1$ which contribute the factors of $\fm$, $\gp$
which will be integrated together. The computation is considerably simplified by the fact that
the remaining factors of $\cZ-1$ can be averaged separately of these two factors, by a simple
extension of identity (3.18), which we used above to factor the leading order terms. Thus

\eq
\lx (\cZ-1)^p \rx_{s\ell,\propto \lxc \psi^2 \rxc} = {p(p-1) \over 2}\lx (\cZ-1)^2
\rx_{s\ell,\propto \lxc \psi^2 \rxc} [ \lx \cZ-1\rx_{\ell} ]^{p-2} ,
\eqno(3.24)
\en
leading to

\eq\eqalign{
\lx \log \cZ \rx_{s \ell, \propto \lxc \psi^2 \rxc} &= \lx (\cZ-1)^2 \rx_{s \ell, \propto \lxc
\psi^2 \rxc} \sum_{p=2}^{\infty} {(-1)^{p-1} \over p} { p (p-1) \over 2}  [ \lx
\cZ-1\rx_{\ell}]^{p-2}\cr &= - \lx (\cZ-1)^2 \rx_{s \ell, \propto \lxc \psi^2 \rxc} \; {1 \over 2}
\left ({1 \over 1 + \lx \cZ-1
\rx_{\ell} }\right )^2   .\cr}
\eqno(3.25)
\en
Using Eqs.~(3.6) and (3.23), and substituting from Eq.~(3.25), we then obtain

\eq\eqalign{
\lx \log \cZ \rx_{s \ell, \propto \lxc \psi^2 \rxc} = &  - {\lxc \psi^2 \rxc \over 4 \lxc \fm
\rxc}
\left [ 1 - \exp(-2 \lambda \lxc \fm \rxc \log n) \right ] \cr  = & -{1 \over 4} {\lxc \psi^2 \rxc
\over
\lxc \fm \rxc} \left (1-n^{-2\lambda \lxc \fm \rxc } \right ) .\cr}
\eqno(3.26)
\en

Recall the definition of $\lxc \fm \rxc$:

\eq
\lxc \fm \rxc \equiv \int_0^{\infty} d \eta \; \eta^{\nu - 1} 
{x^q (\eta) \over y^{\sigma} (\eta)} ,
\eqno(3.27)
\en
where $x(\eta)$ and $y(\eta)$ refer to the unstable manifold for branch competition. Clearly
$\lxc \fm \rxc > 0$, which implies that the correction to $\lx \log \cZ \rx$ displayed in
Eq.~(3.26) approaches a constant as $n \to \infty$. Thus this correction will not affect
the values of the multifractal exponents $\sqq$ (defined in Eq.~(1.6)), provided that they are
computed in this limit, as the terms logarithmic in
$n$ in
$\lx
\log
\cZ \rx$ will still dominate.

\smallskip\noindent
{\it Terms of type 2}
\smallskip

Now we turn to sub-leading terms of type 2, for which a sub-dominant term in one of the
propagators is taken. We first consider such terms involving factors of $\log(1-y)$ or $\log y$
(to be averaged with $\fm$ or $\gp$) arising from the expansion of a leading propagator term,

\eq
\lambda_{N,N} \left \{ \log [(1-y)n] \right \}^N = \lambda_{N,N} [ \log^Nn + N \log(1-y) \log^{N-1}
n + \cdots ] ,
\eqno(3.28)
\en
with a similar result for $\log y$. Since $\lambda_{N,N} = \lambda/N$, the $\log(1-y)$ in
Eq.~(3.28) multiplies $\lambda \log^{N-1}n$, which is comparable to a leading propagator term with
one less vertex. 

We will first consider terms containing the factor
$\lxc
\fm
\log y \rxc$. Looking at such a term arising from $\lx (\cZ-1)^p \rx$, we see that the different
factors of $\cZ-1$ interact only through the various possible orderings of their $\gp$ and one
$\fm$ vertices on the main line; thus we can factor these diagrams, and
consider only the contribution from a single factor of $\cZ-1$ (see Figure 19). This is

\eq\eqalign{
\lx \cZ - 1 \rx_{s \ell, \propto \lxc \fm \log y \rxc} =& \sum_{l=0}^{\infty} \sum_{m=1}^{\infty}
{1 \over (m+l)!} (\lambda \lxc \fm \log y \rxc ) (\lambda \lxc \gp \rxc \log n )^l (\lambda
\lxc \psi \rxc \log n  )^m \cr & + \sum_{l=0}^{\infty} \sum_{m=2}^{\infty}
{m-1 \over (m+l)!} (\lambda \lxc \fm \log y \rxc ) (\lambda \lxc \fm \rxc \log n ) (\lambda
\lxc
\gp \rxc \log n  )^l\cr&\quad\cdot (\lambda \lxc \psi \rxc \log n )^{m-1} . \cr}
\eqno(3.29)
\en
The first term on the right-hand side of Eq.~(3.29) corresponds to the case where the factor
of $\fm$ on the main line is averaged with $\log y$, and the second term corresponds to the
case where a factor off of the main line is averaged with $\log y$. The sums yield

\eq
\lx \cZ - 1 \rx_{s \ell, \propto \lxc \fm \log y \rxc} = \lambda^2 \lxc \psi \rxc \log
n 
\lxc \fm \log y \rxc  \; \exp(\lambda \lxc \psi\rxc \log n ) .
\eqno(3.30)
\en

Turning to $\lx (\cZ - 1)^p \rx$, we see that there are $p$ choices of which factor of $\cZ-1$
contributes the sub-leading term $\propto \lxc \fm \log y \rxc$. Thus

\eq\eqalign{
\lx \log \cZ \rx_{s \ell, \propto \lxc \fm \log y \rxc} & = \lx \cZ - 1 \rx_{s \ell, \propto \lxc
\fm \log y \rxc} \sum_{p=1}^{\infty} {(-1)^{p-1} \over p} p  [\lx \cZ - 1 \rx_{\ell} ]^{p-1}\cr &
= \lx \cZ - 1 \rx_{s \ell, \propto \lxc \fm \log y \rxc} \; \left ( { 1 \over 1 + \lx \cZ-1
\rx_{\ell} }
\right )\cr & =
\lambda^2 \lxc \psi \rxc   \lxc \fm \log y \rxc \log n . \cr}
\eqno(3.31)
\en

By contrast, terms $\propto \lxc \fm \log (1-y) \rxc$ can arise only from a combination of
two or more factors of $\cZ-1$, since a factor of $\fm$ is always associated with a weak branch in
a single partition function, and hence potentially only with a factor of $\log y$. Consider a term
arising from a product of
$p$ factors of
$\cZ - 1$, each containing $l_i$ factors of $\gp$ in the main line, followed by one factor
of $\fm$ and then $m_i$ factors of $\psi$ off of the main line. Let us further suppose that
the factor of $\lxc \fm \log(1-y) \rxc$ arises from the factor of $\fm$ appearing in the
main line of the $i=1$ factor. Then writing $L = \sum_i l_i$ and $M=\sum_i m_i$, we see
that any particular sub-leading contribution proportional to $\lxc \fm \log(1-y \rxc$
(which we term ${\cal D}$) will be given by

\eq
{\cal D} = {\cal N} (\lambda \lxc \gp \rxc \log n)^L (\lambda \lxc \psi \rxc \log n)^M
(\lambda \lxc \fm \log(1-y) \rxc ) \; (\lambda \lxc
\fm
\rxc \log n)^{p-1} ,
\eqno(3.32)
\en
where the factor ${\cal N}$ is determined by the precise ordering of vertices. In fact, it is
easy to see that ${\cal N}$ will be given by a version of Eq.~(3.16) appropriate to $p$
factors of $\cZ-1$, with the one addition that there should be a factor of
$d/dx^{(1)}_{l_1+1}$ inside the integral over the variable $x^{(1)}_{l_1+1}$ (and to the
right the factor of $(x^{(1)}_{l_1+1})^{m_1}/m_1!)$ ). Now the integral of all factors to
the right of this new operator will yield some number times $x^{(1)}_{l_1+1}$ raised to
some power; thus the   $d/d x^{(1)}_{l_1+1}$ operator can be viewed as acting on only one
of the factors of $\cZ-1$ at a time. It follows that the sum over permutations of ${\cal
N}$ can be represented as a sum of averages involving two factors of $\cZ - 1$ only, with
the other factors of $\cZ-1$ averaged separately. More precisely,

\eq
\lx \log \cZ \rx_{\esl, \propto \lxc \fm \log (1-y) \rxc} = \sum_{p=2}^\infty {(p-1) \over
2} (-1)^{p-1}
\lx (\cZ -1)^2 \rx_{\esl, \propto \lxc \fm \log (1-y) \rxc} [\lx \cZ-1 \rx_{\l}]^{p-2} .
\eqno(3.33)
\en
This naturally represents a considerable simplification of the problem.

Now we must compute $\lx
(\cZ -1)^2 \rx_{\esl, \propto \lxc \fm \log (1-y) \rxc}$. This is given by (see Figure 20)

\eq\eqalign{
\lx
(\cZ -1)^2 \rx_{\esl, \propto \lxc \fm \log (1-y) \rxc} = & 2 \sum_{l_1=0}^{\infty}
\sum_{l_2=0}^{\infty} \sum_{m_1=0}^{\infty} \sum_{m_2=1}^{\infty}  \left ( { L
\atop l_1 } \right ) {1 \over m_1!} {1 \over (m_2 -1)!}\cr&\quad\cdot {(M-1)! \over (M+L)!}
(\lambda \lxc \gp \rxc \log n)^L \cr&\quad\cdot (\lambda \lxc \psi \rxc \log n)^M (\lambda \lxc
\fm \log(1-y) \rxc) . \cr} \eqno(3.34)
\en
where the factor of 2 accounts for the distinguishability of the contributions of the two partition
functions. As in the above, $M=m_1 + m_2$, and $L=l_1+l_2$.
The sums may be performed directly, and yield

\eq\eqalign{
\lx
(\cZ -1)^2 \rx_{\esl, \propto \lxc \fm \log (1-y) \rxc} = &  \lambda \lxc \fm \log(1-y)
\rxc { \lxc \psi \rxc \over  \lxc \fm \rxc }\cr&\quad\cdot \left [ \exp ( 2 \lambda \lxc \psi \rxc
\log n ) - \exp ( 2 \lambda \lxc \gp \rxc
\log n ) \right ] .\cr} \eqno(3.35) \en
Since

\eq\eqalign{
\sum_{p=2}^\infty (-1)^{p-1} {(p-1) }    [\lx \cZ-1 \rx_{\l}]^{p-2} = &- \left ({1 \over 1 +\lx
\cZ-1
\rx_{\ell}} \right )^2 
\cr& = -\exp(-2
\lambda
\lxc \psi \rxc \log n ) , \cr}\eqno(3.36) \en
our result is

\eq\eqalign{
\lx \log \cZ \rx_{\esl, \propto \lxc \fm \log (1-y) \rxc} = & - \lambda \lxc \fm \log(1-y)
\rxc { \lxc \psi \rxc \over 2 \lxc \fm \rxc } \left [ 1 - \exp ( -2 \lambda \lxc \fm \rxc \log n )
\right ] \cr = & - \lambda \lxc \fm \log(1-y)
\rxc { \lxc \psi \rxc \over 2 \lxc \fm \rxc } \left ( 1 - n^{-2 \lambda \lxc \fm \rxc} \right )
, \cr}
\eqno(3.37)
\en
i.e., a constant plus a decaying power law in $n$.

Terms proportional to $\lxc \gp \log(1-y) \rxc$ appear at sub-leading order both from one factor
of $\cZ-1$ and from two factors averaged together. The first case arises from the fact that a
$\gp$ factor has a strong branch associated with it downstream, which can provide a factor of
$\log(1-y)$. The combinatorics, and the result, are thus the same as for the term $\propto \lxc \fm
\log y \rxc$, given in Eqs.~(3.29-31). 

The second case involves the averaging of a $\gp$ factor arising from one partition function with
a factor of $\log(1-y)$ arising from the second partition function, leading to the same
combinatorics, and result, as for the term $\propto \lxc \fm \log(1-y) \rxc$ given in
Eqs.~(3.33-37). 
Thus one
obtains

\eq\eqalign{
\lx \log \cZ \rx_{\esl,\propto \lxc \gp \log(1-y) \rxc} = &\lambda \lxc \gp \log(1-y) \rxc \lxc
\psi \rxc \cr&\quad\cdot \left \{ \lambda \log n - { 1 \over 2 \lxc \fm \rxc} \left [ 1 - \exp(-2
\lambda \lxc \fm \rxc \log n) \right ] \right \} . \cr}  
\eqno(3.38)
\en

We have been discussing sub-leading contributions in which a sub-leading propagator term is taken
within the diagram, leading to factors $\log y$ or $\log (1-y)$ being averaged with $\fm$ or
$\gp$. In addition, there are sub-leading contributions in which such factors do not appear, but
in which $\lambda_{N,N-1}$ appears rather than $\lambda_{N,N} \equiv \lambda/N$. In
appendix B we show that

\eq
\lambda_{N,N-1} =  {1 \over 2} {\lxc \log^2 (1-y) \rxc \over \lxc \log (1-y) \rxc^2} \equiv
\lambda^{\prime} , \eqno(3.39)
\en
is independent of $N$. Comparing the expansion of the highest order propagator term,
Eq.~(3.28), with the form of the full propagator, Eq.~(3.4), we note that the independence of
$N$ of $\lambda_{N,N-1} \equiv \lambda^{\prime}$ allows us to directly translate our previous results
to obtain the terms
$\propto
\lambda^{\prime}$, since the two terms in the $N$'th order propagator $\propto \log^{N-1} n$ (one
including a factor of either $\log y$ or $\log(1-y)$, and the other a factor of $\lambda^{\prime}$)
are similar in form. In particular, we substitute
$\lambda
\lxc
\fm
\log y
\rxc
\to
\lambda^{\prime}
\lxc
\fm \rxc$ in Eq.~(3.31), $\lambda \lxc \fm \log (1- y) \rxc \to \lambda^{\prime} \lxc
\fm \rxc$ in Eq.~(3.37), and $\lambda \lxc \gp \log (1- y )\rxc \to \lambda^{\prime} \lxc
\gp \rxc$ in Eq.~(3.38), to obtain

\eq
\lx \log \cZ \rx_{\esl, \propto \lambda^{\prime}} = \lambda^{\prime}  \lxc \psi
\rxc^2 \left [ \lambda \log n - {1 \over 2 \lxc \fm \rxc} \left ( 1 - n^{-2\lambda \lxc
\fm \rxc } \right ) \right ] ,
\eqno(3.40)
\en
which is again a combination of a logarithmic divergence, a constant, and a decaying power
law.

\medskip
\noindent{\bf Summation of sub-leading terms: Non-universal}
\smallskip

At sub-leading order, there are also non-universal terms. These arise from taking one factor of
$a_0 \equiv \lambda_{1,0}$ instead of $\lambda \log n$ in the first propagator sum in a diagram;
taking non-universal terms in any propagator sum upstream of this will lead at most to
sub-sub-leading terms. Reviewing the derivation of Eq.~(3.6), we see that

\eq
\lx \cZ - 1 \rx_{\esl,\propto a_0 } = a_0 \lxc \psi  \rxc \sum_{N=0}^{\infty} {1 \over N!} (\lambda
\lxc \psi \rxc \log n )^N = a_0 \lxc \psi \rxc \exp ( \lambda \lxc \psi \rxc \log n ).
\eqno(3.41)
\en
Note that this summation starts at $N=0$.
It follows immediately, using the methods developed above to prove factorization of
averages of products of $\cZ -1$, that

\eq\
\lx \log \cZ \rx_{\esl,\propto a_0} = \lx \cZ - 1 \rx_{\esl,\propto a_0 } \sum_{p=1}^{\infty} ( - 1
)^{p-1} [ \lx \cZ -1 \rx_{\l} ]^{p-1} = a_0 \lxc \psi \rxc .
\eqno(3.42)
\en

\medskip
\noindent{\bf Summary of leading and sub-leading orders}
\smallskip

To summarize, we have the following result for $\lx \log \cZ \rx$, combining the
leading order, sub-leading universal, and sub-leading non-universal computations
(Eqs.~(3.19), (3.26),(3.31),(3.37-8),(3.40), and (3.42)):

\eq\eqalign{
\lx \log \cZ \rx = & \lambda \lxc \psi \rxc \log n \left [ 1 + \lambda \left ( \lxc \gp \log(1-y)
\rxc + \lxc \fm \log y \rxc \right ) + \lambda^{\prime} \lxc \psi \rxc \right ] \cr&\quad -{1
\over 2 \lxc \fm \rxc } \; \left [ 1 - n^{-2\lambda \lxc
\fm \rxc } \right ] \cr&\quad\quad\cdot \left ( {1 \over 2} \lxc \psi ^2 \rxc  + \lambda \lxc
\psi
\rxc 
\lxc \psi \log (1-y) \rxc +\lambda^{\prime} \lxc \psi \rxc^2 \right ) \cr&\quad+a_0 \lxc
\psi \rxc + O(\delta^3 \log n) , \cr} \eqno(3.43) \en
where the $O(\delta^3 \log n)$ terms are sub-sub-leading (see Table I).

Examining in detail the re-summations performed above, we see that all of the terms contributing to
the coefficient of
$\log n$ in Eq.~(3.43) arise from separate averages of single factors of $\cZ-1$;
averages together of two factors contribute to
the second term on the right-hand side of Eq.~(3.43). Anticipating that this is true to all
orders, we see that\footnote{$^2$}{Using the terminology of
statistical mechanics, averages that  can be reduced to individual
averages $\lx \cZ-1 \rx$ may be termed disconnected, while those involving averages together of more
than one factor of $\cZ-1$ may be termed connected. Thus Eq.~(3.44) corresponds to a sum of
disconnected averages only.} 

\eq \lim_{n \to \infty} { \lx \log \cZ \rx \over \log n} = \lim_{n \to \infty} { \log \lx \cZ \rx
\over \log n} . \eqno(3.44)
\en 
Thus, if we are
concerned only with the term proportional to
$\log n$, which should dominate as $n \to \infty$, this can be obtained from $\log \lx \cZ
\rx$. This quantity is computed by much less elaborate means in appendix D.
 
There are two methods of checking Eq.~(3.43), which resums all terms of the form $\delta^N
\log^N n$ or $\delta^N \log^{N-1} n$, with $N \ge 1$. The power laws in
$n$ may be converted into an expansion in $\log n$, and then compared directly with the $\delta$
expansion, as developed to second order in section 2 above, and to fourth order in appendix B. The
reader may easily satisfy himself that these two series agree, provided that comparison is confined
to terms proportional to $\delta^N \log^N n$ or $\delta^N \log^{N-1} n$, in this case with $1 \le
N \le 4$. 

A second method of checking this result is to recall that
the dependence of the partition function $\cZ$ upon $\sigma$ is quite trivial: $\cZ \propto
n^{\sigma}$ (see Eqs.~(1.7) and (1.10)). However, in our perturbative method this
$\sigma$-dependence is distributed throughout the branched tree, with a factor of
$y^{\sigma}$ or
$(1-y)^{\sigma}$ corresponding to this $\sigma$ dependence appearing at each vertex. This
considerably obscures the $\sigma$ dependence, but may also be used to check the final result for
$\lx
\log \cZ \rx$, Eq.~(3.43). Let us write again Eq.~(1.10),

\eq
\lx \log \cZ(q,\sigma;n) \rx = \sigma \log n + F(q,n) ,
\eqno(3.45)
\en
where $F$ has no dependence upon $\sigma$. This implies that

\eq
{d \lx \log \cZ \rx \over d \sigma} = \log n ,
\eqno(3.46)
\en
and
\eq 
{d^r  \lx \log \cZ \rx \over d \sigma^r} = 0, \quad r > 1 .
\eqno(3.47)
\en
These relations provide powerful constraints on our result. To check these
identities against our result for $\lx \log \cZ \rx$, Eq.~(3.43),
it is necessary to use (see Eqs.~(2.3) and (2.11))

\eq
{d \gp \over d \sigma} = -(\gp+1) \log(1-y) ,
\eqno(3.48)
\en
and
\eq
{d \fm \over d \sigma} = -\fm \log(y) .
\eqno(3.49)
\en
Note that due to Eq.~(3.48), taking the derivative with respect to $\sigma$ effectively
reduces the order in $\delta$ of terms involving $\gp$ by one. Thus we can only compute
derivatives with respect to $\sigma$ of Eq.~(3.43) to $O(\delta)$.

Let us rewrite Eq.~(3.43) as 

\eq
\lx \log \cZ \rx = A(q, \sigma) \log n + B(q,\sigma) (1 - n^{-2\lambda \lxc \fm \rxc}) + C(q,\sigma)
+ \cdots ,
\eqno(3.50)
\en
where the coefficients $A(q,\sigma)$, $B(q,\sigma)$, and $C(q,\sigma)$ can be read from Eq.~(3.43),
and the coefficient
$C=a_o
\lxc
\psi
\rxc$ is the only non-universal one. Taking the derivatives with respect to
$\sigma$, we see after some computation that indeed

\eq
{d A(q, \sigma) \over d \sigma} = 1 + O(\delta^2) ,
\eqno(3.51)
\en
and
\eq
{d B(q, \sigma) \over d \sigma} = O (\delta^2) .
\eqno(3.52)
\en

Thus Eqs.~(3.46-7) are satisfied provided that the non-universal constant $a_0 = 0$. It seems that
we have succeeded in fixing this quantity. However, we remind the reader that many cut-off
procedures will effectively break the invariance expressed by Eqs.~(3.45-47) by introducing a
dependence on the number of particles in an elementary branch $n_e$, so the result
$a_0 = 0$ should not be taken too seriously.

\bigskip
\sect{\bf 4. Interpretation of results}
\medskip

In section 3, we re-summed the leading and sub-leading divergent series to all orders in perturbation
theory. We found a remarkable result. The resummed series consists of three types of terms. In
the first place, there are terms diverging logarithmically with $n$. These terms are all universal
in form. In the second place, there are constant terms--at sub-leading order these include
non-universal constants. Finally, there are power law corrections to $\lx \log \cZ \rx$,  which
involve universal coefficients multiplying $n^{-\Delta}$, with

\eq
\Delta = 2 \lambda \lxc \fm \rxc ,
\eqno(4.1)
\en
a positive quantity. Thus as $n \to \infty$ we expect only the terms $\propto
\log n$ to be relevant in determining the quenched dimensions $\sqq$. 
Recalling the coefficient of this term from Eq.~(3.43), we conclude that

\eq\eqalign{
\lxc \psi (q, \sqq) \rxc \bigg  \{ 1 +   \lambda  \big [ \lxc \gp(q, \sqq) \log(1-y)
\rxc  & + \lxc \fm(q, \sqq) \log y \rxc  \big ]  \cr & + \lambda^{\prime} \lxc
\psi(q,\sqq) \rxc  +O(\delta^2) \bigg \} =0 , \cr}
\eqno(4.2)
\en
which is equivalent to $\lxc \psi (q, \sqq) \rxc = 0$. (Setting the other factor
equal to zero does not lead to an acceptable solution.)\footnote{$^3$} {Numerical evaluation of the
second factor on the right-hand side of Eq.~(4.2) for, e.g., model Z (see section 1 or
Eqs.~(4.3-4)) quickly leads one to the conclusion that setting this factor equal to zero does not
lead to a physical solution for
$\sqq$, or, indeed, to any solution for some values of
$q$.} But this is {\it precisely the same} as the criterion that determines the annealed
multifractal dimensions
$\sqa$! This is unsurprising, as the $\log n$ term in Eq.~(3.43) resulted only from products of
single averages of $\lx \cZ -1 \rx$, as discussed in section 3 or in appendix D. This implies
that the true $n \to \infty$ quenched multifractal dimensions are identical to the
annealed dimensions, negative values of $f$ and all.

However, this conclusion is based upon an overly naive
interpretation of Eq.~(3.43), overlooking crossover effects, which we now discuss. Consider again
the power law term
$\propto n^{-\Delta }$, with
$\Delta$ given by Eq.~(4.1). We can understand the role played by this term by considering model Z,
in which the unstable manifold for branch competition in the $x-y$ plane is simply a straight line
emanating from the unstable fixed point at $(1/2,1/2)$, which turns vertical upon reaching the
$x=0$ or $x=1$ lines (see Figure 3). As a function of $\eta$, we can write$^{(11)}$

\eq
x = \cases{ {1 \over 2} (1 - \eta^{\nu}), & $\eta < 1$ ; \cr 0, & $\eta \ge 1$ . \cr}
\eqno(4.3)
\en
and
\eq
y = \cases{ {1 \over 2} \{ 1 - [\eta^{\nu} / (1 + \nu)]\}, & $\eta < 1$ ; \cr {\bar \eta / \eta}, & $\eta \ge 1$ . \cr}
\eqno(4.4)
\en
with $\bar \eta = [\nu / 2(1 + \nu)]$.

Recalling the definition of $\lxc \fm \rxc$,

\eq
\lxc \fm \rxc = \int_0^{\infty} d \eta \; \eta^{\nu-1} {x^q(\eta) \over y^{\sigma} (\eta)}
\eqno(4.5)
\en
we see that in general, $\lxc \fm \rxc$ is a function of both $q$ and $\sigma$.  This presents us with a
paradox, because any exponent $\Delta$ of a power law correction to $\lx \log \cZ
\rx$ must be independent of $\sigma$, due to the invariances expressed by Eqs.~(3.45-47).  We resolve
this paradox by realizing that Eq.~(4.1) is merely the lowest order in $\delta$ expression for the
exponent $\Delta$; we should write (4.1) more precisely as

\eq
\Delta (q) = 2 \lambda \lxc \fm \rxc + O(\delta^2)
\eqno(4.6)
\en
where $\Delta$ is now the exact exponent of the leading power law correction to $\lx \log \cZ
\rx$. The higher order corrections on the right-hand side of Eq.~(4.6) will cancel the dependence
upon $\sigma$ arising from $\lxc \fm \rxc$.

Let us first compute
compute $\lxc \fm \rxc$ in model Z for $\sigma=0$,

\eq
\lxc \fm(q, \sigma=0) \rxc  =
 {1 \over \nu( q+1) }  \left ({1 \over 2}\right )^q ,
\eqno(4.7)
\en 
obtaining

\eq
\Delta (q) = {2 \lambda  \over
\nu(q+1)} \left ({1 \over 2} \right )^q + O(\delta^2),
\eqno(4.8)
\en
which goes to zero exponentially in $q$ for large $q$. 

Now consider $\lx \log \cZ \rx$ from
Eq.~(3.43). For convenience, we set the non-universal constant $a_0 = 0$. If $\lim_{q
\to \infty} \Delta(q) =0$, we see that (returning to the case of general $\sigma$),

\eq\eqalign{
\lim_{q \to \infty} \lim_{n \to \infty} {\lx \log \cZ \rx_{\ell + s \ell} \over \log n}= 
\lim_{q
\to
\infty} 
\lambda \lxc \psi \rxc\Big [ 1 + \lambda  ( \lxc \gp \log(1-y)
\rxc &+ \lxc \fm \log y \rxc  ) \cr & + \lambda^{\prime} \lxc \psi \rxc
\Big ] 
  , \cr}
\eqno(4.9)
\en
while
\eq\eqalign{
\lim_{n \to \infty} \lim_{q \to \infty} {\lx \log \cZ \rx_{\ell + s \ell} \over \log n} =  & 
\lim_{q \to \infty} 
\bigg \{ \lambda \lxc \psi \rxc   \Big [ 1 + \lambda \left ( \lxc \gp \log(1-y)
\rxc + \lxc \fm \log y \rxc \right ) + \lambda^{\prime} \lxc \psi \rxc \Big ]
\cr& -  \lambda \Big [ {1 \over 2} \lxc \psi ^2 \rxc  + \lambda \lxc \psi
\rxc 
\lxc \psi \log (1-y) \rxc +\lambda^{\prime} \lxc \psi \rxc^2 \Big ]  \bigg \}\cr
=& \lim_{q \to \infty} \left [\lambda \lxc \psi \rxc     + \lambda^2 \lxc \psi \rxc  \lxc \fm \log
({y
\over 1-y})
\rxc - {\lambda  \over 2} \lxc \psi ^2 \rxc \right ], \cr}
\eqno(4.10)
\en
because
\eq
\lim_{n \to \infty} \lim_{q \to \infty} (1-n^{-\Delta(q)}) = 2 \lambda \lxc \fm \rxc \log n
+O(\delta^2). \eqno(4.11)
\en
Thus, for large values of $q$, the {\it apparent} quenched dimensions, which we term
$\sqap$, will be determined by

\eq 
\lambda \lxc \psi \rxc     + \lambda^2 \lxc \psi \rxc  \lxc \fm \log ({y \over 1-y})
\rxc - {\lambda  \over 2} \lxc \psi ^2 \rxc   +
O(\delta^3) = 0 .
\eqno(4.12)
\en
These apparent dimensions are the ones that will be seen, e.g., in numerical studies.
Note that this is simply the naive (non-resummed) result for the dimensions $\sqq$ which
cam be obtained from $\lx \log \cZ \rx = \lx 
\log \cZ \rx_1 + \lx \log \cZ \rx_2  + O(\delta^3)$ (see Eqs.~(2.20) and (2.34)). 

Of course, for sufficiently large $n$, we will always see a crossover back
to the true quenched dimensions. The crossover value of $n$ for which this occurs, $n_c (q)$,
will be determined by

\eq
\Delta(q) \log n_c (q) = \left [ 2 \lambda \lxc \fm \rxc  + O(\delta^2) \right ] \log n_c(q) = 1
,
\eqno(4.13)
\en
or, from Eq.~(4.7), 

\eq
n_c(q) = \exp \Big [  {\nu (q+1) 2^q \over 2 \lambda  }  + O(\delta^0)\Big ] .
\eqno(4.14)
\en
(Note that the first term in the exponential in Eq.~(4.14) is formally $O(\delta^{-1})$.)
Thus, in practice, the true quenched dimensions will not be observable, even for
moderate values of $q$, except for monstrously large systems.
Although we have taken an explicit form for $\Delta$ appropriate for model Z, we
expect the same qualitative results to hold for any branched growth model, and in
particular, for DLA.

Suppose that we had not set $\sigma=0$ at the outset of this calculation. It is simple to compute
$\lxc \fm (q, \sigma) \rxc$ for model Z,

\eq
\lxc \fm (q , \sigma) \rxc =  {1 \over \nu} \left ( {1 \over 2} \right)^{q-\sigma} \left (1 +{1 \over
\nu}\right )^{\sigma} \int_0^1 du \; u^q \left (1 + {u \over \nu} \right)^{- \sigma} 
\eqno(4.15)
\en
and
\eq
\lim_{q \to \infty} \lxc \fm (q , \sigma) \rxc = {1 \over \nu (q+1)} \left ( {1 \over 2}
\right)^{q-\sigma} 
\eqno(4.16)
\en
which is a simple modification of our previous result, Eq.~(4.7). For fixed $\sigma$, Eq.~(4.16)
leads to the same super-exponential dependence of $n_c$ on $q$ displayed in Eq.~(4.14). We expect
the annealed
$\sqa$ to approach a constant as
$q
\to
\infty$, while
$\sqap \to
\beta q, 
\beta \approx 1- \nu < 1$ for $q \to \infty$.$^{(11)}$ It is not, however, legitimate to introduce
these functions $\sigma(q)$ into Eq.~(4.16), because once $\sigma$ becomes large, we must properly
include the effects of the next order term in $\delta$ in Eq.~(4.14), which will offset this
spurious $\sigma$-dependence.

\bigskip \noindent
{\bf Numerical results}
\medskip
In addition to the analytical computation of $\lx \log \cZ \rx$, to which the bulk of
this work is devoted, we have also computed this quantity, for the random branched
growth model, by a Monte Carlo method. For these purposes, it is convenient to use
model Z, which can be defined (see Figure 3) by writing $x$ and $y$ as functions of
$\eta$ as in Eqs.~(4.3-4). The curve in the $x-y$ plane parameterized by
these functions (and their complements $1-x$, $1-y$) is precisely the unstable manifold
shown in Figure 3.

We must also fix the function $\rho( \epsilon )$. We expect our results to be
sensitive only to the $\epsilon \ll 1$ form of $\rho(\epsilon) \propto \rho_0
\epsilon^{\nu-1}$. It is thus convenient to choose

\eq
\rho(\epsilon) = {\epsilon^{\nu-1} e^{-\epsilon} \over \Gamma (\nu)} ,
\eqno(4.17)
\en
as a suitably normalized form for $\rho(\epsilon)$.

Our procedure is now quite simple. Starting at the first branch point, at the root of the tree, we fix
$n$, the total number of particles in the tree. We then choose a value of $\epsilon$ with the
distribution given by Eq.~(4.17), and compute the values of $x$ and $y$ at that branch point
from Eqs.~(4.3-4). We then repeat the process on each of the two sub-branches, and so
forth. If the total probability on a branch is zero (as will be the case for values of $\eta
\ge 1$ on the weaker branch in model Z), then it will not contribute to $\sum_i p_i^q$, and
it can be disregarded. If the total mass of a branch is less than one, then we stop the subdivision
process. We then form the sum either of the probability moments of these ends, which correspond
to the elementary sub-branches (indexed by
$i$),

\eq
Z(q) = \sum_i p_i^q ,
\eqno(4.18)
\en
or else the sum weighted by a simple function of the mass of each elementary sub-branch

\eq
Z^{\prime}(q) = \sum_i { p_i^q \over n_i^{q-1} } .
\eqno(4.19)
\en
We have used the factor $n_i^{1-q}$ on the right-hand side of Eq.~(4.19) because $\sqa = \sqq =
q-1$ for a non-fractal cluster, and the elementary sub-branches are presumably not fractals.
Numerical results for the scaling behavior of these two versions of the partition function were, for
practical purposes, identical.  We can now determine the ``apparent" dimensions $\sqap$ by studying
the scaling of $\lx
\log Z (q)
\rx$ with
$\log n$, where
$\lx
\cdots \rx$ indicates the average over the statistically independent cluster realizations.

Although we performed simulations for a variety of values of $0<\nu<1$, we shall display our results
only for $\nu = 0.6$ (results for other values of $\nu$ are qualitatively similar). We used values of
$n$ varying from $n=10$ to $n=2560$, varying by powers of two, and we averaged over
$10^4$ realizations at each size.  Figure 21 displays $\lx \log Z \rx$ vs. $\log n$ for two
values of $q$. The linear scaling is quite evident in these plots. Extracting the slope of
the linear part, we find $\sqap$ as displayed in Figure 22, with its corresponding $\fa$
(defined as the Legendre transform of $\sqap$, not of $\tau(q)$--see section 1 above
for a discussion of the relation between these two quantities). Note that the $\fa$ corresponding to
the numerical quenched dimensions does not have any negative values of $f$ appearing.

We can implicitly solve Eqs.~(4.2) or (4.12) above in order to find $\sqa$ or the apparent
quenched dimensions $\sqap$. These are displayed with the numerical results in Figure
23. For comparison, we also show the one-propagator result from Eq.~(2.37). It is clear that
the annealed dimensions agree with the numerical results only for $q < 2$, while the
apparent quenched dimensions from Eq.~(4.12) agree over the larger range $q < 5$. From
Eq.~(4.14) for  $n_c (q)$, we already expect that for $q=3$ the crossover from apparent
quenched to annealed dimensions (for model Z with $\nu = 0.6$) in the numerical results
will occur only for $n \gg 10^5$. Since these dimensions are essentially identical for $q
<3$, it is not feasible to see this crossover in the numerical results.

It is possible to understand the fact that for $q$ not too large, the $\sqap \approx
\sqa$, by recasting the expansion in the formal parameter $\delta$ as an
expansion in $q-1$. In this way, one can show that $\sqap = \sqa + O((q-1)^2)$.  This
is discussed in detail in appendix E.  \bigskip
\noindent{\bf 5. Conclusions}
\medskip

We now summarize the main results of this work, which concern the partition function $\cZ(q,\sigma;
n)$ defined in Eq~(2.2).
\smallskip

\item{1.} The multifractal dimensions for the branched growth model may be
defined either by annealed averaging, for which $\lim_{n \to \infty}\lx \cZ (q, \sqa) \rx = 0$, or by
quenched averaging, $\lim_{n \to \infty} \lx \log \cZ (q, \sqq) \rx = 0$.

\item{2.} It is possible to compute $\lx \log \cZ \rx$ perturbatively by
expanding in a formal parameter $\delta$, which counts branching vertices in a
diagrammatic approach. The result is a formal expression

\eq
\lx \log \cZ \rx = \sum_{N=1}^{\infty} \delta^N \lx \log \cZ \rx_N =
\sum_{N=1}^{\infty}\sum_{M=0}^N \beta_{N,M} \log^M n ,
\eqno(5.1)
\en
\item{} where the parameter $\delta =1$ in the physical case and
the coefficients $\{ \beta_{N,M} \}$ can be determined from the unstable
manifold for branch competition in the underlying model.

\item{3.} It is possible to resum the leading $M=N$ and sub-leading $M=N-1$ terms at
all orders $N$ in Eq.~(5.1). The result is an expression for $\lx \log \cZ \rx$ which
combines a logarithmic term in $n$ with a constant and a decaying power law:

\eq
\lx \log \Zqsn \rx = \lxc \psi(q,\sigma) \rxc \Gamma_0 (q, \sigma) \log n + \Gamma_1 (q, \sigma)
(1 - n^{-\Delta}) +  \cdots ,
\eqno(5.2)
\en
\item{} with $\Delta (q) > 0$. 

\item{4.} From Eq.~(5.2), we see that in the limit $n \to \infty$, $\sqq$ is fixed by
requiring $\lxc \psi \rxc =0$, which is also the criterion that fixes $\sqa$. Thus,
in the limit of large $n$, $\sqq = \sqa$.
This limit is attained only for $n \gg n_c = \exp(a \exp b q)$. For $n \ll
n_c$, the apparent multifractal dimensions $\sqap \ne \sqa$. Furthermore, there is
numerical evidence that, unlike the annealed dimensions, these apparent dimensions do
not exhibit negative values of $\fa$.

\item{5.} For appreciable values of $q-1$, $n_c$ is enormous, so that the true
quenched (i.e., annealed) dimensions are virtually unobservable. For $q-1 \sim 1$, the
true quenched dimensions will be observed; but in this case $\sqap \to \sqa$ anyway.
\smallskip

The multifractality of these models is of a quite interesting kind. The
average of the partition function displays true scaling over the entire range of $q$.
The average of its logarithm, on the other hand, contains regions above and below $n_c$ with
different scaling properties. As a function of q, there is a weak function $q_c (n)$
such that for $q< q_c$ and $q > q_c$ essentially different scaling properties are
being explored. Since the essence of multifractality is the smooth variation of
scaling properties with $q$, this represents, in a subtle sense, a failure of
multifractality. 

It is therefore most accurate to say that these branched growth models are multifractal
only as an ensemble; typical members of the ensemble exhibit this weak deviation from
multifractality. Since the branched growth model seems to be a quite adequate theory,
both qualitatively and quantitatively, for diffusion-limited aggregation, we expect
these quite novel properties to hold also for DLA.

\bigskip
\noindent{\bf Acknowledgements}

We are grateful to A. Coniglio and M. Leibig for stimulating discussions. T.C. Halsey
would like to thank the Dipartimento di Fisica, Universit\` a di Napoli, and the
Institute for Theoretical Physics, Santa Barbara for their hospitality. He is
also grateful for the support of the National Science Foundation, through a Presidential
Young Investigator Grant, DMR-9057156, and through grant PHY89-04035, as well as
for the support of an Alfred P. Sloan junior fellowship. K. Honda would like to thank the
Ministry of Education, Science, and Culture of Japan for a travelling grant. B. Duplantier
is grateful for the support of the CNRS through the CNRS-NSF France-U.S. Binational Program.

\vfill \eject \bigskip \noindent{\bf Appendix A: ``Propagator" sums} \medskip

It was claimed above in section 2 that

\eq
\lx \sum_{j \ge 1} { \rho_0 \log^{N-1} n_j \over n_j^{\nu} } \rx =
\sum_{M=0}^{N} \lambda_{N,M} \log^M n ,
\eqno(A.1)
\en
where the sum over $j$ is down a ``main line" with a total of $n$ particles. First we repeat the
computation of $\lambda_{1,1}$. From Eq.~(A.1)

\eq
\lx \sum_{j \ge 1} {\rho_0 \over n_j^{\nu} } \rx =
 \lambda_{1,1} \log n + \lambda_{1,0} .
\eqno(A.2)
\en
Separating out the first term in the sum, we have

\eq
\lambda_{1,1} \log n + \lambda_{1,0} = {\rho_0 \over n^{\nu} } +  \lx  
 \lambda_{1,1} \log
[(1-y ) n ] + \lambda_{1,0} \rx ,
\eqno(A.3)
\en
which simplifies to

\eq
{\rho_0 \over n^{\nu} } +  \lx  
 \lambda_{1,1} \log
(1-y )\rx = 0 ,
\eqno(A.4)
\en
or (recalling Eqs.~(2.7), (2.8))
\eq
\lambda_{1,1} \equiv \lambda = - {1 \over \lxc \log(1-y) \rxc} ,
\eqno(A.5)
\en
The factor of $\rho_0$ has cancelled against that coming
from the integral, so that $\lambda_{1,1}$ is independent of $\rho_0$, as are all of the
$\{\lambda_{N,M}\}$. The constant $\lambda_{1,0}$ (elsewhere called $a_0$) is not
determined by this argument; in fact we expect it to depend on the specific cut-off
procedure in the theory, and thus be non-universal.

To compute the remaining $\{ \lambda_{N,M} \}$, we simply generalize the above argument.

\eq\eqalign{
\lx \sum_{j\ge 1}{\rho_0 \log^{N-1} n_j \over n_j^{\nu} } \rx &=
 \sum_{M=0}^{N} \lambda_{N,M} \log^M n \cr
&= {\rho_0 \log^{N-1} n \over n^{\nu} } +  \lx \sum_{M=0}^{N} \lambda_{N,M}
\log^M [(1-y) n ] \rx \cr
&= {\rho_0 \log^{N-1} n \over n^{\nu} } + \sum_{M=0}^{N} \lambda_{N,M}
\sum_{L=0}^M \left ( { M \atop L } \right ) \log^{L} n \; \lx \log^{M-L} (1-y ) \rx  ,
\cr
}
\eqno(A.6) 
\en
leading immediately to

\eq
\log^{N-1} n  + \sum_{M=1}^{N} \lambda_{N,M} \sum_{L=0}^{M-1} \left (
{ M \atop L } \right ) \log^{L} n \; \lxc \log^{M-L} (1-y ) \rxc 
= 0 .
\eqno(A.7)
\en
The left hand side of Eq.~(A.7) is a sum of terms of the form $a_K \log ^K n$, with $ 0 \le K
\le N-1$. Setting each of the coefficients $a_K$ equal to zero yields $N$
simultaneous linear equations for the $N$ quantities $\{ \lambda_{N,M} , M \ge 1 \}$. The solution
of these equations then gives $\{ \lambda_{N,M} , M \ge 1 \}$. The non-universal coefficients $\{
\lambda_{N,0} \}$ do not appear in Eq.~(A.7), and are left un-determined by this procedure.

As an example, consider $\lambda_{N,N}$. These may be obtained from the coefficient
of the $\log^{N-1} n$ term in Eq.~(A.7),

\eq
\lambda_{N,N} =- { 1 \over N \lxc \log (1-y ) \rxc } = {\lambda_{1,1} \over N} \equiv
{\lambda \over N} .
\eqno(A.8)
\en
It is also simple to determine the $\{ \lambda_{N,N-1} \}$. The coefficient of the $\log^{N-2} n$
term in Eq.~(A.7) is

\eq
\lambda_{N,N-1} \left ( N - 1 \atop N-2 \right ) \lxc \log (1-y) \rxc + \lambda_{N,N} \left ( N \atop
N-2 \right )  \lxc \log^2 (1-y) \rxc = 0 ,
\eqno(A.9)
\en
giving

\eq
\lambda_{N,N-1} = - \lambda_{N,N} {N \over 2} {\lxc \log^2 (1-y) \rxc \over \lxc \log (1-y) \rxc}
= {1 \over 2} {\lxc \log^2 (1-y) \rxc \over \lxc \log (1-y) \rxc^2} \equiv
\lambda^{\prime}. \eqno(A.10)
\en

The next coefficient, that for the $\log^{N-3} n$ term from Eq.~(A.7), gives

\eq
\lambda_{N,N-2} =  (N-1) \left \{ {1 \over 6} {\lxc \log^3 (1-y) \rxc \over \lxc \log (1-y)
\rxc^2} - {1 \over 4} {\lxc \log^2 (1-y) \rxc^2 \over \lxc \log (1-y) \rxc^3} \right \} .
\eqno(A.11)
\en
Clearly, one may iterate to generate any of the $\lambda_{N,M}$ that one needs.

\bigskip \noindent{\bf Appendix B: Series in $\delta$ to fourth order} \medskip

In section 2 above, we demonstrated the rules for computing $\lx \log \cZ \rx$ order by
order in $\delta$, and computed the terms of $O(\delta)$ and $O(\delta^2)$. The diagrams
corresponding to these terms are displayed in Figure 24, the results are

\eq
\lx \log \cZ \rx_1 = (\lambda \log n + a_0 ) \lxc \psi \rxc ,
\eqno(B.1)
\en
and
\eq
\lx \log \cZ \rx_2 = ( \lambda \log n + a_0 ) \left [ \lambda \lxc \psi \rxc \lxc \fm \log
 ( {y \over 1-y } ) \rxc -{1 \over 2} \lxc \psi^2 \rxc
\right ] .
\eqno(B.2)
\en
Recall that in our more formal notation, $\lambda \equiv \lambda_{1,1}$ and the non-universal
constant $a_0 \equiv \lambda_{1,0}$. In the expressions below, we will find it convenient to mix
these two notations, as well as to use $\lambda^{\prime} \equiv \lambda_{2,1}$.

At third order in $\delta$, the diagrams appearing are shown in Figure 25.  The attentive
reader will recall that each diagram will in general involve a number of terms, due to the
different ways in which propagator terms can be averaged with vertices $\fm$ or $\gp$. A
lengthy computation yields the following result for $\lx \log \cZ \rx_3$:

\eq\eqalign{
\lx \log \cZ \rx_3 = & ( \lambda_{2,2} \log^2 n + \lambda_{2,1} \log n + \lambda_{2,0} )
 \cr & \quad \cdot
\left [ 2 \lambda \lxc \fm \rxc \lxc \psi \rxc \left (
\lambda \lxc \psi \log(1-y) \rxc + \lambda^{\prime} \lxc \psi \rxc \right ) + \lambda \lxc \fm
\rxc \lxc \psi^2 \rxc
\right ] \cr & + (\lambda_{1,1} \log n + \lambda_{1,0}) \cr &\quad \cdot \Big \{  \lambda^2 \lxc
\psi \rxc \Big ( {1 \over 2} \lxc \psi \rxc \lxc \fm \log^2 ({y \over 1-y} ) \rxc + \lxc \fm
\log ({y \over 1-y} ) \rxc^2 \cr&\quad\quad + \lxc \fm \log [ y(1-y)] \rxc \lxc \psi \log (1-y) \rxc
\Big )
\cr &\quad + \lambda \lambda^{\prime} \lxc \psi \rxc^2 \lxc \fm \log [ y (1-y) ]\rxc  +
\lambda \lxc \psi^2 \rxc \lxc \fm \log(1-y) \rxc \cr &
\quad\quad  - \lambda \lxc \psi \rxc \lxc \fm \psi \log({y
\over 1-y}) \rxc  + {1 \over 3} \lxc \psi^3 \rxc \Big \} \cr &
+ (\lambda_{1,1} \log n + \lambda_{1,0}) \Big \{ \lxc \psi \rxc [ 2 \lambda a_0 \lxc \fm \rxc \lxc
\psi
\log(1-y) \rxc \cr &\quad + 2 \lambda \lambda_{2,0} \lxc \fm \rxc \lxc \psi \rxc + a_0^2 \lxc \fm
\rxc
\lxc
\psi \rxc ] + a_0 \lxc \fm \rxc \lxc \psi^2 \rxc \Big \}   
,} 
\eqno(B.3)
\en
where the reader will recall that $\lambda_{2,2} = \lambda/2$, the other propagator coefficients
being given after Eq.~(B.2).

The non-universal terms at this order are collected at the end of the right-hand side of
Eq.~(B.2). Unlike the second order in $\delta$, at $O(\delta^3)$ there are non-universal terms
proportional to
$\log n$ appearing. It can be shown by explicit computation that up to $O(\delta^4)$,
all such terms come from decaying power laws in $n$, which in addition vanish when
$\lxc \fm \rxc \to 0$. Thus these non-universal terms survive neither in the limit $n
\to \infty$ nor in the limit $q \to \infty$.

We have also performed the $O(\delta^4)$ computation; we do not show the diagrams. In Eq.~(B.4)
below, we organize the result into four parts: (universal) terms multiplying the third-order
propagator $
\lambda_{3,3}
\log^3 n + \lambda_{3,2} \log^2 n + \lambda_{3,1} \log n +
\lambda_{3,0}$, universal terms multiplying the second-order propagator $\lambda_{2,2} \log^2 n +
\lambda_{2,1} \log n + \lambda_{2,0}$, non-universal terms multiplying this propagator, and terms
both universal and non-universal multiplying the first order propagator $\lambda_{1,1} \log n +
\lambda_{1,0}$, which we omit. The universal results allow the reader to check our re-summation in
section 3. The non-universal terms allow the reader to check our assertion that such terms resum to
simple decaying power laws in
$n$.

\eq\eqalign{
\lx \log \cZ \rx_4 = & ( \lambda_{3,3} \log^3 n + \lambda_{3,2} \log^2 n + \lambda_{3,1} \log n +
\lambda_{3,0} )
 \cr& \quad\cdot \Big [ -2 \lambda^2 \lxc \fm \rxc^2 \lxc \psi \rxc ( \lambda \lxc \psi \log (1-y)
\rxc +
\lambda^{\prime} \lxc \psi \rxc ) - \lambda^2 \lxc \fm \rxc^2 \lxc \psi^2 \rxc \Big ] \cr & + 
(\lambda_{2,2} \log^2 n + \lambda_{2,1} \log n + \lambda_{2,0}) \cr &\quad \cdot
\Bigg \{   2 \lambda^3  \lxc \psi \rxc \Big [ \lxc \fm \rxc \lxc \psi \rxc \lxc \fm \log(1-y)
\log {y
\over 1-y} \rxc \cr &\quad\quad+ \lxc \psi \rxc \lxc \fm \log {y \over 1-y} \rxc \lxc \psi \log(1-y)
\rxc
\cr &\quad -5 \lxc \fm \rxc \lxc \fm \log(1-y) \rxc \lxc \psi \log (1-y) \rxc  \cr&\quad\quad+ \lxc
\fm
\rxc
\lxc
\fm
\log y
\rxc \lxc \psi \log(1-y) \rxc \Big ] \cr & \quad + \lambda^2 \lambda_{3,2} \lxc \psi \rxc^3 \lxc \fm
\log{y \over 1-y} \rxc \cr & \quad\quad+ \lambda^2 \lambda^{\prime} \lxc \psi \rxc \Big [ \lxc \psi
\rxc^2 
\lxc \fm \log {y \over 1-y} \rxc \cr & \quad + 4 \lxc \fm \rxc \lxc \psi \rxc \lxc \fm \log {y \over
1-y}
\rxc - 4 \lxc \fm \rxc^2 \lxc \psi \log(1-y) \rxc \cr & \quad\quad - 8 \lxc \fm \rxc \lxc \psi \rxc
\lxc
\fm
\log(1-y) \rxc \Big ] -4 \lambda \lambda^{\prime \; 2} \lxc \fm\rxc^2 \lxc \psi \rxc^2  \cr&\quad
+  \lambda^2 \Big [ 2 \lxc \fm \rxc \lxc \psi \rxc \lxc \fm \psi \log{y \over 1-y} \rxc
\cr &\quad\quad - \lxc \fm \rxc \lxc \psi \rxc \lxc \psi^2 \log(1-y) \rxc  + \lxc \psi \rxc \lxc
\psi^2 \rxc \lxc
\fm \log{ y \over 1-y} \rxc \cr & \quad -\lxc \psi \log(1-y)
\rxc \Big ( \lxc \fm \rxc \lxc \psi^2 \rxc + 2 \lxc \psi \rxc \lxc \fm \psi \rxc + \lxc \psi \rxc
\lxc \fm^2 \rxc \Big ) 
 \cr & \quad\quad- 4 \lxc \fm \rxc \lxc \psi^2 \rxc \lxc \fm \log(1-y) \rxc \Big
]
\cr & \quad
 - \lambda \lambda^{\prime} \Big ( 2 \lxc \fm \rxc \lxc \psi \rxc \lxc \psi^2 \rxc + 2 \lxc \fm
\rxc^2 \lxc \psi^2 \rxc \cr & \quad\quad + 2\lxc \psi \rxc^2 \lxc \fm \gp \rxc + 3 \lxc \psi \rxc^2
\lxc
\fm^2
\rxc \Big ) \cr & \quad- \lambda \Big ( \lxc \fm \gp \rxc \lxc \psi^2 \rxc + {3 \over 2} \lxc
\fm^2
\rxc \lxc \psi^2 \rxc + \lxc \fm \rxc \lxc \psi^3 \rxc \Big ) \Bigg \} \cr & + 
( \lambda_{2,2}
\log^2 n + \lambda_{2,1} \log n + \lambda_{2,0} ) \cr & \quad \cdot  \Big [ -4 \lambda^2 a_0 \lxc
\fm \rxc^2 \lxc \psi \rxc \lxc \psi \log(1-y) \rxc - 2 \lambda a_0^2 \lxc \fm \rxc^2 \lxc \psi
\rxc^2 \cr & \quad \quad- 4 \lambda^2 \lambda_{2,0} \lxc \fm \rxc^2 \lxc \psi \rxc^2   -2 \lambda a_0
\lxc
\fm
\rxc^2 \lxc \psi^2 \rxc \Big ]  \cr & + (\lambda_{1,1} \log n + \lambda_{1,0} ) \cdot ( \cdots )
 .}
\eqno(B.4)
\en

\bigskip \noindent{\bf Appendix C: Some useful sums} \medskip

Our starting point is the standard exponential sum:

\eq
\sum_{L=0}^{\infty} { x^L \over L!} = \exp(x) .
\eqno(C.1)
\en
and a modified version thereof

\eq
\sum_{L=0}^{\infty} \sum_{l=0}^L \left (L \atop l \right )  { x^L
\over L! } = \exp(2x) .
\eqno(C.2)
\en
Now consider the sum appearing in Eq.~(3.10), which has the form

\eq
{\cal S}^{\prime} (x,y) = \sum_{l_1=0}^{\infty} \sum_{l_2 = 1}^{\infty} \sum_{l=0}^{l_2}
\sum_{m=0}^{\infty} \left ( {L-l \atop l_1} \right ) {(m + l)!
\over m! \; l! } {1 \over (m + L + 1)!} x^{m} y^L .
\eqno(C.3)
\en
where $L=l_1+l_2$. It is convenient to add and subtract the term $l_2=0$ in Eq.~(C.3), and
set

\eq
{\cal S}^{\prime} (x,y) = {\cal S} (x,y) - {\cal S}_0(x,y)
\eqno(C.4)
\en
with
\eq
{\cal S}(x,y) = \sum_{L=0}^{\infty} \sum_{l_1 = 0}^L \sum_{l=0}^{L-l_1}
\sum_{m=0}^{\infty} \left ( {L-l \atop l_1} \right ) {(m + l)!
\over m! \; l! } {1 \over (m + L + 1)!} x^{m} y^L .
\eqno(C.5)
\en
and

\eq
{\cal S}_0(x,y) = \sum_{m=0}^{\infty} \sum_{l_1=0}^{\infty} {1 \over ( m + l_1 + 1)!}
 x^{m} y^{l_1}
\eqno(C.6)
\en

In order to compute ${\cal S}(x,y)$ from Eq.~(C.5) we first
alter the order of summation

\eq
\sum_{l_1=0}^L \sum_{l=0}^{L-l_1} \to \sum_{l=0}^L
\sum_{l_1=0}^{L-l},
\eqno(C.7)
\en
leading immediately to

\eq
{\cal S}(x,y) = \sum_{L=0}^{\infty} \sum_{l = 0}^L 
\sum_{m=0}^{\infty} 2^{-l}{(m + l)!
\over m! \; l! } {1 \over (m + L + 1)!} x^{m} (2y)^L .
\eqno(C.8)
\en
To perform the remaining sums, we change variables to $N=m+L$,
$n=m+l$, and rewrite the sums as

\eq
\sum_{L=0}^{\infty} \sum_{l = 0}^L 
\sum_{m=0}^{\infty} \to \sum_{N=0}^{\infty} \sum_{n = 0}^N 
\sum_{m=0}^{n} .
\eqno(C.9)
\en
We now have

\eq
{\cal S}(x,y) = \sum_{N=0}^{\infty} \sum_{n = 0}^N 
\sum_{m=0}^{n} \left ( {n \atop m} \right ) {1 \over (N+1)! }
2^{m-n} x^m (2y)^{N-m} .
\eqno(C.10)
\en
The remaining sums are now elementary. Summing over $m$, we
obtain

\eq
{\cal S}(x,y) = \sum_{N=0}^{\infty} \sum_{n = 0}^N {1 \over (N+1)!}
\left ( 1 + {x \over y} \right )^n 2^{-n} (2y)^N ,
\eqno(C.11)
\en
and summing over $n$,

\eq
{\cal S}(x,y) = \sum_{N=0}^{\infty} {(2y)^N \over (N+1)!} {
\left ( 1 - \left [ {1 \over 2} \left (1 + { x \over y} \right ) \right
]^{N+1} \right ) \over \left ( 1 - {1 \over 2} \left (1 + { x \over y}
\right )  \right )} .
\eqno(C.12)
\en
Recognizing that

\eq
\sum_{N=0}^{\infty} {x^N \over (N+1)!} = {\exp(x)-1 \over x} ,
\eqno(C.13)
\en
we obtain our final result:

\eq\eqalign{
{\cal S}(x,y) &=  {1 
\over  
\left ( 1 - {1 \over 2} \left (1 + { x \over y}
\right )  \right )}
 \left \{ {\exp(2y) - 1 \over 2y} 
- {1 \over 2} \left (1 + { x \over y}
\right )  
{\exp \left (2y \cdot {1 \over 2} \left (1 + { x
\over y} \right )  \right ) -1 \over 2y \cdot {1 \over 2} \left
(1 + { x \over y} \right )  } \right \} \cr
&= e^y \left ( {e^y-e^x \over y-x}  \right )  .\cr} \eqno(C.14)
\en
A similar but simpler algebra yields the sum ${\cal S}_0(x,y)$ from Eq.~(C.6),

\eq
{\cal S}_0 (x,y) = {e^y - e^x \over y-x}
\eqno(C.15)
\en
so that

\eq
{\cal S}^{\prime} (x,y) = (e^y - 1) \left ( {e^y - e^x \over y-x} \right )
\eqno(C.16)
\en

This result illustrates the necessary tricks for performing the types of sums appearing in this
paper.  To give a further example, the sum appearing in Eq.~(3.9) is of the form

\eq
{\cal S}^{\prime \prime} (x,y) =
\sum_{l_1=0}^{\infty}
\sum_{l_2=0}^{\infty}
\sum_{m_1=0}^{\infty}
\sum_{m_2=0}^{\infty}
\sum_{l=0}^{l_2} 
\left ( {L-l \atop l_1} \right ) 
{ (M + l + 1)! \over m_1! \; (m_2 + l + 1)!} 
{1 \over ( M + L+ 2)!} 
x^L y^M ,
\eqno(C.17)
\en
with $L=l_1 + l_2$ and $M=m_1+m_2$. Once again, 

\eq
\sum_{l_1=0}^{\infty}
\sum_{l_2=0}^{\infty}
\sum_{l=0}^{l_2} \to \sum_{L=0}^{\infty}
\sum_{l=0}^{L}
\sum_{l_1=0}^{L-l} ,
\eqno(C.18)
\en
so that

\eq
{\cal S}^{\prime\prime} (x,y) = \sum_{m_1=0}^{\infty}
\sum_{m_2=0}^{\infty} \sum_{L=0}^{\infty}
\sum_{l=0}^{L} 2^{L-l} { (M + l + 1)! \over m_1! \; (m_2 + l + 1)!} 
{1 \over ( M + L+ 2)!} 
x^L y^M  .
\eqno(C.19)
\en
In addition to $M=m_1+m_2$ and $L=l_1 + l_2$, we take $N=M+L$, and $n=M+l$. We then
rearrange the sums

\eq
\sum_{m_1=0}^{\infty}
\sum_{m_2=0}^{\infty} \sum_{L=0}^{\infty}
\sum_{l=0}^{L} \to \sum_{N=0}^{\infty}
\sum_{n=0}^{N} \sum_{m_1=0}^{n}
\sum_{M=m_1}^{n} ,
\eqno(C.20)
\en
we find that the sums can now be performed to yield

\eq
{\cal S}^{\prime\prime} (x,y) = {1 \over 2} \left ( { e^x - e^y \over x-y } \right )^2 .
\eqno(C.21)
\en

The other sums appearing in section 3 can all be performed using this repertoire
of tricks.

\bigskip \noindent{\bf Appendix D: Computation of $\log \lx \cZ \rx$} \medskip

Although, as we have seen, the computation of $\lx \log \cZ \rx$ is quite intricate, the computation
of $\lx \cZ \rx$ is by contrast straightforward. This is quite useful, since we have seen that the
terms in $\lx \log \cZ \rx$ that are $\propto \log n$ as $n \to \infty$ are identical to $\log \lx
\cZ \rx$. 

Consider the initial branching point in the cluster. At this point, we can write the relation

\eq
\lx \cZ (q, \sigma; n) \rx = \int_0^{\infty} d \epsilon \rho( \epsilon ) \left \{ \fm (\epsilon n)
\lx \cZ(q, \sigma; y(\epsilon n) n) \rx + \fp (\epsilon n) \lx \cZ (q, \sigma; (1-y(\epsilon n)) n )
\rx \right \} ,
\eqno(D.1)
\en
where the averages inside the integral on the right-hand side now do not include the first average
over
$\epsilon$. In Ref.~11, we showed how this integral equation can be expanded into
differential equations of increasing order, whose solution determines the annealed
dimensions $\sqa$. For general $\sigma$ these differential equations have solutions of the
form $\lx \cZ \rx = n^{\mu(q, \sigma)}$. Using this {\it Ansatz} in Eq.~(D.1), we
obtain

\eq
n^{\mu} = \int_0^{\infty} d \epsilon \rho( \epsilon ) \left \{ \fm(\epsilon n) [y(\epsilon n) n
]^{\mu} +
\fp(\epsilon n)
\left [\left (1-y(\epsilon n) \right ) n \right ]^{\mu} \right \} .
\eqno(D.2)
\en
Factoring out $n^{\mu}$, substituting $\fp \equiv 1 + \gp$, and then transforming to the
$\eta
\equiv
\epsilon n$ variable, we see that

\eq\eqalign{
\int_0^{\infty} d \eta \; \eta^{\nu - 1} \big \{ \fm(\eta) y^{\mu} (\eta)  
+ (1+\gp(\eta)) & [1-y(\eta)]^{\mu} - 1
\big \}
\cr &
\equiv
\lxc
\fm y^{\mu} + (1+\gp) (1-y)^{\mu} - 1 \rxc = 0 , \cr } 
\eqno(D.3)
\en
which implicitly determines $\mu(q,\sigma)$. We can expand Eq.~(D.3) so as to obtain $\mu$
order by order in $\delta$, the formal parameter that counts the number of powers of $\fm$
or $\gp$. If $\fm \to 0$ and $\gp \to 0$, we see that $\mu \to 0$ as well. To implement the
expansion, we write $y^{\mu} \equiv
\exp (\mu \log y)$ and $(1-y)^{\mu} = \exp(\mu\log(1-y))$; we can then expand the exponentials order
by order in $\mu$ in Eq.~(D.3), thereby obtaining solutions for $\mu$ to any desired order in
$\delta$. To second order in $\delta$, we can write

\eq
\lxc \psi + \mu [ \log (1-y) + \gp \log (1-y) + \fm \log y   ] + {\mu^2 \over 2} \log^2 (1-y) \rxc +
O(\delta^3) = 0 ,
\eqno(D.4)
\en
which gives

\eq
\mu = -{\lxc \psi \rxc \over \lxc \log (1-y) \rxc} \Big [ 1 - { \lxc \fm \log y + \gp
\log (1-y) \rxc 
\over
\lxc
\log (1-y) \rxc} + {\lxc \psi \rxc \lxc \log^2 (1-y) \rxc \over 2 \lxc \log(1-y) \rxc^2} \Big ] +
O(\delta^3) ,
\eqno(D.5)
\en
or, recalling the definitions of $\lambda$ and $\lambda^{\prime}$ (Eqs.~(A.5) and (A.10)),

\eq
\mu = \lambda \lxc \psi \rxc [ 1 + \lambda \lxc \fm \log y + \gp \log (1-y) \rxc + \lambda^{\prime}
\lxc \psi \rxc  + O(\delta^2) ] ,
\eqno(D.6)
\en
which agrees with the term $\propto \log n$ in Eq.~(3.43) above. Note that $\lxc
\psi \rxc = 0$ implies that $\mu = 0$ to any order in $\delta$.
Clearly terms of higher
order in
$\delta$ can be obtained from Eq.~(D.3) with considerably less labor than by direct perturbative
computation.
\bigskip
\noindent{\bf Appendix E: $\delta$ expansion as an expansion in $q-1$}
\medskip

The perturbation expansion derived above can be formally written in the following form:

\eq
\lx \log \cZ \rx = \lxc \psi \rxc F_1 (q,\sigma; n) + \lxc \psi^2 \rxc F_2 (q, \sigma; n) +
\cdots ,
\eqno(E.1)
\en
because the last vertex (at the far right) of any diagram is always averaged without
any powers of $\log y$ or $\log(1-y)$, and can always be either $\fm$, $\gp$, or a
product of factors of these two. Thus this last vertex can always be regarded as being of the form
$\lxc
\psi^n \rxc$. (This property is satisfied by Eqs.~(3.43) and (B.1-4).) Recall the definition of
$\psi$:

\eq
\psi (\eta; q, \sigma)  = \fm (\eta)  + \gp (\eta ) = {x^q (\eta) \over y^{\sigma}
(\eta)} + {(1-x(\eta))^q \over (1-y(\eta))^{\sigma} } - 1 .
\eqno(E.2)
\en
At $q=1$, for $\sigma=0$ we have $\psi = 0$, which implies that both $\sqa$ and the apparent $\sqap$
are equal to zero. Thus it is clear that 

\eq
\psi(q, \sqap; \eta) = O(q-1) .
\eqno(E.3)
\en
Since by definition $\lxc \psi(q,\sqa) \rxc = 0$ exactly, Eq.~(E.1) together with (E.3) implies that
$\sqap =
\sqa + O((q-1)^2)$, with

\eq
\lxc \psi(q, \sqap) \rxc = O((q-1)^2) ,
\eqno(E.4)
\en
and

\eq
\lxc \psi^2 \rxc = O((q-1)^2) .
\eqno(E.5)
\en

\vfill \eject 
\sect{\bf
References} \medskip

\item{1.} T.A. Witten, Jr. and L.M. Sander, Phys. Rev. Lett. {\bf 47}, 1400
(1981).

\item{2.} P. Meakin, Phys. Rev. A {\bf 27}, 1495 (1983).

\item{3.} P. Meakin, in {\it Phase Transitions and Critical Phenomena, Vol.~12} eds. C. Domb
and J. Lebowitz, (Academic Press, New York, 1988).

\item{4.} T. Vicsek, {\it Fractal Growth Phenomena}, (World Scientific, Singapore, 1989)

\item{5.} H. Gould, F. Family, and H.E. Stanley, Phys. Rev. Lett. {\bf 50}, 686
(1983); T. Nagatani, Phys. Rev. A {\bf 36}, 5812 (1987); J. Phys. A {\bf 20}, L381
(1987); P. Barker and R.C. Ball, Phys. Rev. A {\bf 42}, 6289 (1990).

\item{6.} X.R. Wang, Y. Shapir and M. Rubinstein,  Phys. Rev. A {\bf 39}, 5974 (1989);
J. Phys. A {\bf 22}, L507 (1989).

\item{7.} L. Pietronero, A. Erzan, and C.J.G. Evertsz, Phys. Rev. Lett.
{\bf 61}, 861 (1988); Physica A {\bf 151}, 207 (1988).

\item{8.} P. Meakin, H.E. Stanley, A. Coniglio, and T.A. Witten, Jr., Phys. Rev. A {\bf 32}, 2364
(1985); C. Amitrano, A. Coniglio, and F. di Liberto, Phys. Rev. Lett. {\bf 57}, 1016
(1986); C. Amitrano, Phys. Rev. A {\bf 39}, 6618 (1989).

\item{9.} T.C. Halsey, P. Meakin, and I. Procaccia, Phys. Rev. Lett. {\bf 56}, 854
(1986); T.C. Halsey, Phys. Rev. Lett. {\bf 59}, 2067 (1987); Phys.
Rev. A {\bf 38}, 4749 (1988). 

\item{10.}  Y. Hayakawa, S. Sato, and M. Matsushita, Phys. Rev. A {\bf 36}, 1963
(1987); R.C. Ball and O. Rath Spivack, J. Phys. A {\bf 23}, 5295 (1990).

\item{11.} T.C. Halsey and M. Leibig, Phys. Rev. A {\bf 46}, 7792
(1992).

\item{12.} T.C. Halsey, Phys. Rev. Lett. {\bf 72}, 1228 (1994).

\item{13.} R. Blumenfeld and A. Aharony, Phys. Rev. Lett. {\bf 62}, 2977 (1989); R.C. Ball and
R. Blumenfeld, Phys. Rev. A {\bf 44}, 828 (1991); S. Schwarzer, J. Lee, A. Bunde, S. Havlin,
H.E. Roman, and H.E. Stanley, {\bf 65}, 603 (1990); B. Mandelbrot and C.J.G.
Evertsz, Nature {\bf 348}, 143 (1990); S. Schwarzer, J. Lee, S. Havlin, H.E.
Stanley, and P. Meakin, Phys. Rev. A {\bf 43} 1134 (1991).

\item{14.} B.B. Mandelbrot, H. Kaufman, A. Vespinani, I. Yekutieli, and C.H. Lam,
Europhys. Lett. {\bf 29}, 599 (1995).

\item{15.} P. Ossadnik, Physica A {\bf 195}, 319 (1993).

\item{16.} M. Muthukumar, Phys. Rev. Lett. {\bf 50}, 839 (1983);
S. Tolman and P. Meakin, Phys. Rev. A {\bf 40}, 428 (1989).

\item{17.} R. Brady and R.C. Ball, Nature (London) {\bf 309}, 225; J. Nittmann, G. Daccord, and
H.E. Stanley, Nature (London) {\bf 314}, 141 (1985); D. Kessler, J. Koplik, and H. Levine,
Advances in Physics {\bf 37}, 255 (1988).

\item{18.} P. Meakin, Phys. Rev. Lett. {\bf 51}, 1119 (1983); M. Kolb, R. Botet, and R.
Jullien, {\it ibid}, 1123; J. Nittmann, G. Daccord, and H.E. Stanley, Nature (London) {\bf
314}, 141 (1985); P. Meakin, G. Li, L.M. Sander, H. Yan, F. Guinea, O. Pla, and E. Louis,
in {\it Disorder and Fracture}, J.C. Charmet, S. Roux, and E. Guyon, eds. (Plenum Press,
New York, 1990).

\item{19.} A. Renyi, {\it Probability Theory} (North-Holland, Amsterdam, 1970);  B.B.
Mandelbrot, Ann.~Isr.~Phys.~Soc. {\bf 225} (1977); H.G.E. Hentschel and I. Procaccia,
Physica {\bf 8D}, 435 (1983);  U. Frisch and G. Parisi, in {\it Turbulence and Predictability in Geophysical Fluid Dynamics
and Climate Dynamics,  Proc.~of Int.~School of Physics ``Enrico Fermi'' LXXXVIII}, eds. M.
Ghil, R. Benzi, and G. Parisi (North-Holland, Amsterdam,1985) p. 84.

\item{20.} T.C. Halsey, M.H. Jensen, L.P. Kadanoff, I. Procaccia, and B. Shraiman,
Phys. Rev. A {\bf 33}, 1141(1986).

\item{21.} M.E. Cates and T.A. Witten, Phys. Rev. A {\bf 35}, 1809 (1987); T.C. Halsey,
in {\it Fractals: Physical Origin and Properties}, ed. L. Pietronero, (Plenum
Publishing Co., London, 1989).

\item{22.} S. Redner, Am. J. Phys. {\bf 58}, 267 (1990).

\item{23.} B.B. Mandelbrot, in {\it Fractals: Physical Origin and Properties}, ed. L. Pietronero,
(Plenum Publishing Co., London, 1989); A.B. Chhabra and K.R. Sreenivasan, Phys. Rev. A {\bf 43},
1114 (1991).

\item{24.} T.C. Halsey, K. Honda, B. Duplantier, unpublished.

\vfill \eject 
\headline{ \hfill}
\sect{\bf
Table caption} \medskip

\item{1.} This table shows the orders of the terms divergent in $\log n$ appearing in the
$\delta$-expansion for $\lx \log \cZ \rx$. A solid cross indicates that a term of that order in
$\delta$ and
$\log n$ appears; a dashed cross indicates that while such terms appear in some diagrams,
they cancel between diagrams. The leading and sub-leading resummations discussed in
section 3 correspond to sums down the diagonals of this table.

\vfill \eject 
\sect{\bf
Figure captions} \medskip

\item{1.} A typical two dimensional DLA cluster, grown using off-lattice random walkers.
There are approximately $35,000$ particles in this cluster (courtesy of M. Leibig).

\item{2.} Multifractal scaling functions $\fa$ for (a) a typical non-stochastic
system, and (b) a typical stochastic system, with annealed averaging. Note
the appearance of negative values of $f$ in the latter case. These
functions are obtained by Legendre transformation of $\tau(q)$ or
$\sigma(q) \propto \tau(q)$.

\item{3.} The unstable manifold for the dynamics of branch competition in
``model Z". The two branches move along the indicated diagonal line until either $x=0$
or $x=1$. This corresponds to the complete screening of one or the other of
the two sibling branches. From this point only the parameter $y$ reflecting
the relative masses of the two branches changes.

\item{4.} The unstable manifold for the dynamics of branch competition for
DLA in two dimensions, as calculated by a renormalization method in Ref.~12. Comparing
with model Z, we see that screening is a more gradual process in DLA.

\item{5.} Two sibling elementary sub-branches. The
last node in the branching cluster is indexed by $J$, has a total of $n_J$ descendants, and has a
total growth probability of $p_J$. The stochastic variable controlling the branching at
this node is $\epsilon_J$. The stronger descendant sub-branch has a number of particles
$n_e=(1-y(\epsilon_J n_J))n_J$, and a total growth probability $p_e = (1-x(\epsilon_J n_J))
p_J$

\item{6.} The sub-branch found at the bottom of the main branch, defined by taking at
every node the stronger of the two paths down the branch. The index $j$
indexes the nodes along this main branch.

\item{7.} The terms of first order in $\delta$ containing one factor of $\fm$ arise from
the elementary sub-branches at the end of the ``first-order" sidebranches off of the main
branch. The terms of first order in $\delta$ containing one factor of $\gp$ arise from
corrections to the partition function contribution coming from the elementary sub-branch
at the end of the main branch.

\item{8.} Diagrams at $O(\delta)$ in perturbation theory. The solid circle indicates an
$\fm$ vertex, and the open circle indicates a $\gp$ vertex. A solid and an open circle
connected by a short vertical line indicate a vertex of $\psi = \fm + \gp$. A horizontal
segment indicates a branch over which $n^{-\nu}$ is summed, possibly with a
logarithmic factor in $n$.

\item{9.} A diagram at $O(\delta^2)$ that is $\propto \gp^2$. This diagram comes from
$\lx \cZ-1 \rx$. The two horizontal bars indicate the two propagator sums. Note that
including a factor of $\gp$ does not lead one onto a weaker side-branch.

\item{10.} A diagram at $O(\delta^2)$ that is $\propto \fm^2$. This diagram comes from
$\lx \cZ-1 \rx$. The two horizontal bars indicate the two propagator sums. The vertical bar
indicates that the rightmost $\fm$ vertex is off of the main
branch. Taking a factor of $\fm$ in a diagram arising from $\lx \cZ-1 \rx$ always moves
one onto a weaker sidebranch.

\item{11.} Mixed terms $\propto \fm \cdot \gp$ arising from $\lx \cZ-1 \rx$. These diagrams
are $O(\delta^2)$.

\item{12.} All terms of $O(\delta^2)$ arising from $-(1/2) \lx (\cZ-1)^2 \rx$. Note the
symmetry factors.

\item{13.} All diagrams containing only one propagator in the full expression for $\lx
\log \cZ \rx$. Note that the joined vertices indicate factors of $\fm$ or $\gp$ averaged
together, e.g. $\lxc \fm \gp \rxc$ for the last diagram. The sum of these diagrams gives the
``one-propagator" approximation $\lx \log \cZ \rx_{1p}$.

\item{14.} A typical contribution to $\lx \cZ -1 \rx$. At leading order, the leading
logarithm in each propagator is taken. Each vertical segment indicates a move to a
side-branch.

\item{15.} A typical contribution to $\lx ( \cZ -1)^2 \rx$, arising from a product of a
diagram with $l_1$ factors of $\gp$ upstream from the first factor of $\fm$, with $m_1$
factors of $\psi$ off of the main branch, with a diagram with 
$l_2$ factors of $\gp$ upstream from the first factor of $\fm$, with $m_2$
factors of $\psi$ off of the main branch. For simplicity, we have drawn all factors of
$\psi$ along the same branch; terms involving more than one factor of $\fm$ will properly
have further vertical segments. (Note that we use ``upstream" to mean closer to the root, at the
left, not closer to the elementary sub-branches.)

\item{16.} One term arising from the product indicated in Figure 15. Defining $L=l_1 +
l_2$, we have $L-l \ge l_1$ factors of $\gp$ upstream of the first factor of $\fm$,
followed by $l \le l_2$ factors of $\gp$, and then the second factor of $\fm$. The diagram
includes $m_1$ factors of $\psi$ coming off of the first $\fm$ vertex, and $m_2$ factors
of $\psi$ coming off of the second $\fm$ vertex.

\item{17.} This diagram indicates a term in the product indicated in Figure 15 that is
$\propto \lxc \fm^2 \rxc$. Now all $L=l_1 + l_2$ factors of $\gp$ are to the left of the
vertex containing the two factors of $\fm$ averaged together. Note that at the sub-leading
order, which we are considering here, the two separate side-branches emanating from these two factors
of $\fm$ (one containing $m_1$ factors of $\psi$, and the other containing $m_2$ factors of $\psi$)
can be treated as independent.

\item{18.} This diagram indicates a term $\propto \lxc \fm^2 \rxc$ arising from $\lx (
\cZ-1)^3 \rxc$. At sub-leading order, the factor of $\cZ -1$ whose first $\fm$
vertex appears separately can be averaged independently of the other two factors of
$\cZ-1$.

\item{19.} At sub-leading order, terms $\propto \lxc \fm \log y \rxc$ can be obtained by
averaging a single factor of $\cZ -1$; a typical diagram is shown. The arrows indicate the
$\fm$ vertices, each of which is a potential source of a factor of $\lxc \fm \log y \rxc$
except the last, which has no propagator to the right to provide a factor of $\log y$.

\item{20.} At sub-leading order, terms $\propto \lxc \fm \log(1-y) \rxc$ require at least
two factors of $\cZ -1$. This diagram arises from a product of the form shown. The
arrow in the diagram corresponding to a single factor of $\cZ-1$ indicates the vertex (with a factor
of one rather than a factor of
$\fm$ or
$\gp$) at which the term
$\log(1-y)$, arising from the propagator, is taken. To the right of this arrow, it is not necessary
to keep track of whether vertices are
$\fm$ or $\gp$, provided at least one such vertex appears. In the combination of the two factors
of $\cZ-1$, this $\log(1-y)$ is averaged with a factor of $\fm$ contributed by the other partition
function.

\item{21.} Numerical results for $\lx \log \sum_i p_i^q \rx$ vs. $\log n$ for a Monte Carlo
realization of model Z random branched growth, for $q=2$ and $q=5$. We used a parameter
$\nu=0.6$ to specify model Z. The linear slope indicates apparent multifractality, the value
of the slope giving the apparent dimensions $\sqap$.

\item{22.} (a) The apparent dimensions $\sqap$ vs. $q$ for $0<q<10$, and (b) the Legendre
transform of this function, $\fa$, computed numerically for model Z with $\nu=0.6$. No negative values of $f$ appear. In addition, since we compute $\sqa$ only for $q>0$,
we show only the left side of the multifractal spectrum. Note that model Z has the
pathological feature that $\lim_{q \to 0} \sigma(q) \ne -1$, due to the fact that much of
the cluster surface has growth probability zero.

\item{23.} $\sigma$ vs. $q$ for model Z with $\nu=0.6$ computed in various manners. (a)
solid line: numerical results, as in Figure 22(a); (b) dashed line: annealed dimensions
$\sqa$; these are also the apparent dimensions $\sqap$ at $O(\delta)$; (c) dot-dashed
line: apparent dimensions, computed to $O(\delta^2)$, and (d) dotted line: the
one-propagator result for $\sigma(q)$, from Eq.~(2.37). We note that the computation for apparent
dimensions gives excellent results for $q<5$, but fails at higher values of $q$. The
one-propagator result gives qualitatively correct results over the entire range of $q$.

\item{24.} Diagrams at $O(\delta)$ and $O(\delta^2)$. These correspond to the terms
appearing in the text in Eqs.~(B.1-2). They also summarize Figs.~(8-12).

\item{25.} Diagrams at $O(\delta^3)$. These correspond to the terms appearing in Eq.~(B.3).
\vfill\eject

\def\dofig#1{\epsfysize = 10truein \epsfbox{#1}\eject}
\hoffset=-0.9 true in
\voffset=-0.7 true in

\dofig{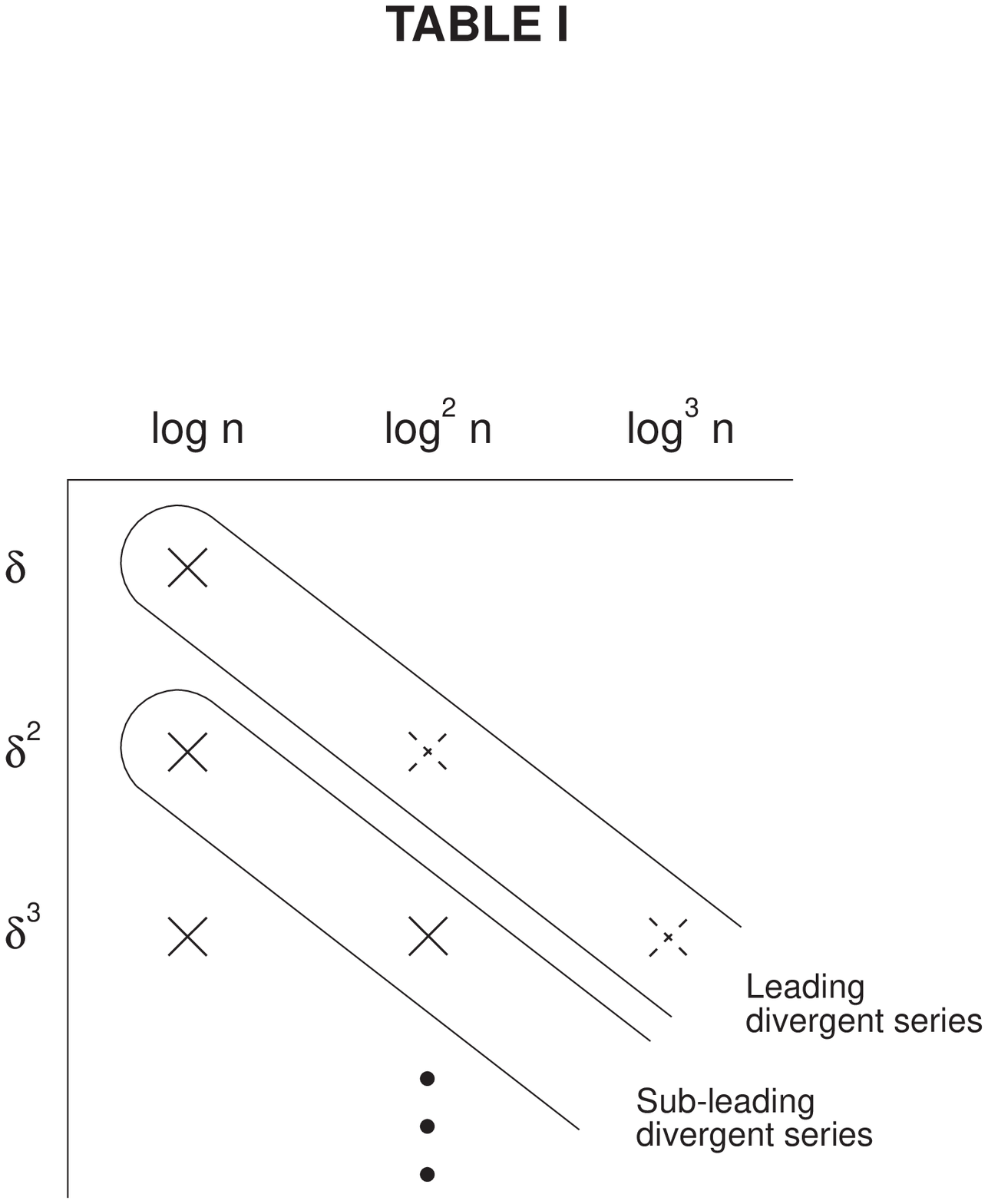}
\dofig{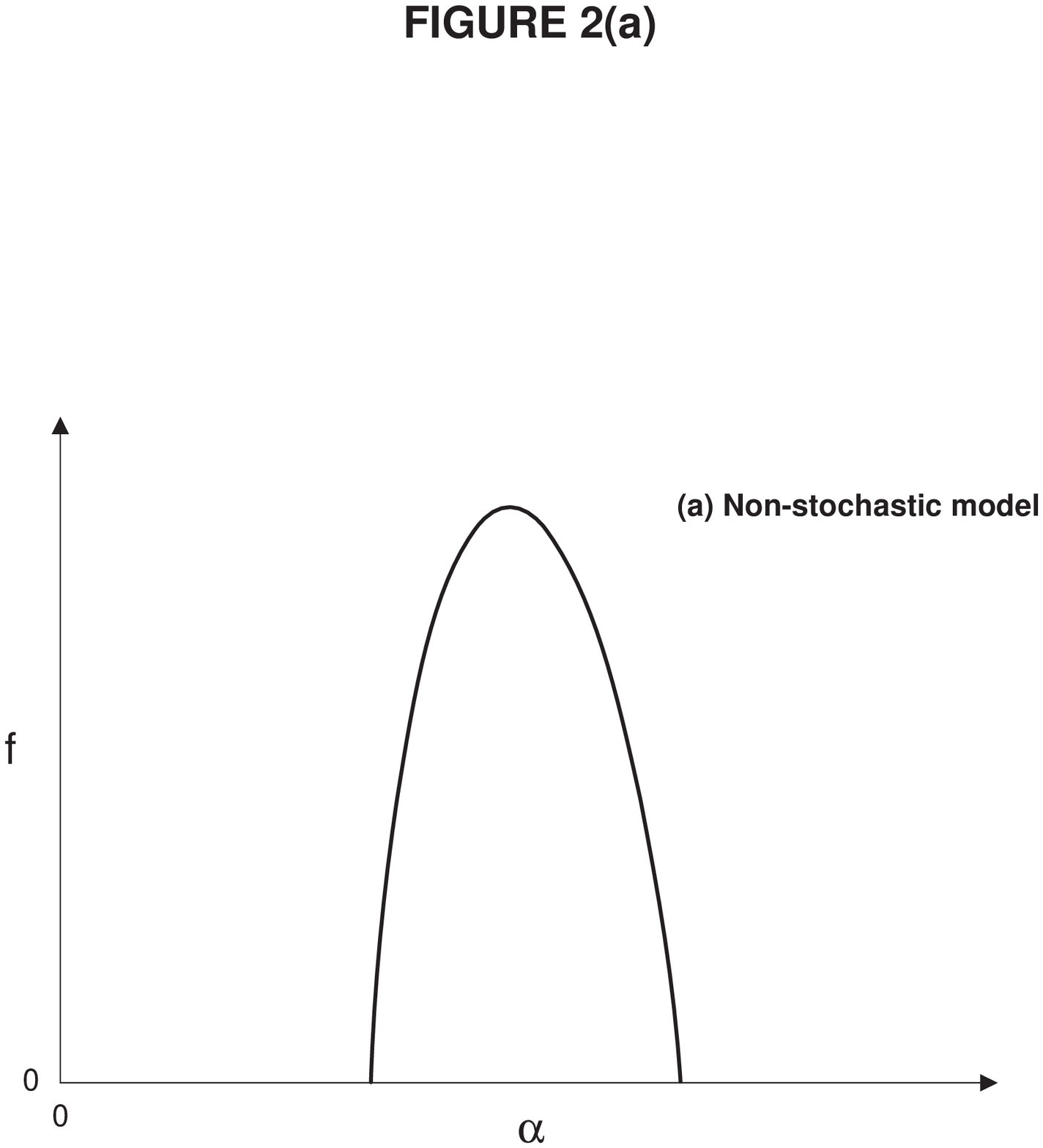}
\dofig{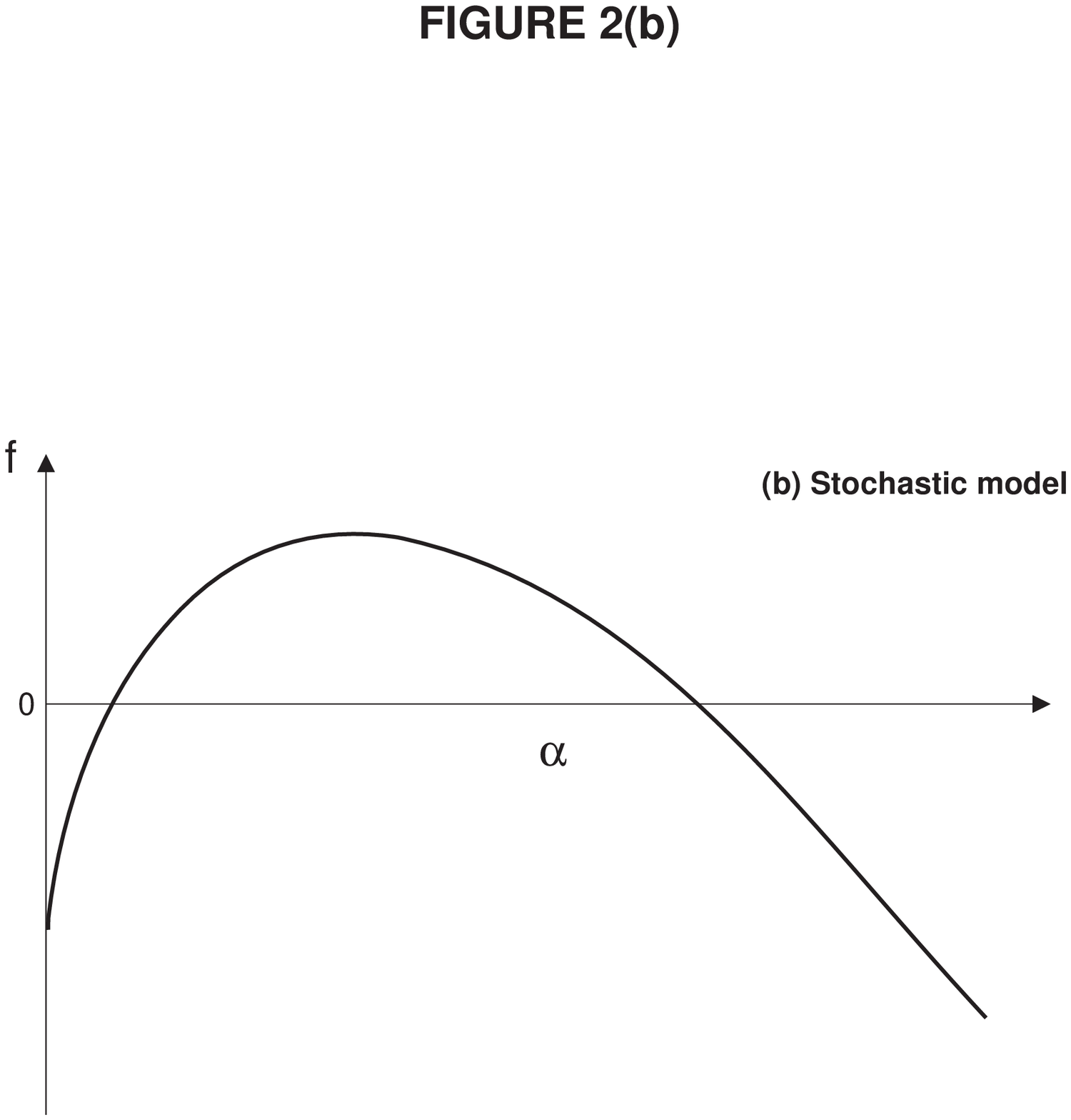}
\dofig{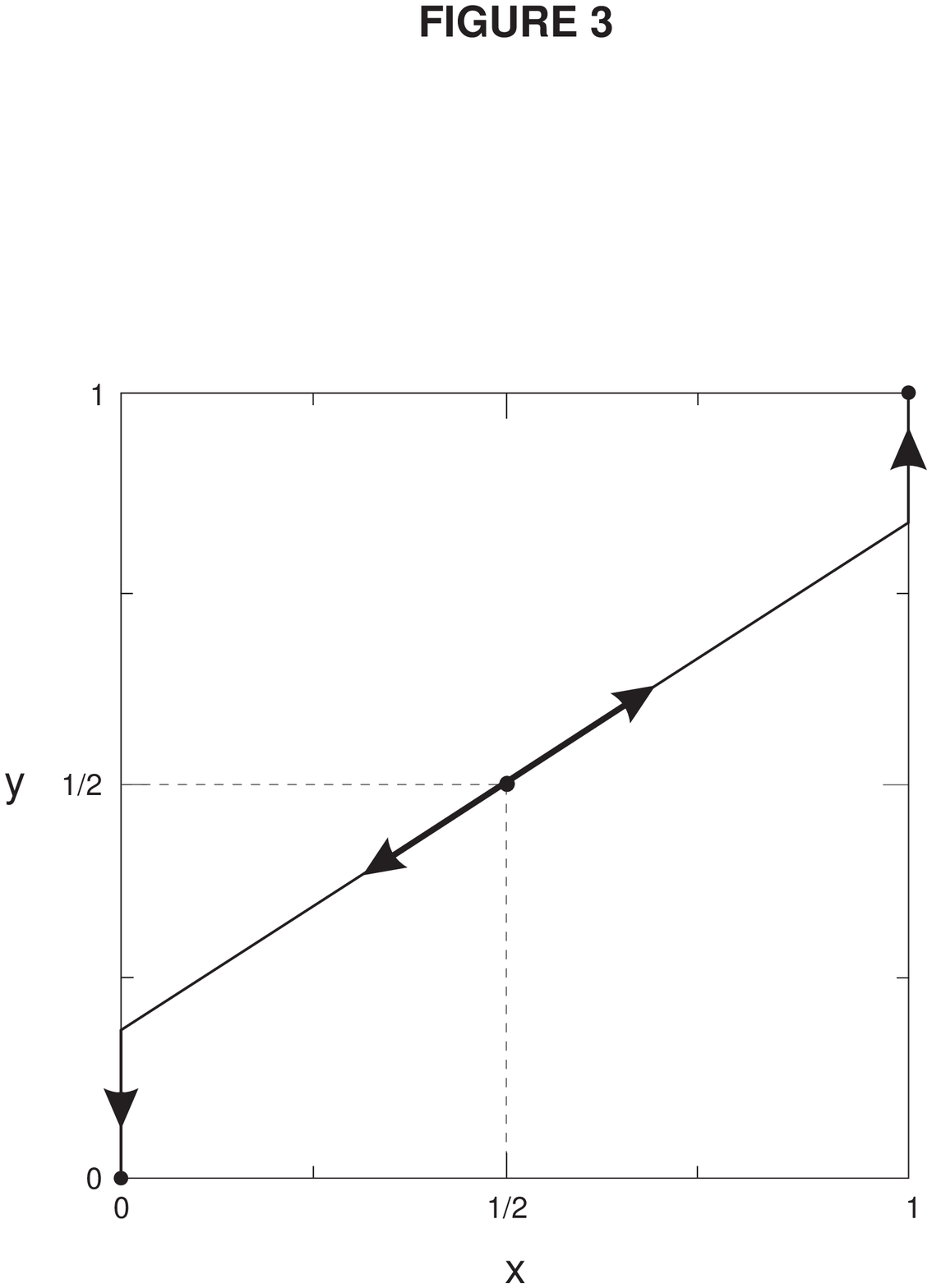}
\dofig{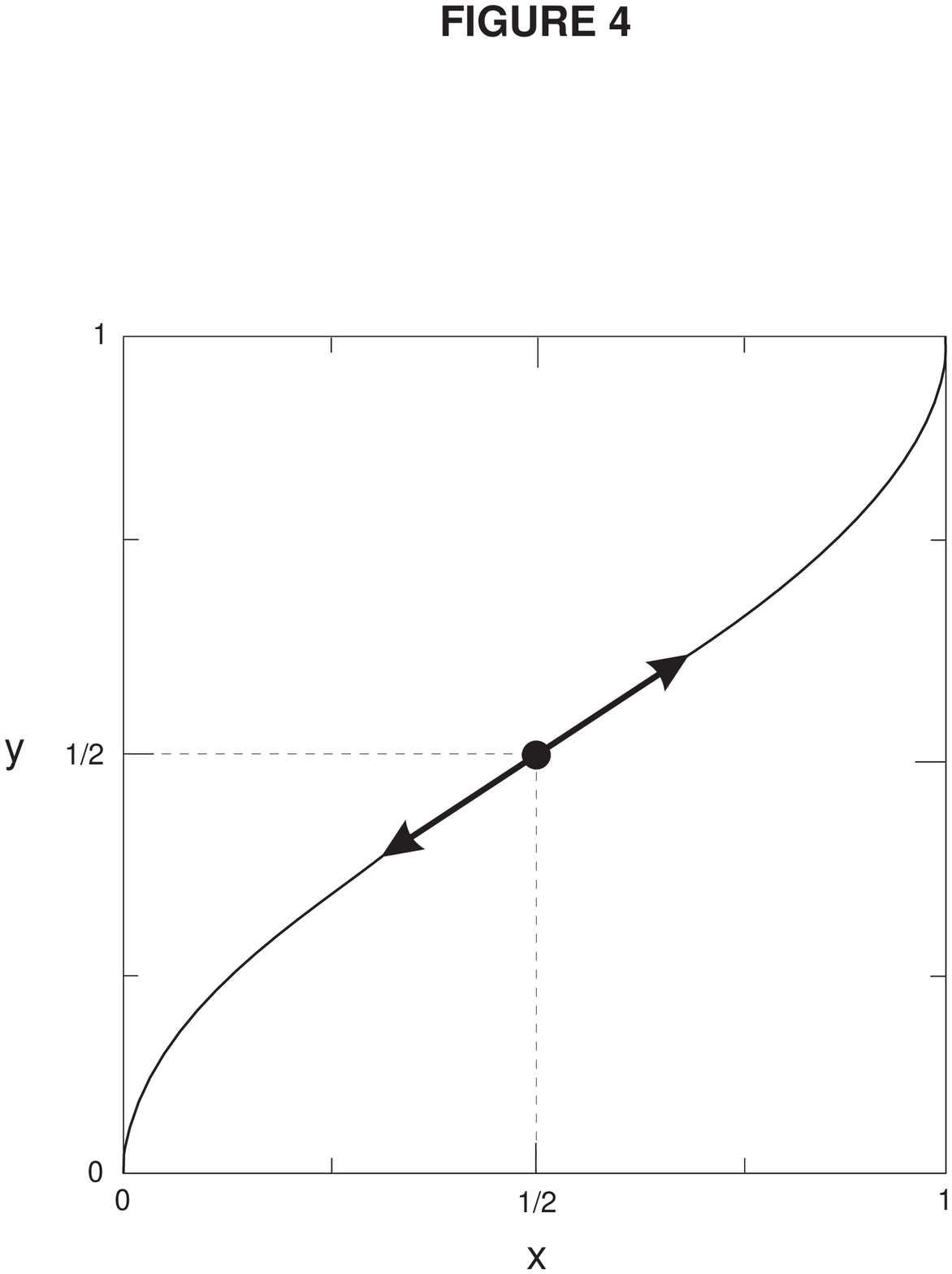}
\dofig{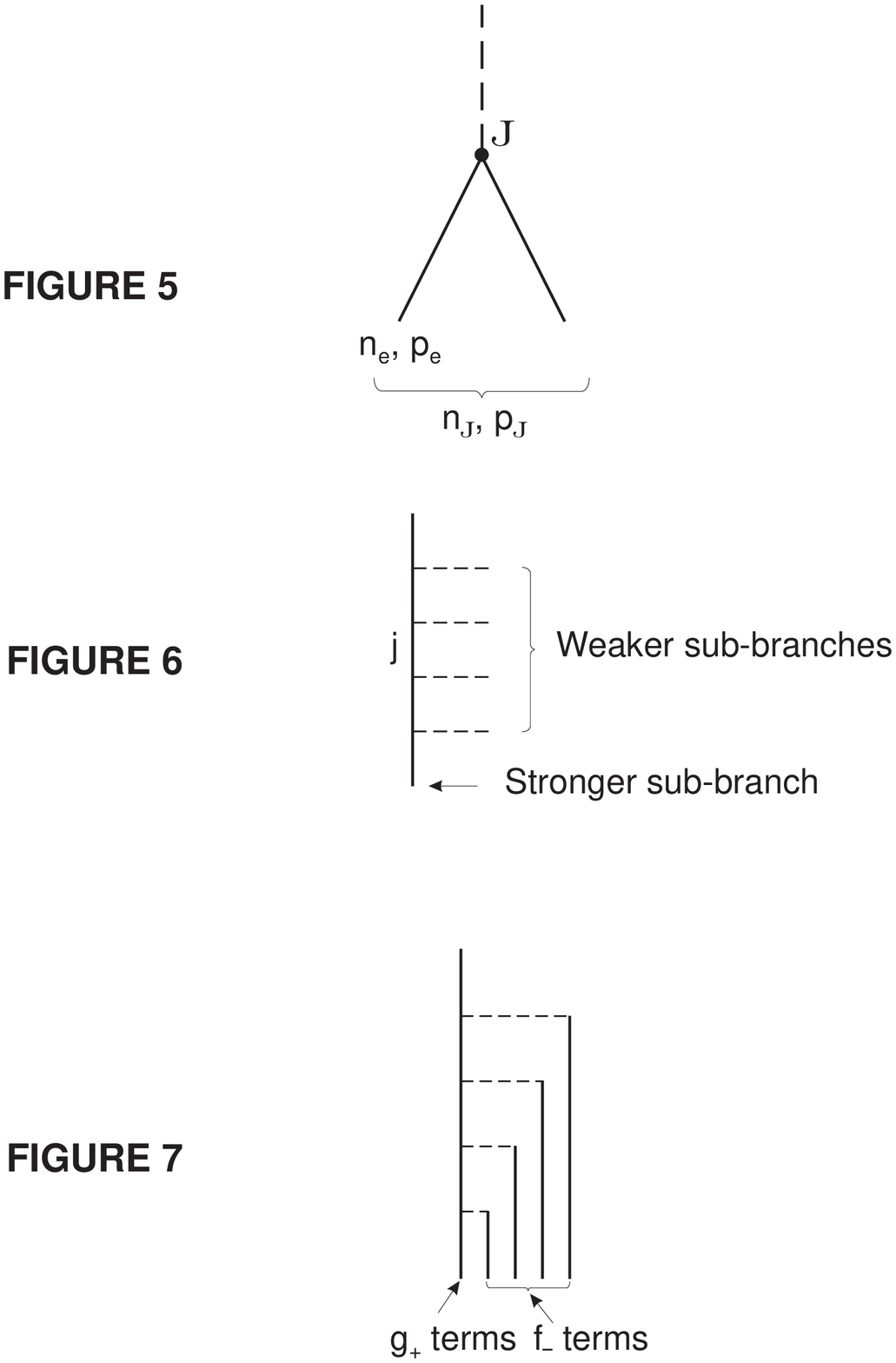}
\dofig{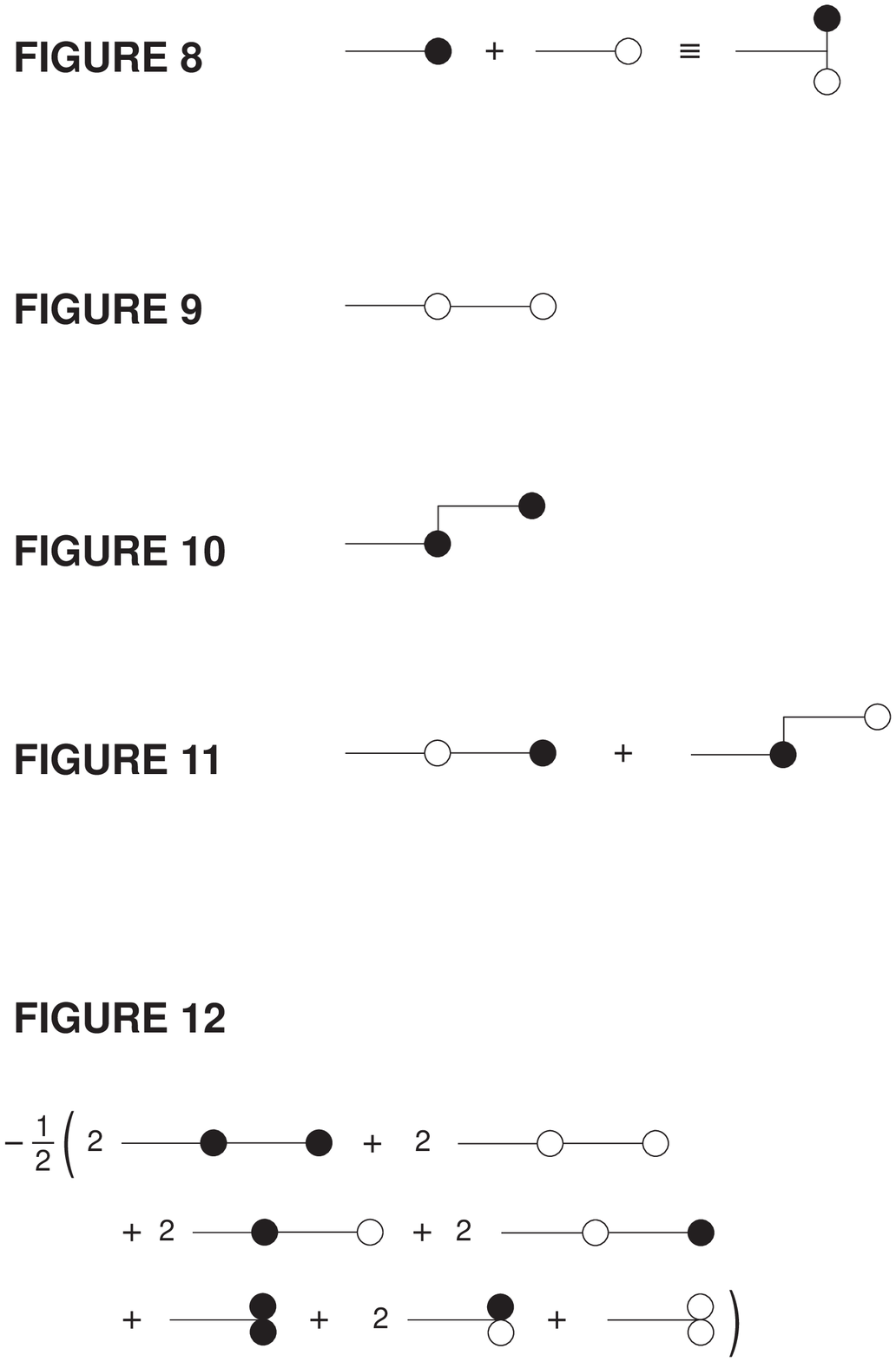}
\dofig{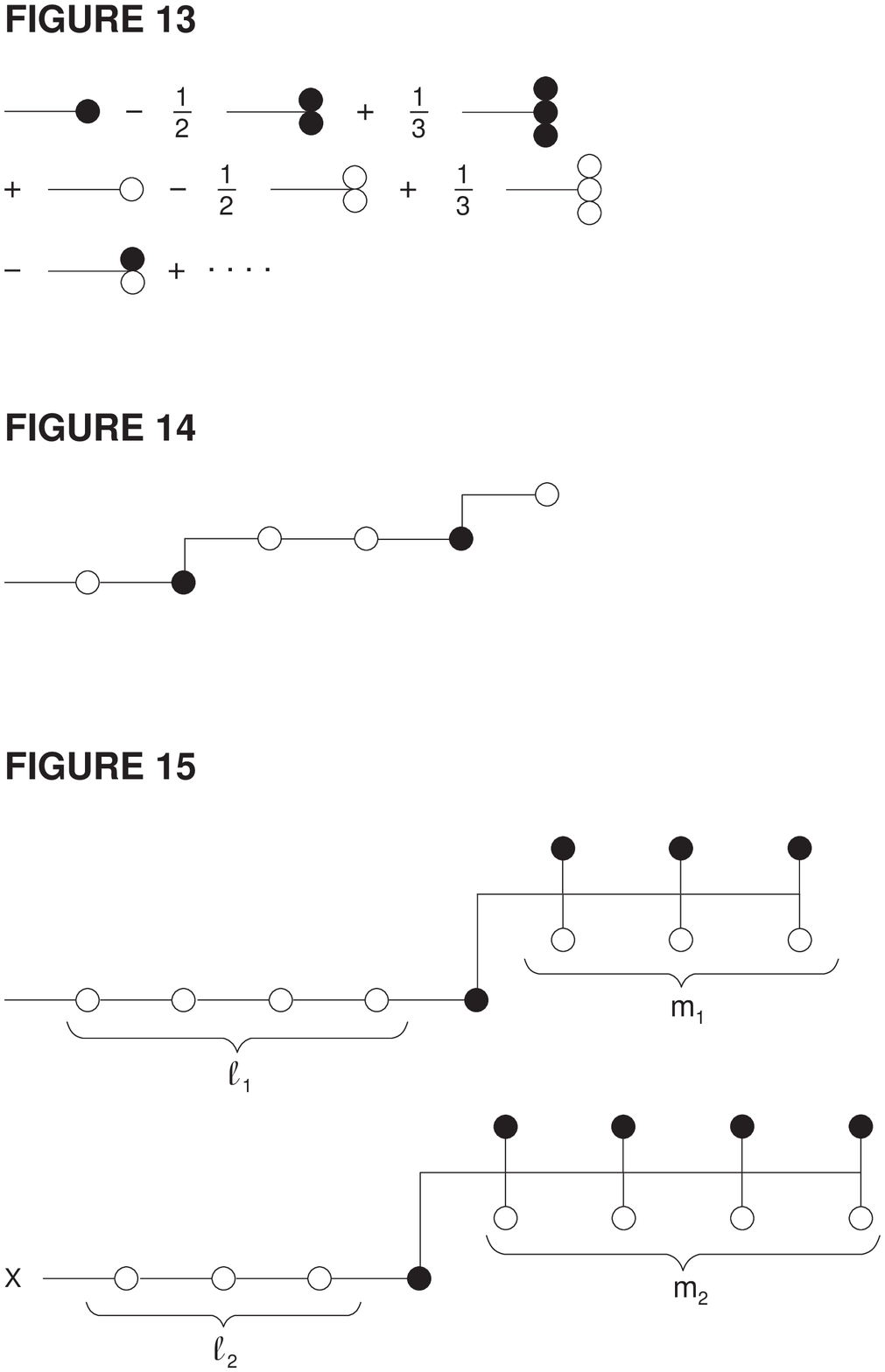}
\dofig{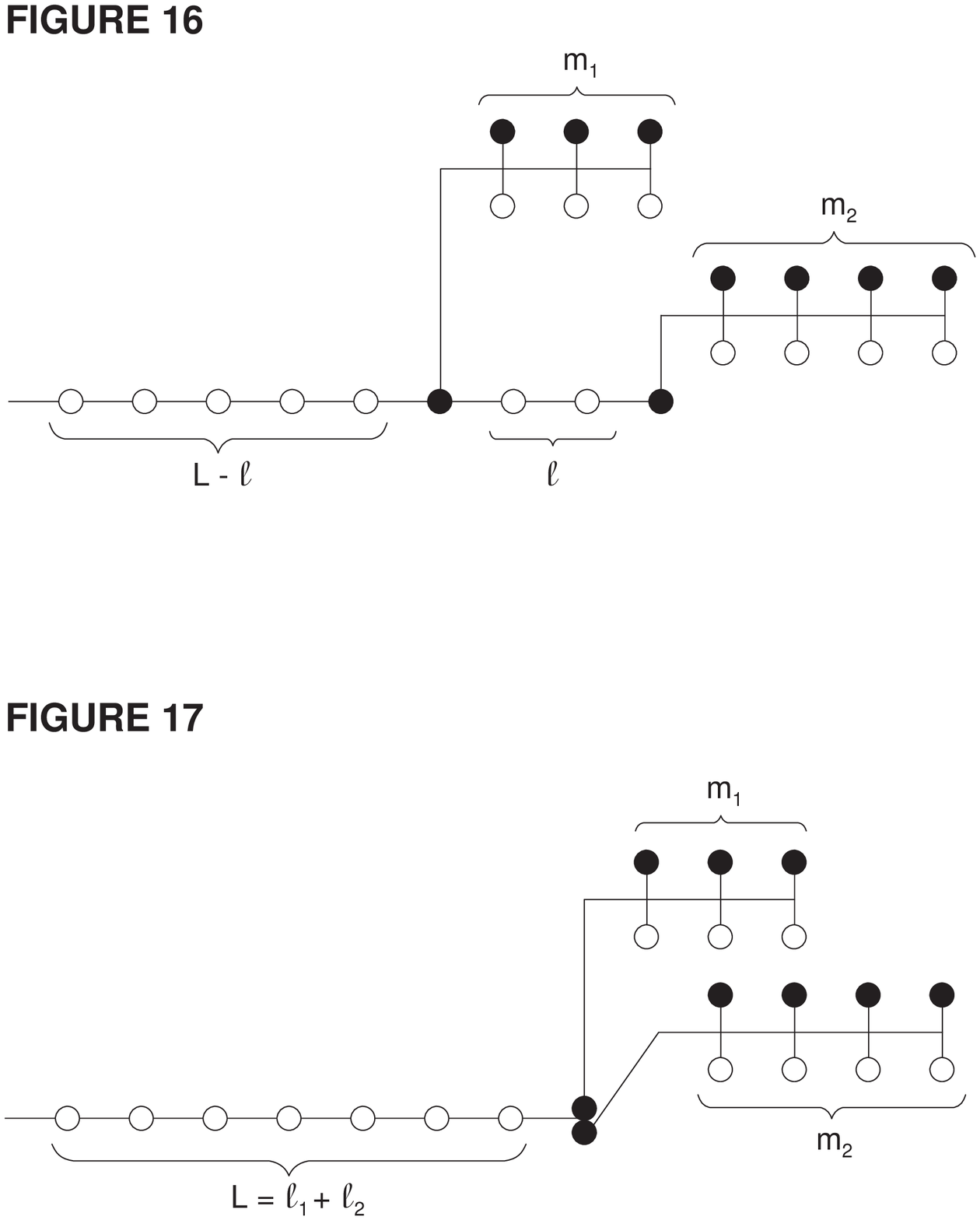}
\dofig{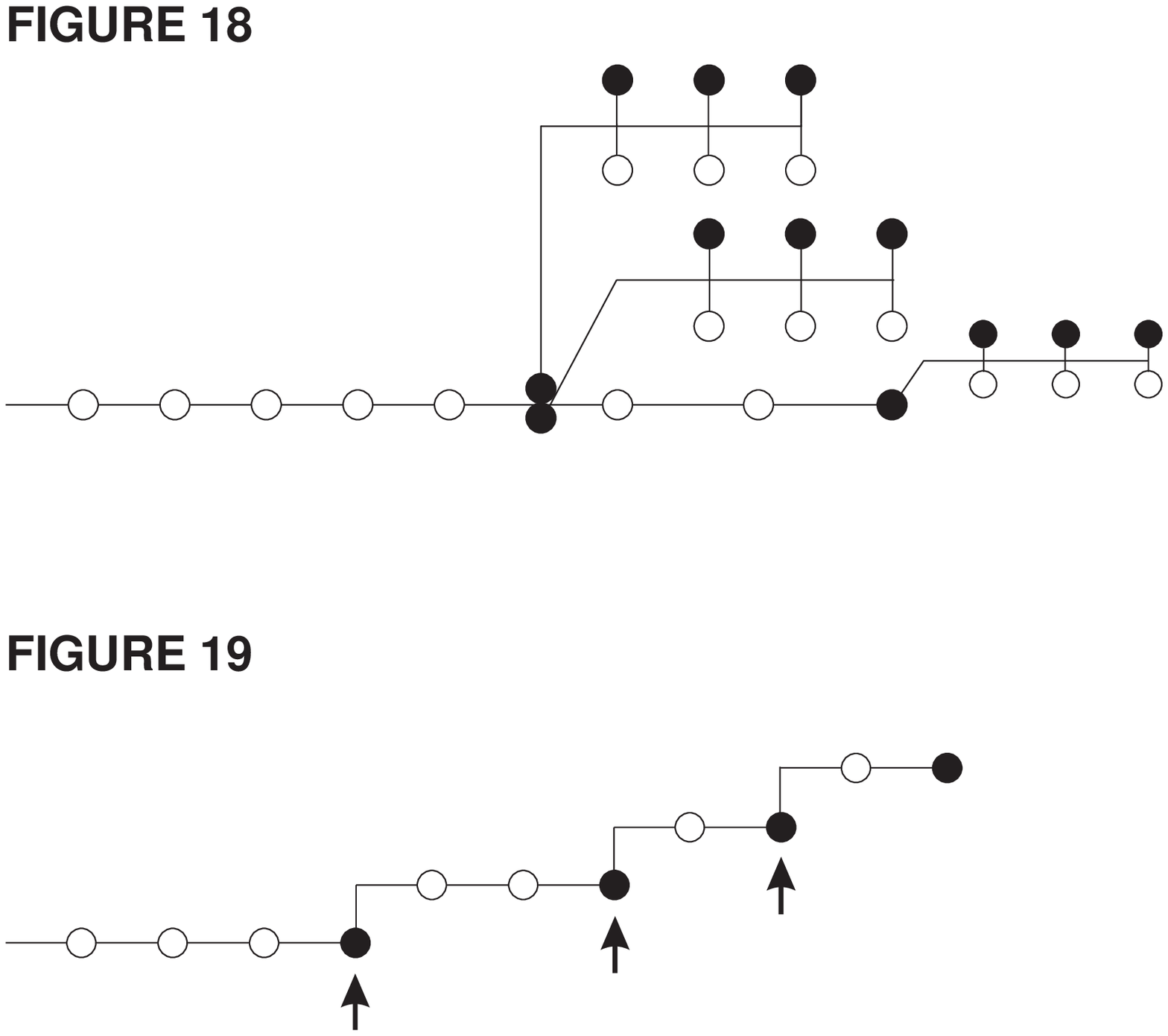}
\dofig{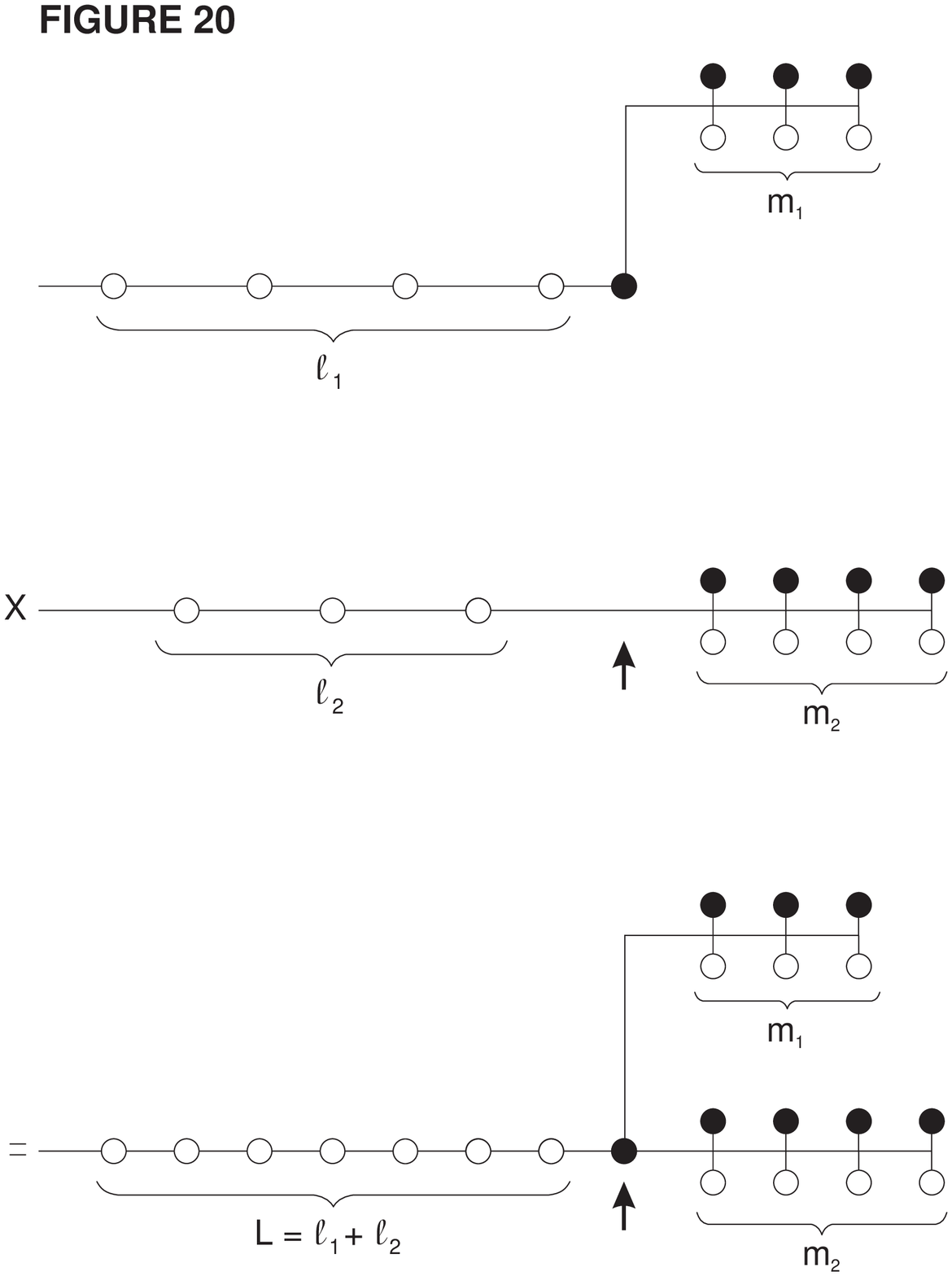}
\dofig{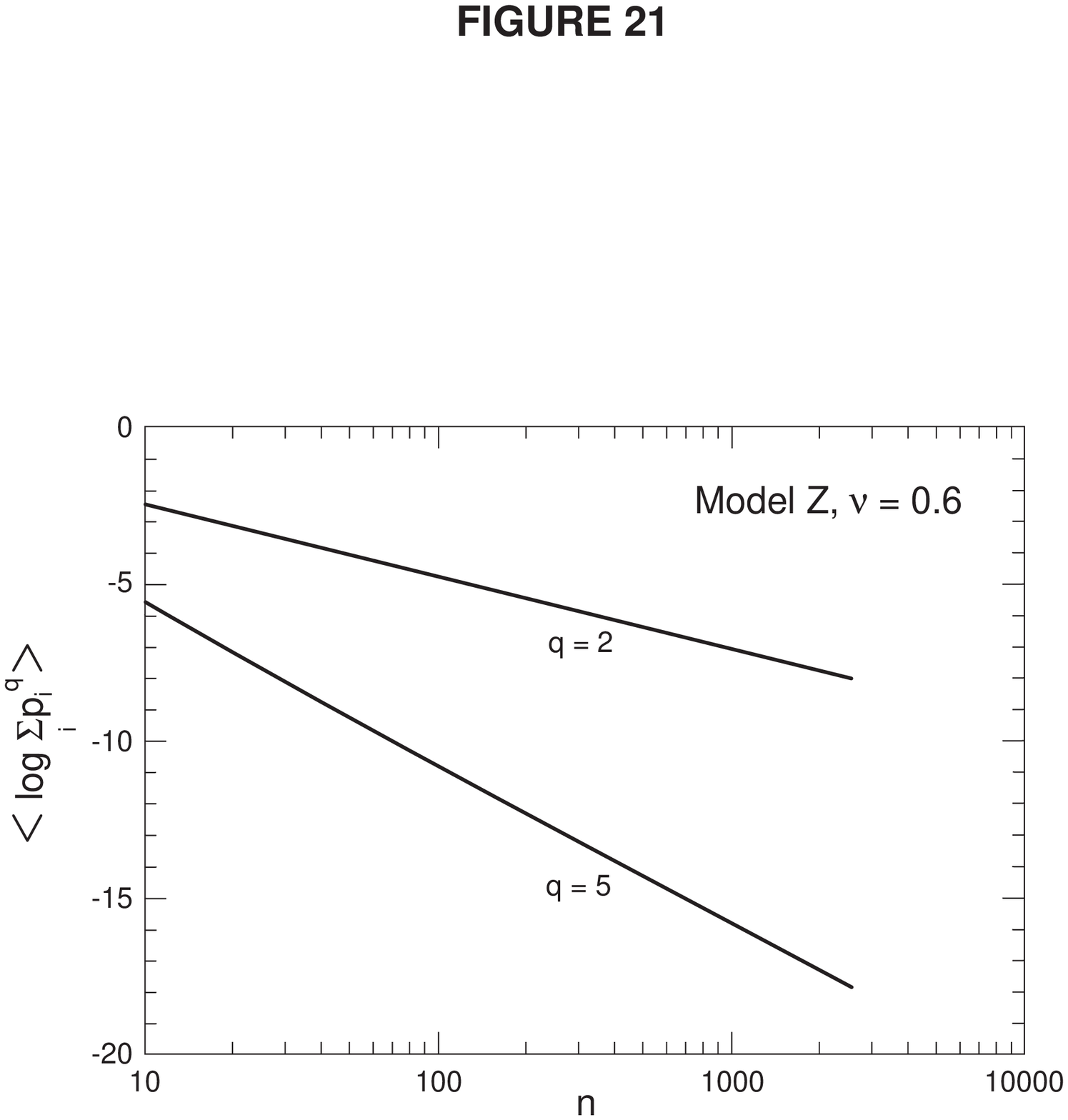}
\dofig{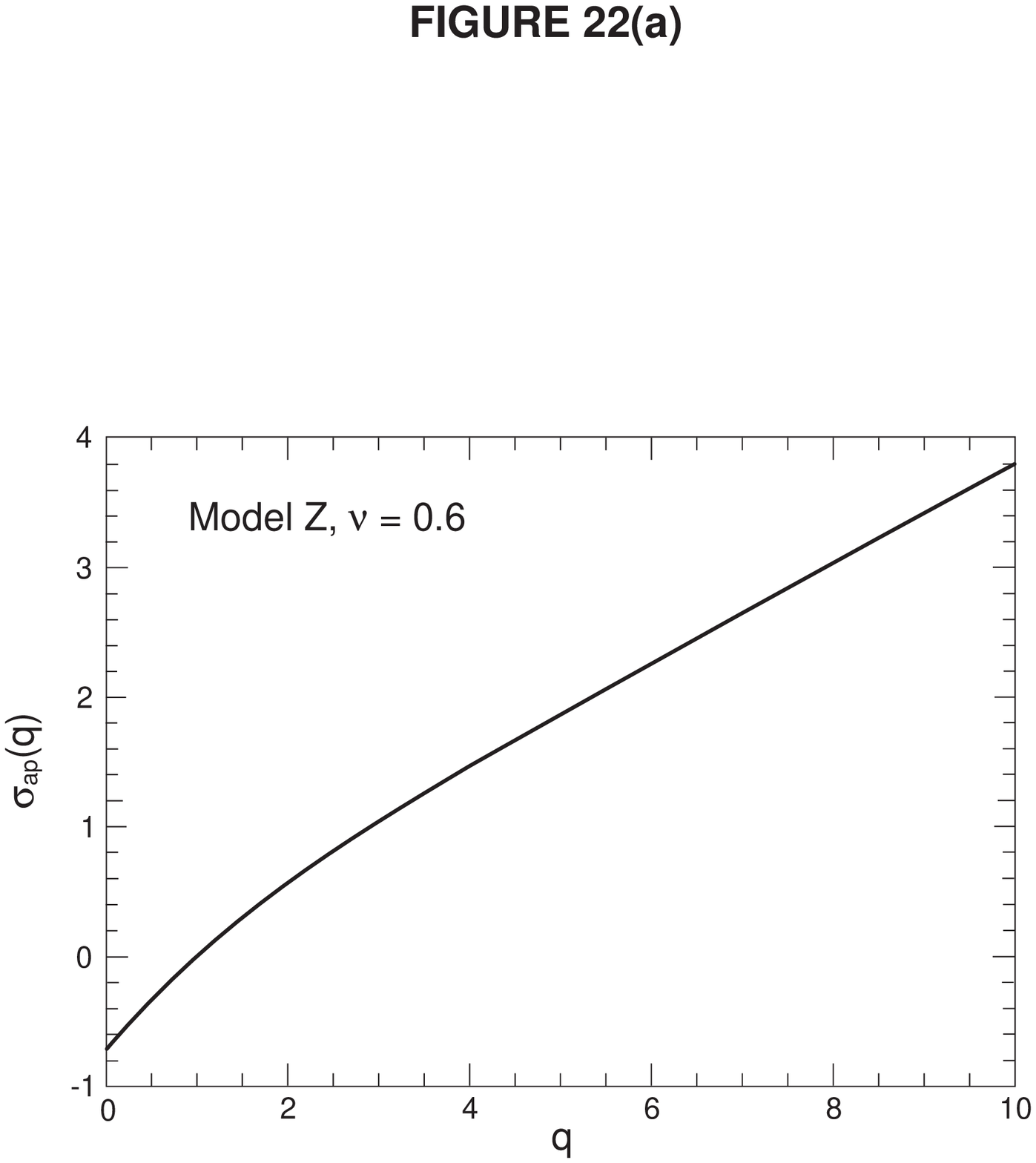}
\dofig{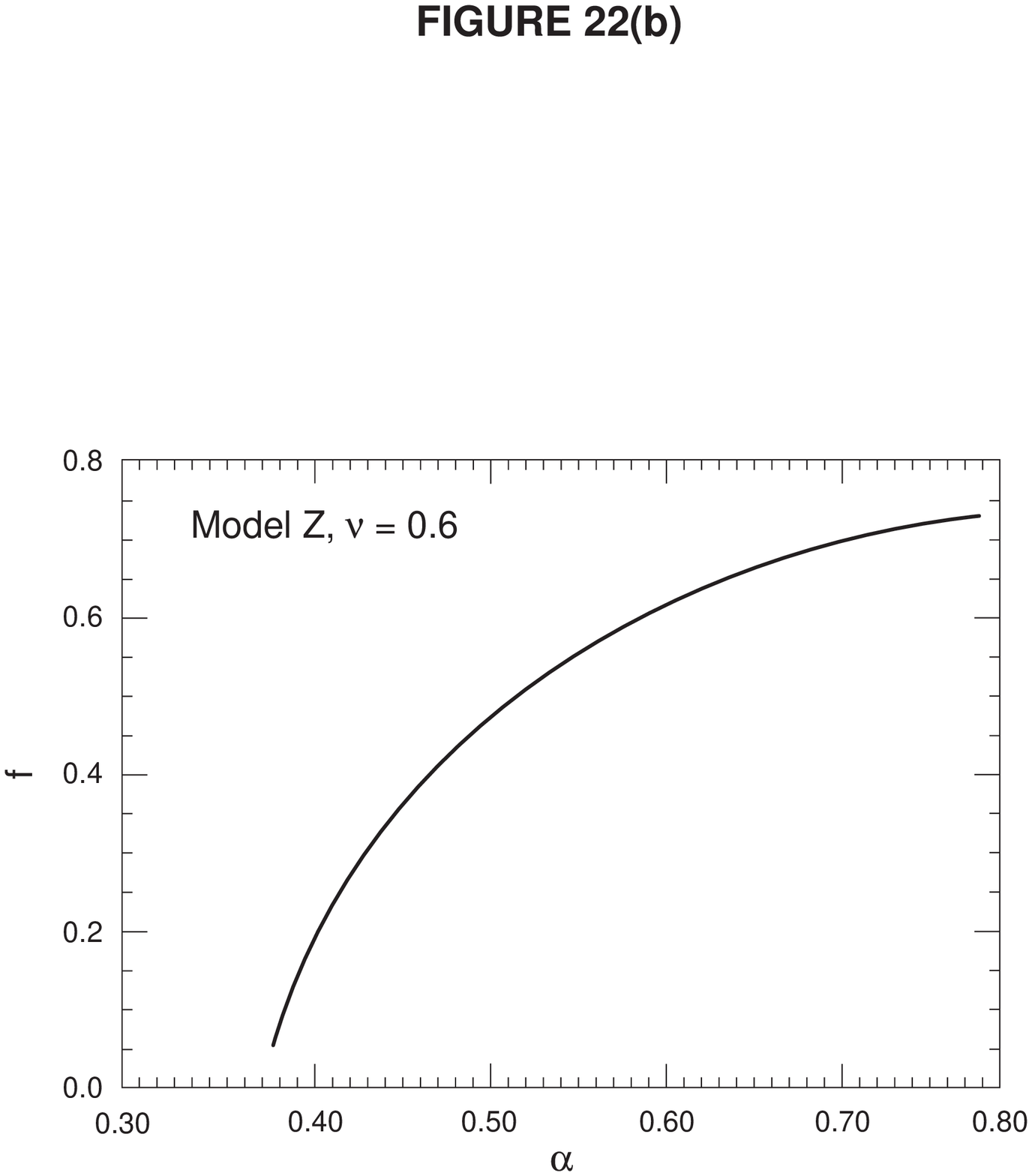}
\dofig{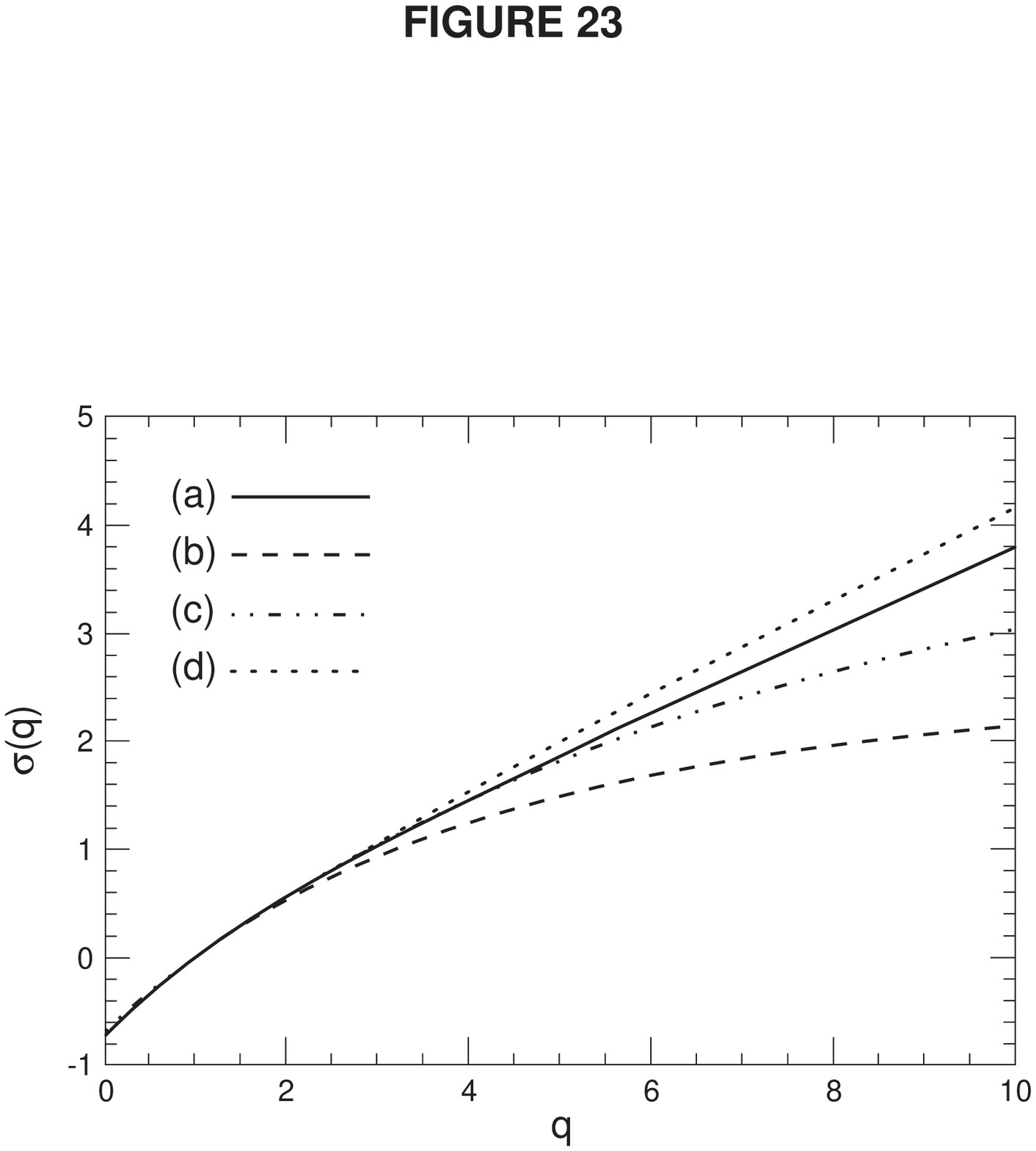}
\dofig{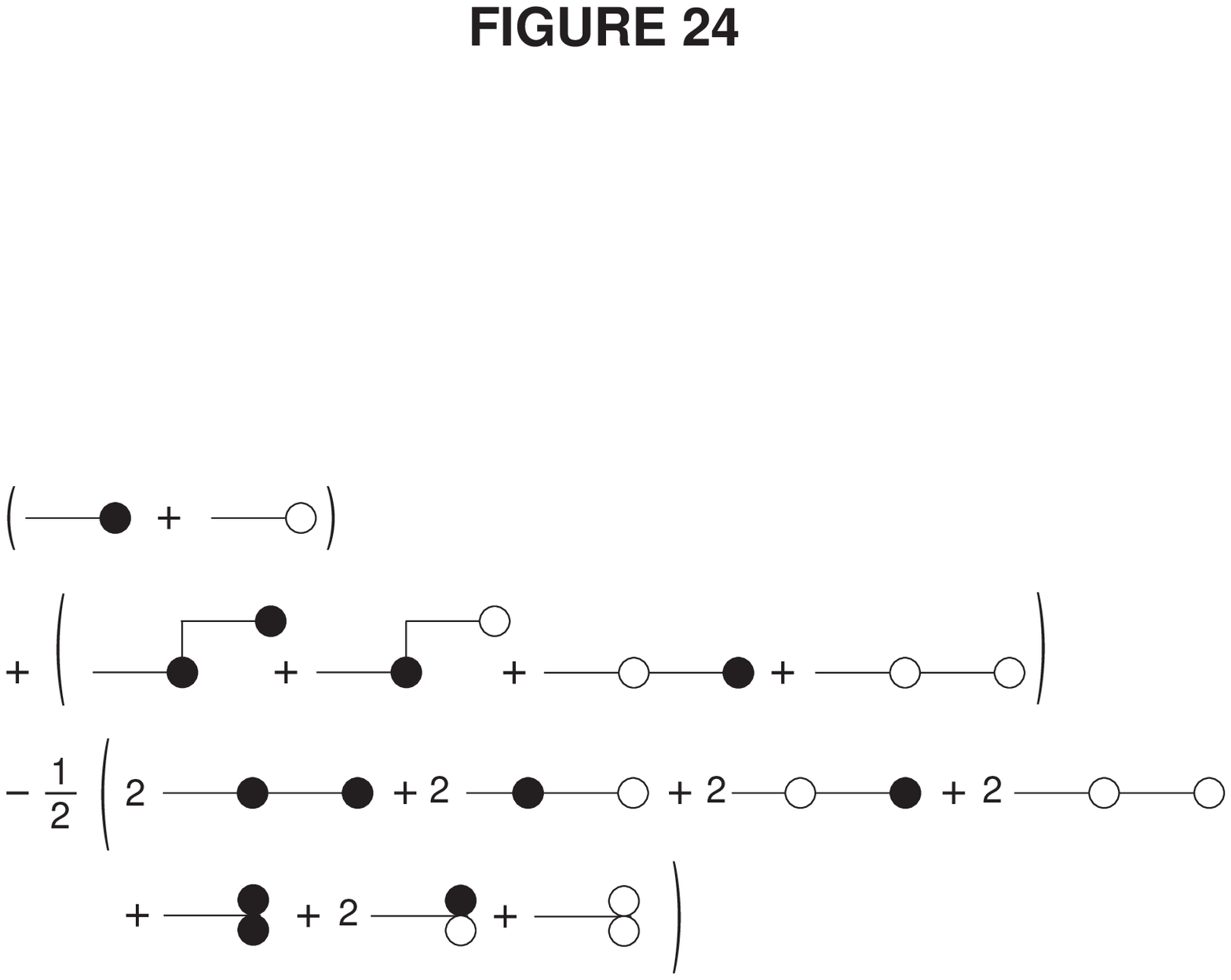}
\dofig{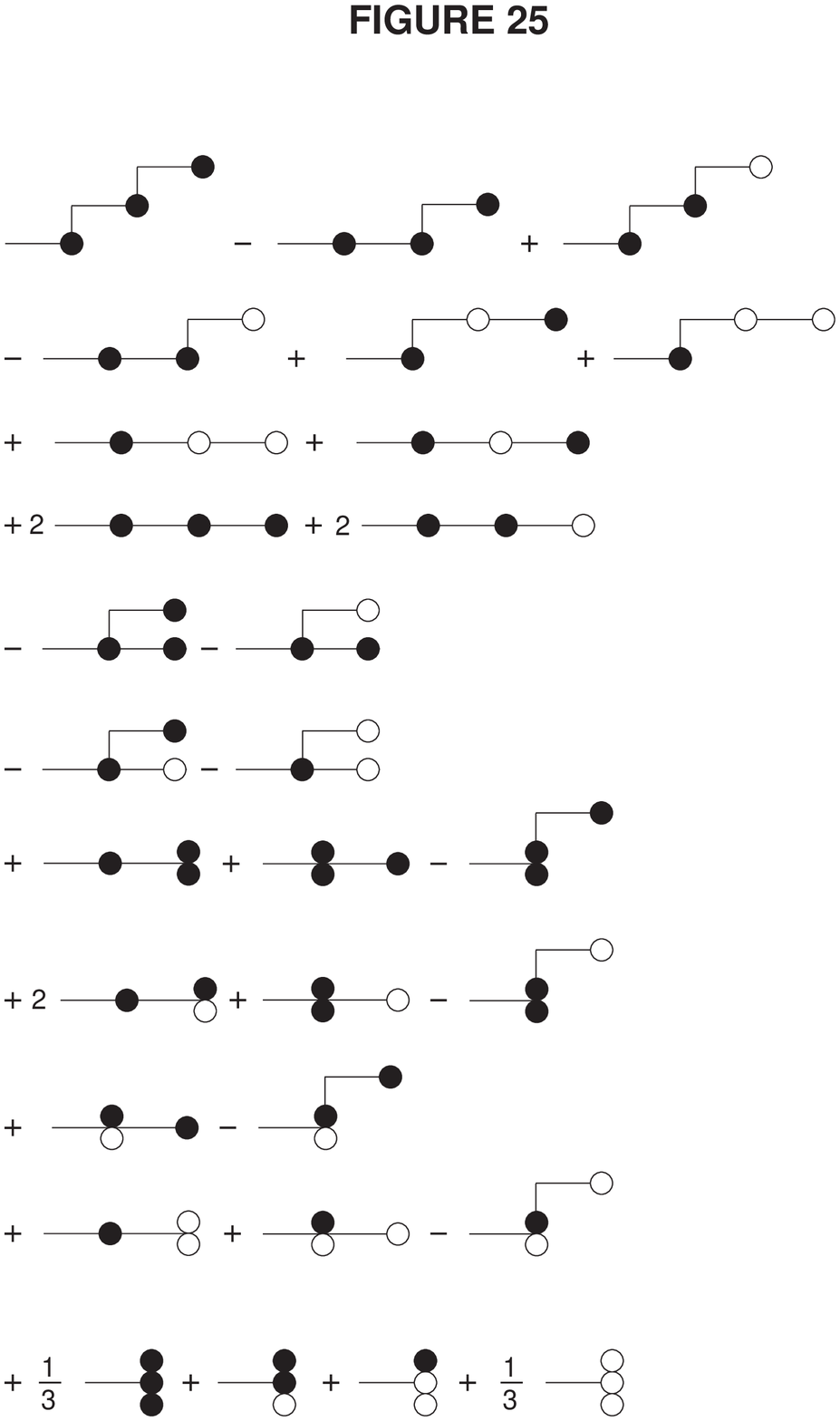}
\vfill\eject\end 

\item{8.} M. Matsushita, Y. Hayakawa, S. Sato, and K. Honda, Phys. Rev. Lett. {\bf
59}, 86 (1987).

 For a recent experimental study, see F. Mas and F.
Sagu\' es, Europhysics Lett. {\bf 17}, 541 (1992).

\item{15.} L. Turkevich and H. Scher, Phys. Rev. Lett. {\bf 55}, 1026 (1985); Phys. Rev. A {\bf 33},
786 (1986).

\item{18.} J. Lee and H.E. Stanley, Phys. Rev. Lett. {\bf 61}, 2945 (1988).

\item{20.} J. Lee and H.E. Stanley, unpublished.

\item{21.} N.G. Makarov, Proc. London Math. Soc. {\bf 51}, 369 (1985).

\item{22.} P. Jones, unpublished.

\item{23.} F. Argoul, A. Arneodo, G. Grasseau, and H.L. Swinney, Phys. Rev. Lett. {\bf
61}, 2558 (1988).

\item{24.} G. Li, L.M. Sander, and P. Meakin, Phys. Rev. Lett. {\bf
63}, 1322(C) (1989); R.C. Ball and O. Rath Spivack, J. Phys. A {\bf 23}, 5295 (1990).

\item{25.} A number of groups have recently studied the hierarchical structure of DLA
clusters. We note in particular B. Derrida, V. Hakim, and J. Vannimenus, Phys. Rev. A
{\bf 43}, 888 (1991); P. Ossadnik, Phys. Rev. A {\bf 45}, 1058 (1992).

\item{26.} There is considerable qualitative evidence that the scaling structure of
the low growth probability regions of DLA differs from that of the high growth
probability regions. See, e.g., B. Mandelbrot and C.J.G. Evertsz, Nature {\bf 348},
143 (1990).

\item{27} R.C. Ball and M. Blunt, Phys. Rev. A {\bf 39}, 3591 (1989).

\item{29.} Y. Sawada, A. Dougherty, and J.P. Gollub, Phys. Rev.
Lett. {\bf 56}, 1260 (1986); D. Grier, E. Ben-Jacob, Roy Clarke, and L.M. Sander, Phys
Rev. Lett. {\bf 56}, 1264. For a review, see  

\item{33.} R.C. Ball, Physica A {\bf 140}, 62 (1986), and references therein. This paper
also presents an argument for $D \approx 1.71$ based on the macroscopic stability of DLA
clusters.

\item{34.} J.-P. Eckmann, P. Meakin, I. Procaccia, and R. Zeitak, Phys. Rev. A {\bf
39}, 3185 (1989); and references therein.

\item{35.} R.C. Ball and T.A. Witten, Phys. Rev. A {\bf 29}, 2966 (1984).
